\documentclass[english,reprint,aps,prd,superscriptaddress,hidelinks,longbibliography]{revtex4-2}
\usepackage[utf8]{inputenc}
\usepackage[english]{babel}
\usepackage{mathrsfs}
\usepackage{bm}
\usepackage{amsmath}
\usepackage{amsthm}
\usepackage{amssymb}
\usepackage{stmaryrd}
\usepackage{graphicx}
\usepackage{multirow}

\usepackage[bookmarks=false,
 breaklinks=false,pdfborder={0 0 1},backref=false]
 {hyperref}
\hypersetup{
 colorlinks,linktocpage=true,linkcolor={red!50!black},citecolor={blue!50!black},urlcolor={blue!80!black}}

\makeatletter

\usepackage{array}
\usepackage{tabularray}
\usepackage{amsfonts}

\usepackage{newtxtext}
\usepackage[dvipsnames]{xcolor}

\definecolor{DarkGreen}{RGB}{1,140,1}
\definecolor{DarkBlue}{RGB}{1,1,185}
\definecolor{DarkRed}{RGB}{185,1,1}

\newif\ifarxiv
\arxivtrue
\ifarxiv
\global\long\def\smref#1{\ref{app:#1}~\cite{sm}}
\global\long\def\appref#1{Appendix~\ref{app:#1}}
\global\long\def\newstuff#1{#1}

\newenvironment{newstuffenv}
 {}
 {}
\else
\newcommand{\smref}[1]{\MakeUppercase{#1}~\cite{sm}}
\global\long\def\appref#1{Appendix~\ref{app:#1}}
\global\long\def\newstuff#1{{\color{DarkBlue}#1}}

\newenvironment{newstuffenv}
 {\color{DarkBlue}}
 {}
\fi

\DeclareMathOperator*{\argmin}{\arg\!\min}
\DeclareMathOperator*{\argmax}{\arg\!\max}

\global\long\def\II{\bm{I}}
\global\long\def\numstate{d}%
\global\long\def\numedge{m}%
\global\long\def\epr{\sigma}%
\global\long\def\zz{\bm{0}}%
\global\long\def\oneone{\bm{1}}%
\global\long\def\ff{\bm{f}}%
\global\long\def\rr{\rho}%
\global\long\def\jj{\bm{j}}%
\global\long\def\jjrr{j_{\rr}}%
\global\long\def\negedge{\tilde{\rr}}%
\global\long\def\jjR{\tilde{\bm{j}}}%
\global\long\def\jjRrr{\tilde{j}_{\rr}}%
\global\long\def\ffrr{f_{\rr}}%
\global\long\def\ffrrR{f_{\negedge}}%
\global\long\def\DD{\mathcal{D}}%
\global\long\def\BBbase{\nabla}%
\global\long\def\BBdiv{\BBbase^{\top}}%
\global\long\def\xx{i}%
\global\long\def\yy{j}%
\global\long\def\BBgrad{\BBbase}%
\global\long\def\negBBgrad{{-\BBbase}}%
\global\long\def\pot{\bm{\phi}}%
\global\long\def\potx{\phi_{\xx}}%

\global\long\def\pp{\bm{x}}%
\global\long\def\ppx#1{x_{#1}}%
\global\long\def\dtpp{\dot{\pp}}%
\global\long\def\ppeq{\pp^{\eq}}%
\global\long\def\ppeqx#1{x_{#1}^{\eq}}%
\global\long\def\ppss{\pp^{\sss}}%
\global\long\def\ppssx#1{x_{#1}^{\sss}}%

\global\long\def\ppp{\bm{p}}%
\global\long\def\pppx#1{p_{#1}}%
\global\long\def\dtppp{\dot{\ppp}}%
\global\long\def\dtpppx#1{\dot{p}_{#1}}
\global\long\def\pppeq{\ppp^{\eq}}%
\global\long\def\pppeqx#1{p_{#1}^{\eq}}%
\global\long\def\pppss{\ppp^{\sss}}%
\global\long\def\pppssx#1{p_{#1}^{\sss}}%

\global\long\def\ccc{\bm{c}}%
\global\long\def\cccx#1{c_{#1}}%
\global\long\def\dtccc{\dot{\ccc}}%
\global\long\def\dtcccx#1{\dot{c}_{#1}}
\global\long\def\ccceq{\ccc^{\eq}}%
\global\long\def\cccss{\ccc^{\sss}}%
\global\long\def\cccssx#1{c_{#1}^{\sss}}%

\DeclareMathOperator{\diag}{diag}
\DeclareMathOperator{\grad}{grad}
\DeclareMathOperator{\divop}{div}
\global\long\def\nnn{n}%
\global\long\def\potspace{\mathbb{R}^{\numstate}}%
\global\long\def\fluxspace{\mathbb{R}_{+}^{\numedge}}%
\global\long\def\distspace{\mathbb{R}_{+}^{\numstate}}%
\global\long\def\forcespace{\mathbb{R}^{\numedge}}%
\global\long\def\ft{T}%

\global\long\def\hk{\text{hk}}%
\global\long\def\ex{\text{ex}}%
\global\long\def\ons{\text{ons}}%

\global\long\def\eprHK{\epr_{\hk}}%
\global\long\def\eprHKons{\sigma_{\hk}^{\ons}}%
\global\long\def\eprEXons{\sigma_{\ex}^{\ons}}%
\global\long\def\ep{\Sigma}%
\global\long\def\epEX{\ep_{\ex}}%
\global\long\def\epHK{\ep_{\hk}}%
\global\long\def\eprEX{\epr_{\ex}}%
\global\long\def\eprHKhs{\epr_{\hk}^{\text{HS}}}%
\global\long\def\eprEXhs{\epr_{\ex}^{\text{HS}}}%
\global\long\def\epEXHS{\ep_{\textrm{\ensuremath{\ex}}}^{\textrm{HS}}}%
\global\long\def\ee{\bm{\theta}}%
\global\long\def\eerr{\theta_{\rr}}%
\global\long\def\eerrR{\theta_{\negedge}}%

\global\long\def\eq{\textrm{eq}}%
\global\long\def\poteq{\bm{\phi}^{\eq}}%
\global\long\def\poteqx#1{\phi_{#1}^{\eq}}%
\global\long\def\potxx#1{\phi_{#1}}%

\global\long\def\chempot{\bm{\mu}}%
\global\long\def\FreeEnergy{\mathcal{F}}%

\global\long\def\potopt{\pot^{\star}}%
\global\long\def\potoptx{\phi_{\xx}^{\star}}%
\global\long\def\potoptxx#1{\phi_{#1}^{\star}}%

\global\long\def\sss{\textrm{ss}}%
\global\long\def\potss{\bm{\phi}^{\sss}}%
\global\long\def\potssx{\phi_{\xx}^{\sss}}%

\global\long\def\cgfLeg{\mathcal{L}}%

\global\long\def\bath{\alpha}%
\global\long\def\suchthat{\;\;\text{where}\;\;}%
\global\long\def\Rija{R_{ij}^{\bath}}%
\global\long\def\Rjia{R_{ji}^{\bath}}%

\global\long\def\potlag{\pot}%

\global\long\def\nproj{P_\varnothing}%
\global\long\def\HHH{\mathcal{H}}%
\global\long\def\HH{H}%

\global\long\def\DDexpFam{\mathscr{D}}%

\global\long\def\ldpProb#1{\mathrm{Pr}\big[#1\big]}%

\makeatother

\begin{document}
\title{
\newstuff{Generalized free energy and excess/housekeeping decomposition in nonequilibrium systems:\\
from large deviations to thermodynamic speed limits}
}
\author{Artemy Kolchinsky}
\affiliation{ICREA-Complex Systems Lab, Universitat Pompeu Fabra, 08003 Barcelona, Spain}
\affiliation{Universal Biology Institute, The University of Tokyo, 7-3-1 Hongo, Bunkyo-ku, Tokyo 113-0033, Japan}
\author{Andreas Dechant}
\affiliation{Department of Physics No. 1, Graduate School of Science, Kyoto University, Kyoto 606-8502, Japan}
\author{Kohei Yoshimura}
\affiliation{Department of Physics, The University of Tokyo, 7-3-1 Hongo, Bunkyo-ku, Tokyo 113-0033, Japan}
\author{Sosuke Ito}
\affiliation{Universal Biology Institute, The University of Tokyo, 7-3-1 Hongo, Bunkyo-ku, Tokyo 113-0033, Japan}
\affiliation{Department of Physics, The University of Tokyo, 7-3-1 Hongo, Bunkyo-ku, Tokyo 113-0033, Japan}
\begin{abstract}

\newstuff{
In genuine nonequilibrium systems that undergo continuous driving, the thermodynamic forces are nonconservative, meaning they cannot be described by any free energy potential. Nonetheless, we show that the dynamics of such systems are governed by a ``generalized free energy'' that is derived from a large-deviations variational principle. This variational principle also yields a decomposition of fluxes, forces, and dissipation (entropy production) into a conservative ``excess'' part and a nonconservative ``housekeeping'' part. 
Our decomposition is universally applicable to stochastic master equations, deterministic chemical reaction networks, and open systems. We also show that the excess entropy production obeys a thermodynamic speed limit (TSL), a fundamental thermodynamic constraint on the rate of state evolution and/or external fluxes. We demonstrate our approach on several examples, including real-world metabolic networks, where we derive fundamental dissipation bounds and uncover ``futile'' metabolic cycles. Our generalized free energy and decomposition are empirically accessible to thermodynamic inference in both stochastic and deterministic systems. We discuss important connections to several theoretical frameworks, including information geometry and Onsager theory, as well as previous excess/housekeeping decompositions.
}
\end{abstract}
\maketitle

\section{Introduction}

Thermodynamics is one of the most far-reaching and useful theoretical
frameworks in the physical sciences. Traditionally, this framework
was mainly used to study ``passive'' systems, which tend toward equilibrium 
when allowed to relax freely. \newstuff{When a passive system is brought out of equilibrium,  
many of its thermodynamic, dynamical,
and statistical properties are governed by the \emph{\newstuff{nonequilibrium} free energy} $\mathcal{F}$. As we discuss below, the nonequilibrium free energy controls relaxation dynamics~\citep{maas_gradient_2011,mielke2011gradient}, equilibrium fluctuations~\citep{schloglFluctuationsThermodynamicNon1971,qian2001relative}, and extractable work~\citep{procaccia1976potential,esposito2011second}. %
In addition, the thermodynamic forces that describe a passive system are ``conservative'', being determined by the gradient of the nonequilibrium free energy. For this reason, we refer to passive systems as \emph{conservative systems}.} %

In nonequilibrium thermodynamics, \emph{entropy production rate} (EPR) refers to the increase of the entropy of a system and its environment, and it serves as the fundamental measure of thermodynamic dissipation. In conservative systems, EPR is proportional to the loss of the nonequilibrium free energy over time, vanishing 
at equilibrium, where the state remains constant. 
Recently, the connection between entropy production and nonstationarity has been used to derive \emph{thermodynamic
speed limits} (TSL)~\citep{shiraishi_speed_2018,ito2018stochastic,van2020unifiedCSLTUR,yoshimura2021thermodynamic,shiraishi2024wasserstein}. TSLs bound the time and dissipation needed to transform
a system from one state to another, and they quantify the thermodynamic efficiency of finite-time transformations~\cite{lee2022speed}. 
Moreover, the minimal dissipation required to go between two states
in a finite time defines a thermodynamic distance between states. 
In the
near-equilibrium regime, this distance is called ``{thermodynamic
length}''~\citep{salamon1985length,salamon_thermodynamic_1983,crooks_measuring_2007,sivak_thermodynamic_2012}.
Far from equilibrium, it is the distance based on thermodynamic 
optimal transport~\citep{aurell2011optimal,aurell2012refined,dechant2019thermodynamic,nakazato2021geometrical,van2022thermodynamic,kohei2022,nagayama2023geometric,ito2024geometric}.

Nowadays, there is growing interest in ``genuine nonequilibrium'' systems that undergo continuous driving. Such driving may arise due to nonequilibrium boundary conditions, applied mechanical
forces, or internal fuel sources (as in active matter
\citep{fodor2022irreversibility}). When allowed to relax freely, such systems tend toward nonequilibrium steady states, or possibly limit cycles and chaos in the case of nonlinear dynamics. 
Many examples are found in molecular biology, where driving is provided by nonequilibrium concentrations of ATP.

We refer to continuously driven systems as \newstuff{\emph{nonconservative}}, because they exhibit nonconservative thermodynamic forces that cannot be expressed as the gradient of any free energy potential. 
\newstuff{Nonconservative systems exhibit cyclic fluxes that contribute to dissipation but do not influence state evolution. 
For example, a nonconservative system may remain near a steady state while parameters slowly change, incurring arbitrarily large entropy production despite negligible state change. 
In nonconservative systems, vanishing speed of evolution does not imply small EPR. Therefore, TSLs that relate EPR and speed may become arbitrarily loose in nonconservative systems, and they typically do not provide useful measures of efficiency for finite-time transformations. }

\begin{table*}

\begin{tblr}{
width=\linewidth,
 colspec = {l c c c},
rows={m},
 stretch = 0,
 rowsep = 6pt,
 hlines = {gray, 0.5pt},
 vlines = {gray, 0.5pt},
}
& {\textbf{Our generalized free energy $\pot^{*}$} }
& {\textbf{Section}} 
& {\textbf{Steady-state free energy $\potss$ }} 
\\

\textbf{Definition} 
& {${\displaystyle \BBdiv \jj =-\BBdiv(\jj\circ e^{\BBgrad\pot^{*}})}$ }
& §\ref{sec:variationaldb}-\ref{sec:variational-ctive}
& {${\displaystyle \potss=\ln\frac{\pp}{\ppss}}$ } 
\\
{\textbf{Excess/housekeeping}\\\textbf{decomposition}} 
& {Information-geometric\\decomposition}
& \SetCell[r=3]{c} §\ref{sec:ex-vs-hk} 
& {Hatano-Sasa, also called\\adiabatic/nonadiabatic}
\\
{\qquad\newstuff{\emph{Excess part}}}
& {\newstuff{Conservative}}
& %
& {\newstuff{Nonstationary}}
\\
{\qquad\newstuff{\emph{Housekeeping part}}} 
& {\newstuff{Nonconservative}}
& %
& {\newstuff{Stationary}}
\\
{\textbf{Linear response}\\\textbf{\newstuff{coefficients}}}
& {\newstuff{Short-time}\\\newstuff{diffusion coefficients}}
& {§\ref{subsec:lr}}
& {\newstuff{Steady-state}\\\newstuff{diffusion coefficients}}
\\
{\textbf{Large deviations}} 
& {Dynamical fluctuations} 
& §\ref{sec:Physical-interpretations}
& {Steady-state fluctuations} 
\\
{\textbf{Thermodynamic speed limit}} 
& {Wasserstein speed}
& §\ref{sec:tsl}
& {Total variation speed (MJPs only)} 
\end{tblr} %

\caption{Summary of our results and comparison with
the steady-state/Hatano-Sasa (HS) approach (discussed in more detail in Sections~\ref{subsec:ss} and \ref{subsec:hs-comparison}). Top row: $\jj$ indicates the vector of one-way fluxes, $\BBdiv$ the stoichiometric matrix, and $\pp$ the concentration vector or probability distribution. 
}\label{tab:1}
\end{table*}

\begin{newstuffenv}
In this paper, we consider the problem of defining a ``generalized'' free energy for nonconservative systems. We also consider the related problem of decomposing nonconservative systems into \emph{excess} and \emph{housekeeping} parts. 
The excess part is the conservative contribution associated with the generalized free energy, and it vanishes once the potential becomes ``equilibrated'' and the internal dynamics become stationary. The excess part should discount the contribution of cyclic fluxes to dissipation, thus giving rise to meaningful TSLs in nonconservative systems. The housekeeping part is the nonconservative contribution that cannot be associated with the generalized free energy, and it does not vanish even when the internal dynamics are stationary. As we show, this decomposition may be considered at the level of thermodynamic forces (conservative/nonconservative forces), %
fluxes (gradient/cyclic fluxes), %
or dissipation (excess/housekeeping EPR). %

There has been extensive work on generalizations of free energy~\citep{graham_nonequilibrium_1986,freidlin_random_1998,derridaFreeEnergyFunctional2001,ao_potential_2004,wangPotentialLandscapeFlux2008,geDissipationGeneralizedFree2013,liFreeActionNonequilibrium2015,fang_nonequilibrium_2019} and  excess/housekeeping decompositions~\cite{glansdorffNonequilibriumStabilityTheory1970,oono1998steady,hatano2001steady,speck2005integral,esposito2007entropy,komatsu2008steady,sagawa2011geometrical,maes2014nonequilibrium,smith2020intrinsic,kohei2022} in nonconservative systems. 
The best known existing approach is based on nonequilibrium steady states~\cite{keizer1985heat,oono1998steady,fang_nonequilibrium_2019,falascoMacroscopicStochasticThermodynamics2023a}. Here, the generalized free energy is defined as the large-deviation rate function of steady-state fluctuations, which in many cases is the relative entropy to the steady state distribution, sometimes called the  ``quasipotential''~\cite{falascoMacroscopicStochasticThermodynamics2023a}. 
In this approach, the excess EPR is defined as the decrease of the quasipotential over time, and it captures the nonstationary component of dissipation~\citep{hatano2001steady,esposito2007entropy,esposito2010three,rao2016nonequilibrium,ge2016nonequilibrium}. The housekeeping EPR is defined as the remainder, and it captures the stationary component of dissipation. In the literature, this excess/housekeeping decomposition is  termed the ``Hatano-Sasa'' (HS)~\cite{hatano2001steady} or the ``nonadiabatic/adiabatic'' decomposition~\citep{esposito2007entropy,esposito2010three,rao2016nonequilibrium,ge2016nonequilibrium}.

The quasipotential is very useful for studying stability and fluctuations of stationary systems. However, as we discuss below, it is not as useful for studying transient systems away from steady state, nor for deriving transient relations such as TSLs. Fundamentally, this is because the quasipotential is defined in terms of steady-state statistics, rather than  ``local in time'' statistics of a transient system.

Our definition of the generalized free
energy and the excess/housekeeping decomposition is based on a variational principle, and it makes no explicit reference to steady states. This principle is universally applicable to stochastic and deterministic discrete systems, including 
stochastic master equations as well as open and closed deterministic CRNs with arbitrary kinetics. As we show, this variational principle can be understood in terms of dynamical large deviations of local-in-time  flux fluctuations. From this perspective, the excess EPR  quantifies the statistical irreversibility of state dynamics, while the generalized free energy is the ``most irreversible'' observable --- the least likely to evolve backward in time due to a stochastic fluctuation.  The connection to large deviations means that our generalized free energy and excess EPR are directly related to fluctuation statistics, allowing for practical thermodynamic inference~\cite{seifert2019stochastic}. 
A brief summary of our results and comparison with the steady-state approach is provided in Table~\ref{tab:1}.

In addition, we derive a tight TSL that relates our excess EPR to dynamical activity and speed of evolution. Importantly, speed is measured in terms of  optimal-transport distance (1-Wasserstein)~\citep{dechant2019thermodynamic,van2023topological,shiraishi2024wasserstein}. This distance is sensitive to the system's internal topology, thus capturing important dynamical constraints. 
In open systems, including stationary ones, our TSL bounds internal dissipation in terms of activity and the velocity of the external fluxes.

Along the way, we highlight fundamental connections to several  theoretical frameworks. In particular, we relate our excess/housekeeping decomposition to the celebrated Pythagorean theorem from information geometry~\cite{amari2016information}, and we relate the linear-response regime of our approach to Onsager theory and thermodynamic length~\cite{crooks_measuring_2007}.

Our results are illustrated on three examples. The first is a simple unicyclic master equation, where we demonstrate important differences between our approach and the steady-state HS approach. The second example is the Brusselator, a nonlinear chemical oscillator, where we derive our decomposition in closed form and demonstrate a general TSL for limit cycle oscillators. The third example is a set of real-world metabolic networks, modeled as open CRNs in steady state. Using our TSL, we quantify the efficiency of several metabolic pathways, and we use our decomposition to identify  futile metabolic cycles.
\end{newstuffenv}

The paper is laid out as follows. 

In the next section, we describe our physical setup and formalism. Section~\ref{sec:db} provides background on nonequilibrium free energy in conservative systems and its generalization based on nonequilibrium steady states. 
We introduce our variational principle for conservative systems in Section~\ref{sec:variationaldb}. In Section~\ref{sec:variational-ctive}, we extend this variational principle to nonconservative systems, introduce the excess/housekeeping decomposition, and discuss its behavior under coarse-graining. We consider the linear-response regime and make connections to information geometry, Onsager theory, and thermodynamic length.
In Section~\ref{sec:Physical-interpretations}, we relate our approach to  dynamical large deviations and derive a thermodynamic uncertainty relation. 
In Section~\ref{sec:tsl}, we derive a TSL for excess entropy production based on Wasserstein distance. 
Section~\ref{sec:Examples} illustrates our approach on several examples, including real-world metabolic networks. 
\newstuff{In Section~\ref{sec:comparison2}, we compare our proposal to several previous approaches, including ones based on Euclidean-Onsager geometry~\citep{kohei2022}, Hessian geometry~\citep{Kobayashi2022,kobayashi2024information}, and the steady-state approach.} We finish with the Discussion and suggestions for future work in Section~\ref{sec:Discussion}. 

Appendices~\ref{sec:nonlinear_mjps} and \ref{app:odd} discuss nonlinear master equations and systems with odd variables. Supporting derivations and additional numerical comparisons are found in the Supplemental Material (SM)~\cite{sm}.

\section{Setup and preliminaries}
\label{sec:setup}

Before proceeding, we fix notation. Vectors are written
in bold, $\bm{a}=(a_{1},a_{2},\dots)\in\potspace$, including
the special vectors $\zz=(0,0,0,\dots)$ and $\bm{1}=(1,1,1,\dots)$.
We use $e^{\bm{a}},\bm{a}^{2},\vert\bm{a}\vert\dots$ to indicate
elementwise operations, for example $e^{\bm{a}}:=(e^{a_{1}},e^{a_{2}},\dots)$.
The notation $\bm{a}\circ\bm{b}:=(a_{1}b_{1},a_{2}b_{2},\dots)$
indicates elementwise multiplication and 
$\bm{a}/\bm{b}:=(a_{1}/b_{1},a_{2}/b_{2},\dots)$
indicates elementwise division. As discussed below, $\BBbase$
refers to the stoichiometric matrix (which acts like the discrete
gradient operator). To avoid confusion, we write the usual gradient of
a function $f:\potspace\to\mathbb{R}$ as $\grad_{\pp}\,f=(\partial_{\ppx 1}f,\dots,\partial_{\ppx {\numstate}}f)$.

\subsection{States, reactions, and dynamics}
\label{subsec:setup-1}
We focus on discrete Markovian systems with general (linear or nonlinear) dynamics.
As special cases, this includes Markov jump processes (MJP) that
represent the transport of probability between microstates in stochastic systems. 
It also
includes deterministic chemical reaction networks (CRNs) used
to model chemical systems in the large-volume limit
~\citep{feinbergFoundationsChemicalReaction2019,kondepudi2014modern,rao2016nonequilibrium}. 
Continuous-state
systems are briefly mentioned in Section~\ref{sec:comparison-variational} and non-Markovian
systems in Section~\ref{sec:Discussion}.

The system's thermodynamic \emph{state} at time $t$ is specified
by a nonnegative vector $\pp(t)=(\ppx 1(t),\dots,\ppx{\numstate}(t))\in\distspace$. We always use the term \emph{state} to refer to the system's
thermodynamic state $\pp(t)$. When referring to the microstates, we do so explicitly. The term \emph{state function} 
refers to functions $f:\distspace\to\mathbb{R}$.
The term \emph{state observable} refers to functions over microstates/species,
represented by vectors $\pot\in\potspace$.

In an MJP, the state is a probability distribution over
$\numstate$ microstates (or coarse-grained ``mesostates''). \newstuff{For notational clarity, when referring to probability distributions, we sometimes write the state $\pp$ as $\ppp$.} 
In
a deterministic chemical reaction network (CRN), the state is an 
unnormalized vector of concentrations of $\numstate$ chemical species. \newstuff{For notational clarity, when referring to a deterministic concentration vector, we sometimes write the state $\pp$ as $\ccc$.}

The system is associated with $\numedge$ one-way \emph{transitions} or \emph{reactions}
indexed by $\rr\in\{1,\dots,\numedge\}$, corresponding
to jumps between microstates in an MJP or one-way chemical
reactions in a CRN. We use the term \emph{reaction observable} to
refer to functions of individual reactions, represented by a vector 
$\ee\in\forcespace$. \newstuff{As an example, the reaction observable $\ee$ may indicate heat exchanges; then, $\eerr$ indicates the number of joules released to (or absorbed from) a heat bath during reaction $\rr$. Except where otherwise noted, we make no assumptions about the nature of reaction observables $\ee$ (like antisymmetry).}

The one-way \emph{flux} of reaction $\rr$ is
the expected number of reactions per unit time and
volume, written $\jjrr(\pp(t),t)$. The flux depends on
state $\pp(t)$ through system-specific kinetics, and on time $t$ due to external driving (i.e., changing control parameters). Except where otherwise noted, we make
no assumptions about the kinetics or external driving. The entire
set of one-way fluxes is represented by the vector $\jj=(j_{1},\dots,j_{\numedge})\in\fluxspace$.

\newstuff{The system may also be open to exchange matter with the environment via {inflows and outflows}, as in a flow reactor. The exchanges are encoded in the vector $\II(\pp(t),t)\in\mathbb{R}^\numstate$, where $I_i(\pp(t),t)$ is the net outflow (outflow minus inflow) of species $i$ for state $\pp(t)$ and time $t$. We will refer to exchanges encoded by $\II$ as the \emph{external fluxes}.
As we discuss below, chemical systems may also be open to exchanges with external ``chemostated'' species that are kept at constant concentrations.} 

The system's state evolves according to
\begin{align}
\dtpp=\BBdiv\jj -\newstuff{\II}\,.\label{eq:cont-equation}
\end{align}
where $\BBgrad\in\mathbb{Z}^{\numedge\times\numstate}$ is a matrix
whose entries $\BBgrad_{\rr\xx}$ indicate the amount of $\xx$ created
or destroyed by reaction $\rr$. This matrix is called the \emph{incidence
matrix} in MJPs and the \emph{stoichiometric matrix} in CRNs. 
Observe that we often leave dependence on time $t$ and/or state $\pp$ implicit, as in Eq.~\eqref{eq:cont-equation}. 
The quantity $\BBdiv \jj$, which we sometimes call the \emph{net production}, is the change of species due to reaction fluxes. \newstuff{Eq.~\eqref{eq:cont-equation} expresses the state evolution as a combination of net production and external fluxes. In closed systems, the state evolution is equal to net production.}

Eq.~\eqref{eq:cont-equation} can be interpreted as a discrete
continuity equation, where $\BBdiv$ %
acts as the negative discrete divergence operator. In turn, the stoichiometric matrix $\BBgrad$ acts like the discrete
gradient operator: the net increase of any state observable $\pot$
due to reaction $\rr$ is $[\BBgrad\pot]_{\rr}=\sum_{\xx}\BBgrad_{\rr\xx}\potx$.

\subsection{Thermodynamic forces and entropy production}

Each one-way reaction
$\rr$ is associated with a \emph{reverse flux} $\jjRrr$,
where $\jjR=(\tilde{j}_{1},\dots,\tilde{j}_{m})$ indicates the vector
of all reverse fluxes. We sometimes refer to $\jj$ as the \emph{forward
fluxes} to distinguish them from the reverse fluxes $\jjR$. Each reaction $\rr$ is associated with a \emph{(thermodynamic) force},
sometimes called ``reaction affinity'' or ``affinity'' in the literature, defined as the log-ratio
of the forward and reverse fluxes,
\begin{align}
f_{\rr}:=\ln\frac{\jjrr}{\jjRrr}.\label{eq:forcedef}
\end{align}
We represent the forces across all reactions using the vector 
\[
\ff=\ln\frac{\jj}{\jjR} = (f_{1},\dots,f_{m})\in\forcespace \,.
\]
Like the fluxes, the
forces generally depend on the state $\pp$ and time $t$, although
we leave this implicit in our notation.

For simplicity, in the main text, we focus on systems without odd variables (such
as velocities or momenta). 
However, many of our results generalize to systems with odd variables; see \appref{odd}
for more details. 
In systems without odd variables, each reaction $\rr\in\{1,\dots,\numedge\}$
is associated one-to-one with a reverse reaction $\negedge\in\{1,\dots,\numedge\}$ such that
\begin{align}
\BBgrad_{\rr\xx}=-\BBgrad_{\negedge\xx}\qquad\qquad \jjRrr=j_{\negedge}.\label{eq:antisymmetry}
\end{align}
Observe that $\sum_{\rr}\jjrr=\sum_{\rr}j_{\negedge}$ since reversal is one-to-one.

Our main thermodynamic quantity of interest is the entropy production
rate (EPR). For systems without odd variables, the EPR is written
as
\begin{align}
\epr=\sum_{\rr}\jjrr\ln\frac{\jjrr}{\jjRrr}=\jj^{\top}\ff.\label{eq:eprn2}
\end{align}
Using Eq.~\eqref{eq:antisymmetry}, we may also write it as
\begin{align}
\epr=\frac{1}{2}\sum_{\rr}(\jjrr-j_{\negedge})\ln\frac{\jjrr}{j_{\negedge}}\ge0.\label{eq:eprn3}
\end{align}
The total entropy production (EP) over time $0\le t\le T$
is 
\begin{align}
\ep=\int_{0}^{\ft}\epr(t)\,dt\,,\label{eq:integratedEP}
\end{align}
where $\epr(t)$ is the EPR incurred by the fluxes $\jj(t)$ at time $t$. 
We always write thermodynamic quantities (free energy, forces, chemical potentials, etc.) in dimensionless units. 

EPR vanishes only
when the reverse and forward fluxes are equal ($\jj=\jjR$), equivalently
when the forces vanish ($\ff=\zz$), thus it quantifies the dynamical
irreversibility of the system's fluxes.
EPR acquires additional thermodynamic meaning when
the condition of 
\emph{local detailed balance} (LDB) holds, which says that the force
$\ffrr$ associated with each reaction $\rr$ is equal to the increase
of thermodynamic entropy of the system and its environment due to
that reaction~\citep{kondepudi2014modern,maes2021local,rao2016nonequilibrium}.
Assuming LDB, EPR is the expected rate of increase of the
thermodynamic entropy of the system and its environment.

\begin{figure}
\begin{centering}
\includegraphics[totalheight=8cm]{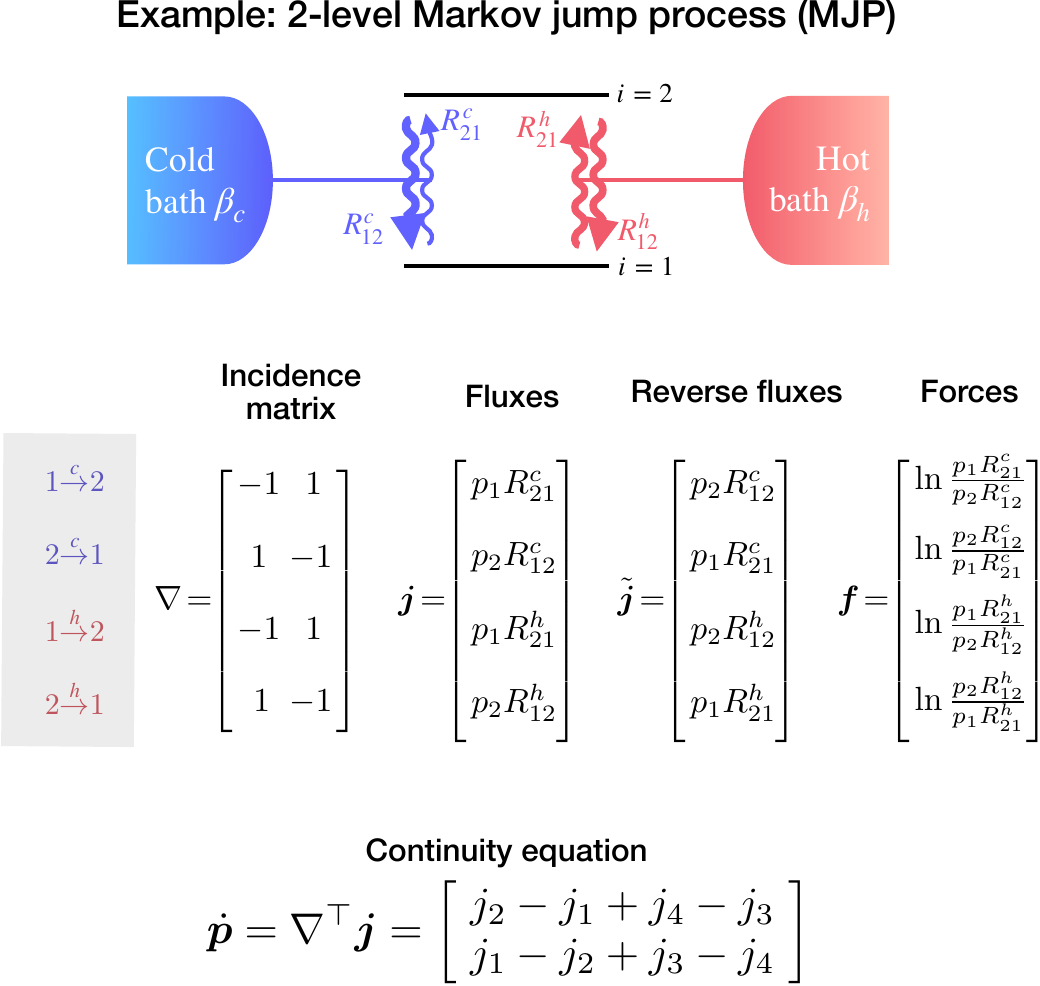} 
\par\end{centering}
\caption{\textbf{Formalism illustrated on a two-level MJP 
coupled to a pair of heat baths, with distribution $\ppp=(\pppx{1},\pppx{2})$.} Transitions 
occur with rates $R_{21}^{c}$ and $R_{12}^{c}$ when exchanging energy
with the cold bath at inverse temperature $\beta_{c}$, and with
rates $R_{21}^{h}$ and $R_{12}^{h}$ when exchanging energy with
the hot bath at inverse temperature $\beta_{h}$. The four
one-way transitions are characterized by the incidence matrix, forward
and reverse flux vectors, and force vector.}\label{fig:twolevel}
\end{figure}

\subsection{Example: Markov jump process (MJP)}

To illustrate our formalism, we consider an MJP that represents a stochastic
system with $\numstate$ microstates, coupled to one or more thermodynamic
reservoirs indexed by $\bath$. The probability distribution
$\ppp$ evolves according to a master equation,
\begin{align}
\dtpppx{\xx}=\sum_{\bath}\sum_{\yy=1}^{\numstate}(\pppx{\yy}\Rija-\pppx{\xx}\Rjia),\label{eq:g2}
\end{align}
where $\Rjia$ is the transition rate from microstate $\xx$ to microstate
$\yy$ mediated by reservoir $\bath$~\cite{esposito2010threefaces}.

In our formalism, there is a one-way reaction $\rr$ for each transition $\xx\shortrightarrow\yy$ mediated
by reservoir $\bath$. It has incidence matrix entries
\begin{align}
\BBgrad_{\rr k}=\delta_{k\yy}-\delta_{k\xx},
\label{eq:MJPincidence}
\end{align}
and flux $\jjrr=\pppx{\xx}\Rjia$. The reverse flux $\jjRrr=\pppx{\yy}\Rija$
is the forward flux across the reverse reaction $\negedge$ (transition
$\yy\to\xx$ mediated by $\bath$). \newstuff{In a standard MJP, there are no inflows or outflows, so $\II=\zz$.} Given these definitions, the continuity equation~\eqref{eq:cont-equation}
is equivalent to the master equation~\eqref{eq:g2}. The force across each reaction is 
\begin{align}
\ffrr=\ln\frac{\jjrr}{\jjRrr}=\ln\frac{\Rjia\pppx{\xx}}{\Rija\pppx{\yy}},\label{eq:mjpforce}
\end{align}

In Figure~\ref{fig:twolevel}, we illustrate our
formalism on a simple MJP that represents a two-level system coupled
to a pair of heat baths. We consider this example again in Section~\ref{subsec:2level}.

\newstuff{As discussed in Appendix~\ref{sec:nonlinear_mjps}, our formalism can also be applied to ``nonlinear MJPs'', sometimes used to model many-body systems with mean-field interactions. %
}

\subsection{Example: chemical reaction network (CRNs)}

\label{subsec:crn}

\global\long\def\numrevreactions{k}%
\global\long\def\revrr{r}%
\global\long\def\jjrrCRN{j_{\revrr}^{+}}%
\global\long\def\jjRevrrCRN{j_{\revrr}^{-}}%
\global\long\def\kkrrCRN{k_{\revrr}^{+}}%
We illustrate our formalism using a CRN with $\numstate$ chemical species and $\numrevreactions$ reversible reactions, 
\begin{align}
\sum_{\xx=1}^{\numstate}\nu_{\revrr\xx}X_{\xx}\rightleftharpoons\sum_{\xx=1}^{\numstate}\kappa_{\revrr\xx}X_{\xx}\qquad\forall\revrr\in\{1,\dots,\numrevreactions\},\label{eq:crndef}
\end{align}
where $\nu_{\revrr i}$ and $\kappa_{\revrr i}$ specify the stoichiometry
of species $i$ as reactant and product in reaction $\revrr$. The
forward and reverse fluxes across each reversible reaction $\revrr$
are indicated as $\jjrrCRN$ and $\jjRevrrCRN$. These fluxes generally 
depend on the concentration vector $\ccc$, possibly in a nonlinear
manner. For instance, for mass-action kinetics, the forward flux for
reaction $r$ is 
\begin{align}
\jjrrCRN=\kkrrCRN\prod_{\xx=1}^{\numstate}\cccx{\xx}^{\nu_{\revrr\xx}},\label{eq:massaction}
\end{align}
where $\kkrrCRN$ is the forward rate constant of reaction $\revrr$.

In our formalism, each reversible reaction $\revrr\in\{1,\dots\numrevreactions\}$
is treated as two one-way reactions $(\rr,\negedge)$ with
fluxes $\jjrr=\jjrrCRN$, $\jjRrr=\jjRevrrCRN$ and stoichiometry
$\BBgrad_{\rr\xx}=\kappa_{\revrr\xx}-\nu_{\revrr\xx}=-\BBgrad_{\negedge\xx}$. 
\newstuff{As mentioned above, a CRN can be open %
due to inflows and outflows, as specified by the flow vector $\II$.} 
The concentrations $\ccc$ evolve according to Eq.~\eqref{eq:cont-equation}, sometimes called the ``reaction rate equation'' in the CRN literature. 

\begin{figure}
\begin{centering}
\includegraphics[totalheight=8cm]{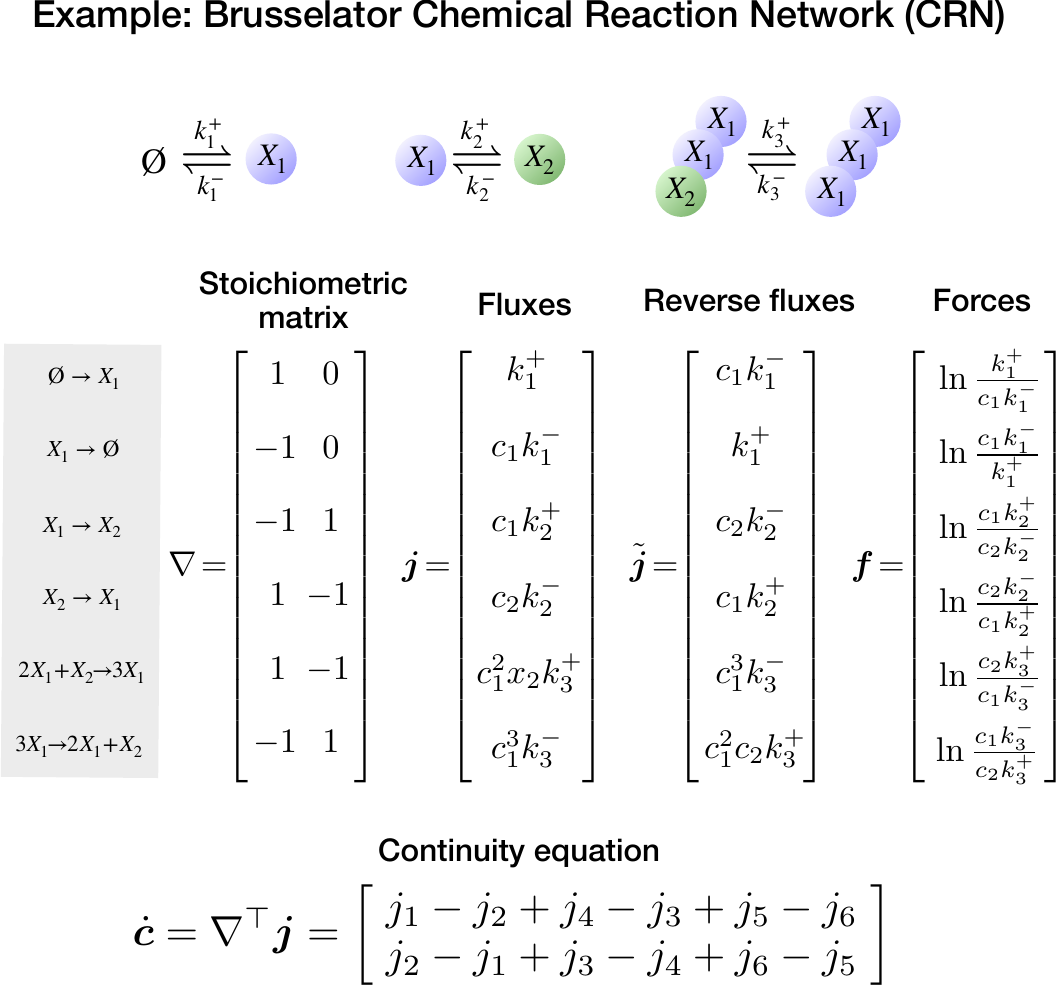}
\par\end{centering}
\caption{\label{fig:crn}
\textbf{Formalism illustrated on the Brusselator CRN, a nonlinear chemical oscillator, with concentrations $\ccc=(\cccx{1},\cccx{2})$.} There are two species $X_1,X_2$ and three reversible reactions:
$\varnothing\rightleftharpoons X_{1}$ (inflow), $X_{1}\rightleftharpoons X_{2}$
(conversion), and $2X_{1}+X_{2}\rightleftharpoons3X_{1}$ (second-order
autocatalysis). The six one-way reactions are characterized
by the stoichiometric matrix, forward and reverse flux vectors, and
thermodynamic forces.}
\end{figure}

In Figure~\ref{fig:crn}, we illustrate our formalism
on the Brusselator~\citep{prigogine1968symmetry}, a simple chemical oscillator. We also consider this example
in Section~\ref{sec:brusselator}.

\newstuff{In addition to inflows and outflows $\II$, a CRN may be open due to presence of ``external'' species. Such species 
are not included in the concentration vector $\ccc$ or the stoichiometric matrix $\BBgrad$, although their concentrations may still affect kinetics of reactions (via pseudo-rate constants). 
This is often used to represent ``chemostatting'', where the concentrations of some species are externally controlled~\cite{rao2016nonequilibrium}. For example, Prigogine's original model of the Brusselator~\cite{prigogine1968symmetry} has four additional species coupled to synthesis/degradation of $X_1$ and conversion $X_{1}\rightleftharpoons X_{2}$. The simplified Brusselator model in Figure~\ref{fig:crn} is derived by assuming these four species have constant concentrations and thus can be eliminated from the model.}

\newstuff{Our general formalism applies to arbitrary CRNs, but the condition of LDB requires further physical  assumptions. LDB is valid for reversible elementary reactions, including some non-ideal systems with non-mass-action kinetics~\citep{avanzini2021nonequilibrium}. LDB is also valid for some types of non-elementary reactions, including reversible Michael-Menten kinetics~\cite{beard2007relationship}, under appropriate definitions of forward and reverse fluxes~\cite{wachtel2018thermodynamically,peng2020universal,avanzini2020thermodynamics}.}

\subsection{Relative entropy}

\emph{Relative entropy} is an information-theoretic measure of divergence that plays a central role in our work. 
The relative entropy between a pair of states $\pp,\bm{y}\in\distspace$
is defined as
\begin{align}
D(\pp\Vert\bm{y}):=\sum_{\xx=1}^{\numstate}\big(\ppx{\xx}\ln\frac{\ppx{\xx}}{y_{\xx}}-\ppx{\xx}+y_{\xx}\big).\label{eq:klstate}
\end{align}
It is always nonnegative and vanishes only when $\pp=\bm{y}$.
We use a generalized version of relative entropy, appropriate
for unnormalized states that may not sum to unity (such as concentration
vectors in CRNs). For normalized probability distributions, it reduces
to the well-known Kullback-Leibler divergence, $D(\pp\Vert\bm{q})=\sum_{\xx}\pppx{\xx}\ln({\pppx{\xx}}/{q_{\xx}})$.

We also consider the relative entropy between pairs of flux vectors $\jj,\jj^{\prime}\in\fluxspace$, written as
\begin{align}
\DD(\jj\Vert\jj^{\prime}):=\sum_{\rr=1}^{\numedge}\big(\jjrr\ln\frac{\jjrr}{j_{\rr}^{\prime}}-\jjrr+j_{\rr}^{\prime}\big).\label{eq:klflux}
\end{align}
We use the calligraphic $\DD$ (rather than $D$) to distinguish the relative entropy
between fluxes.

Importantly, the EPR can be written as the relative
entropy between the forward and reverse fluxes,
\begin{align}
\epr=\DD(\jj\Vert\jjR)=\sum_{\rr}\big(\jjrr\ln\frac{\jjrr}{\jjRrr}-\jjrr+\jjRrr\big)\ge0.\label{eq:eprn}
\end{align}
For systems without odd variables, this expression follows from Eq.~\eqref{eq:eprn2}
and $\sum \jjrr=\sum j_{\negedge}$. 
However, Eq.~\eqref{eq:eprn} also correctly captures EPR for systems with odd variables, where Eq.~\eqref{eq:eprn2} may not apply (see \appref{odd}). 

\newstuff{The generalized relative entropy~\eqref{eq:eprn} induces a decomposition of EPR into nonnegative contributions from individual one-way reactions:  
\begin{align}
\epr = \sum_\rr \sigma^{(\rr)},\quad\quad\sigma^{(\rr)} &:= \jjrr\ln\frac{\jjrr}{\jjRrr}-\jjrr+\jjRrr\,.\label{eq:oneway}
\end{align}
The contribution from reaction $\rr$ can also be written as
\begin{align}
\sigma^{(\rr)} =\jjrr(\ffrr - 1 + e^{-\ffrr})\ge 0\,.
\label{eq:epr-rr}
\end{align} 
Our EPR decomposition~\eqref{eq:oneway} is finer-grained than the usual one, which considers reversible reaction pairs~\cite[Eq.~(68)]{falascoMacroscopicStochasticThermodynamics2023a}:%
\begin{align*}
\epr=\frac{1}{2}\sum_\rr \epr^{(\rr)}_{\text{rev}},\quad\quad \epr^{(\rr)}_{\text{rev}}:=\epr^{(\rr)}+\epr^{(\negedge)}=(\jjrr-\jjRrr)\ffrr. 
\end{align*}
}

\section{Background on nonequilibrium free energy}

\label{sec:db}

In this section, we review the concept of nonequilibrium free energy in conservative systems.
We then discuss a previously proposed generalization to nonconservative systems based
on steady states.

\subsection{Nonequilibrium free energy in conservative systems}

\label{subsec:db}

We use the term \emph{conservative system} to refer to a system 
governed by a nonequilibrium free energy $\FreeEnergy(\pp,t)$. In such systems, 
the thermodynamic forces can be expressed in terms
of the chemical potential $\chempot=\grad_{\pp}\FreeEnergy(\pp,t)$ as
\begin{align}
\ff=\negBBgrad\chempot\,,\label{eq:cons00}
\end{align}
where $\BBbase\in\mathbb{R}^{\numedge\times\numstate}$ is the discrete
gradient operator defined above (i.e., stoichiometric matrix). 
Forces that have the gradient form
\eqref{eq:cons00} are termed {conservative forces}.

The internal dynamics of a conservative system can be expressed as a gradient
flow for $\FreeEnergy$. 
To show this, we write the continuity
equation~\eqref{eq:cont-equation} as
\begin{align}
\dtpp=\frac{1}{2}\BBdiv(\jj-\jjR)-\newstuff{\II}=\BBdiv L\ff-\newstuff{\II}.
\label{eq:cont00}
\end{align}
Here we used $\BBdiv\jj=\frac{1}{2}\BBdiv(\jj-\jjR)$, since reverse reactions have antisymmetric stoichiometry, and defined $L$ as a diagonal matrix with entries $L_{\rr\rr}=\frac{1}{2}\jjrr(1-e^{-\ffrr})/\ffrr\ge0$. 
Using Eq.~\eqref{eq:cons00},  
we can further write %
\begin{align}
\dtpp=-(\BBdiv L\BBgrad)\grad_{\pp}\FreeEnergy-\newstuff{\II}.\label{eq:gradflow}
\end{align}
\newstuff{The (first) net production term is a gradient flow for $\FreeEnergy$ (the positive-semidefinite matrix $\BBdiv L\BBgrad$ defines a metric). 
The relationship between conservative systems and gradient flows was first derived for MJPs in Ref.~\citep{maas_gradient_2011} and CRNs with mass-action kinetics in Ref.~\citep{mielke2011gradient}. Since then, the connection between gradient flows and large deviations has inspired an active research program~\cite{adamsLargeDeviationsGradient2013,mielke_relation_2014,kraaij2020fluctuation}.}

Consider a conservative system that relaxes without time-dependent driving ($\FreeEnergy$ does
not depend explicitly on time) \newstuff{or external fluxes ($\II=\zz$)}. In this case, $\FreeEnergy$ decreases
monotonically over time until the system reaches a stationary equilibrium state
$\ppeq$ of minimal nonequilibrium free energy.  
At this state,  
the chemical potential reaches its equilibrium value and the thermodynamic forces vanish: 
\[
\chempot^{\eq}=\grad_{\pp}\FreeEnergy(\ppeq)\qquad\ff=\negBBgrad\chempot^{\eq}=\zz.
\]
By the definition of the thermodynamic forces, %
forward and reverse fluxes in equilibrium must obey
the condition of \emph{detailed balance},
\begin{align}
\jj(\ppeq)=\jjR(\ppeq)\,.\label{eq:db}
\end{align}

The condition of conservative forces~\eqref{eq:cons00}
is invariant if $\chempot$ is shifted by any null vector of $\BBgrad$.
Such null vectors represent conserved quantities that do not affect
thermodynamic forces. For our purposes, it will be convenient to consider
the equilibrium chemical potential as a reference null vector (recall $\negBBgrad\chempot^{\eq}=\zz$) and define $\poteq=\chempot-\chempot^{\eq}$ as the purely nonequilibrium contribution
to the chemical potential,
\begin{align}
\ff=\negBBgrad\poteq\label{eq:cons}
\end{align}
We usually refer to $\poteq$ as the ``free energy potential''. 

The preceding statements hold for any conservative system. However, things become less abstract 
for MJPs~\cite{esposito2011second} and ideal CRNs with mass-action kinetics~\cite{rao2016nonequilibrium}. In these cases, 
$\poteq$ has an explicit form, as the gradient of the relative entropy
between the current state and the equilibrium state:
\begin{align}
\poteq=\grad_{\pp} D(\pp\Vert\ppeq)\qquad \poteqx{\xx}=\ln\frac{\ppx{\xx}}{\ppeqx{\xx}}.\label{eq:poteqclosedform}
\end{align}
As a concrete example, consider an MJP that evolves according to Eq.~\eqref{eq:g2}. 
Suppose there is a single reservoir and an equilibrium distribution $\pppeq$ that obeys detailed balance, $\pppeqx{\xx}R_{\yy\xx}=\pppeqx{\yy}R_{\xx\yy}$. Then, given any other distribution $\ppp$, the forces $\ffrr:=\ln(\pppx{\xx}R_{ji}/\pppx{\yy}R_{ij})=\poteqx{\xx}-\poteqx{\yy}$ are conservative for the potential $\poteqx{i}=\ln(\pppx{\xx}/\ppeqx{\xx})$.

The relative entropy $D(\pp\Vert\ppeq)$ in Eq.~\eqref{eq:poteqclosedform} quantifies
the purely nonequilibrium contribution to $\mathcal{F}$~\citep{esposito2011second}, 
\begin{align}
D(\pp\Vert\ppeq)=\FreeEnergy(\pp)-\FreeEnergy(\ppeq)\,.
\label{eq:fediff}
\end{align}
It appears under many names in the literature, including ``pseudo-Helmholtz function''~\citep{rao2016nonequilibrium},
``shear Lyapunov function''~\citep{rao2016nonequilibrium}, negative
``nonequilibrium Massieu potential''~\citep{rao2018conservation}, or simply ``free energy''~\citep{qian2001relative,spohnLargeScaleDynamics1991,mielkeNonequilibriumThermodynamicalPrinciples2017,derridaFreeEnergyFunctional2001}. It plays a central role in the thermodynamics of MJPs
and closed ideal CRNs with conservative forces. For instance, it is proportional to the maximal work
that can be extracted while bringing such systems from $\pp$ to $\ppeq$~\citep{procaccia1976potential,esposito2011second,rao2016nonequilibrium}. Conversely, its decrease under free relaxation is
equal to EPR~\footnote{To derive this result, observe that $\epr=\jj^\top \ff =-\dtpp^\top \poteq$, since $\ff=\negBBgrad\poteq$ and $\dtpp = \BBdiv \jj$.},
\begin{align}
\epr= -\partial_{t}D(\pp(t)\Vert\ppeq(s))\vert_{s=t}=-\dtpp^\top \poteq\ge 0.\label{eq:eprLossFE}
\end{align}

The relative entropy has an important statistical interpretation in large deviations theory.
\newstuff{Consider an ensemble of $n$ %
independent and identical fluctuating systems, and let $\bm{P}$ be the empirical distribution of their microstates at a given point in time. Due to stochastic fluctuations, $\bm{P}$ will itself be a random variable. 
Under the equilibrium distribution $\pppeq$, the
probability of observing empirical distribution $\bm{P}\approx \ppp$ due to a fluctuation scales as~\citep{schloglFluctuationsThermodynamicNon1971,qian2001relative,ge2016mesoscopic} 
\begin{align}
\ldpProb{ \bm{P} \approx \ppp}\asymp e^{- \nnn \, D(\ppp\Vert\pppeq)}
\label{eq:ldp} 
\end{align}
where $\asymp$ indicates equality up to sub-exponential
factors in scale parameter $\nnn$. This type of expression is known as a large deviations principle, with the relative entropy $ D(\pp\Vert\ppeq)$ playing the role of the ``rate function''~\citep{touchette2009large}.  
Eq.~\eqref{eq:ldp} implies that distributions with larger nonequilibrium free
energy are exponentially less likely to emerge from equilibrium fluctuations. A similar result can be derived for stochastic chemical systems. Consider an ideal well-mixed system in a large reactor volume $V$, and let the random variable $\bm{C}$ indicate the empirical concentration in an average sub-volume. The probability that concentration vector $\ccc$ emerges as a fluctuation, given true equilibrium concentrations $\ccceq$, scales as~\citep{andersonLyapunovFunctionsStationary2015} 
\begin{align}
\ldpProb{\bm{C} \approx \ccc}\asymp e^{-V\, D( \ccc\Vert\ccceq)}\,. 
\label{eq:ldpccc}
\end{align}}
 
\subsection{The steady-state approach to nonconservative systems}
\label{subsec:ss}

Recent work in thermodynamics has focused on \emph{nonconservative systems}, whose forces cannot be expressed in the form of Eq.~\eqref{eq:cons}
for any $\pot$. Such systems are not governed by a nonequilibrium free energy and they have nonequilibrium steady states %
that do not satisfy detailed balance. 

Nonetheless, it has been suggested that it may be possible to define a ``generalized'' free energy for nonconservative systems. A related idea is that the fluxes, forces, and EPR may be decomposed into excess and housekeeping contributions. In particular, EPR may be decomposed as 
\begin{align}
\epr=\eprEX+\eprHK\,.\label{eq:exhk}
\end{align}
The excess EPR is associated with the generalized free energy and vanishes in steady state, while the housekeeping EPR is the genuine nonequilibrium contribution arising from cyclic fluxes.

\newcommand{\quasip}{{\Phi}_{\text{ss}}}

\begin{newstuffenv}
The best-known generalized free energy and excess/housekeeping decomposition
is based on nonequilibrium steady states~\citep{wangPotentialLandscapeFlux2008,fang_nonequilibrium_2019,falascoMacroscopicStochasticThermodynamics2023a}. 
In this approach, the generalized free energy is defined as the large-deviations rate function of steady-state fluctuations, written $\quasip$, sometimes called the \emph{quasipotential}. 
Let random variable $\bm{X}$ indicate the empirical state averaged across a stochastic system of size $V$ (e.g., number of independent copies as in Eq.~\eqref{eq:ldp} or reactor volume as in Eq.~\eqref{eq:ldpccc}). Then, the quasipotential governs the probability that some state ${\pp}$ (distribution or concentration vector) emerges as a steady-state fluctuation,
\begin{align}
\ldpProb{ \bm{X} \approx \pp }\asymp e^{-V\, \quasip( \pp)}\,.\label{eq:ldpcccnc} 
\end{align} 
Importantly, the quasipotential decreases monotonically under free relaxation, thus it serves as a Lyapunov function~\cite{falascoMacroscopicStochasticThermodynamics2023a}. However, the dynamics of nonconservative systems are not a gradient flow for the quasipotential, nor any other state function.

For systems described at the microscopic fluctuating level,  
the quasipotential is the relative
entropy between the system's actual distribution and the steady-state distribution, $\quasip( \ppp)=D(\ppp\Vert\pppss)$. 
This quasipotential is also used to define an excess/housekeeping decomposition of EPR called the Hatano-Sasa (HS)~\cite{hatano2001steady} or the adiabatic/nonadiabatic~\cite{esposito2010threefaces} decomposition. In analogy to Eq.~\eqref{eq:eprLossFE}, the HS excess EPR is the decrease of the quasipotential due to relaxation,
\begin{align} 
\eprEXhs :=-\partial_{t}D(\ppp(t)\Vert\pppss(s))\vert_{s=t}\ge 0\,,%
\label{eq:eprexhs}
\end{align}
where $\pppss(s)$ is the steady-state distribution specified by the system's parameters at time $s$. 
It can be written in slightly more explicit form as
\begin{align} 
\eprEXhs =-\jj^\top \BBgrad \potss = -\dtppp^\top \potss.\label{eq:eprexhs2}
\end{align}
Here $\potss:=\grad_{\ppp} D(\ppp\Vert\pppss)=\ln(\ppp/\pppss)$ is the generalization of the free energy potential, which we sometimes refer to as the \emph{steady-state potential}. (This terminology should not be confused with the term ``quasipotential'', which refers to a function $\quasip(\pp)$ over thermodynamic states.) 
The HS housekeeping EPR is the remainder EPR, 
\begin{align}
\eprHKhs:=\epr-\eprEXhs.\label{eq:eprhkhs}
\end{align}
This housekeeping contribution is nonnegative for MJPs without odd variables. (With odd variables, it can take on unphysical negative values, see \appref{odd}.)

In systems described at the macroscopic level, including deterministic CRNs, the situation is more complicated. For CRNs that obey \emph{complex balance}, the quasipotential is simply the relative entropy $\quasip( \ccc)= D(\ccc\Vert \cccss)$, where $\cccss$ is the steady-state concentrations of the deterministic dynamics~\eqref{eq:cont-equation}~\citep{andersonLyapunovFunctionsStationary2015,geMathematicalFormalismNonequilibrium2017}. 
Excess and housekeeping EPR are defined in direct analogy to Eqs.~\eqref{eq:eprexhs} and \eqref{eq:eprhkhs}~\cite{ge2016mesoscopic,geMathematicalFormalismNonequilibrium2017}. 
However, complex balance is a highly restrictive condition that excludes most CRNs, including those that exhibit rich nonlinear phenomena like oscillations and chaos~\citep{ge2016nonequilibrium}. 

Without complex balance, 
the quasipotential $\quasip(\ccc)$ is no longer the relative entropy $D(\ccc\Vert\cccss)$~\citep{hugangStationarySolutionMaster1987a,falascoMacroscopicStochasticThermodynamics2023a}. 
In fact, there may not even exist 
stable fixed point $\cccss$ for the 
the deterministic dynamics~\eqref{eq:cont-equation}, because the large-volume limit (used to derive the deterministic CRN description) may not commute with the long-time limit (used to derive steady-state behavior). This phenomenon, known as Keizer's paradox~\cite{keizer_statistical_1987,ge2011non}, highlights the implicit choice of timescales involved in the large-volume and long-time limits. 
In general deterministic CRNs, %
computing $\quasip$ is challenging, although sophisticated numerical methods have been developed~\cite{weinan2004minimum,dahiya2018ordered,li2022machine,zakine2023minimum,gagrani2023action}. 
However, once it is found, macroscopic HS excess and housekeeping EPR can be defined as in Eq.~\eqref{eq:eprexhs} and Eq.~\eqref{eq:eprhkhs}, with the relative entropy replaced by $\quasip$~\cite{falascoMacroscopicStochasticThermodynamics2023a}.

The steady-state approach has 
found numerous applications in physics, chemistry, and biology. For example, it has been used to study transition rates between stochastic attractors, quantify stability and fluctuations in stationarity, and to visualize potential landscapes that govern long-term relaxations~\cite{falascoMacroscopicStochasticThermodynamics2023a,nolting_balls_2016,fang_nonequilibrium_2019,wangPotentialLandscapeFlux2008,ge2009thermodynamic}. 

On the other hand, the steady-state approach is less useful for studying transient systems that never approach stationarity, including systems observed over short timescales and systems driven by changing external parameters and/or flows. 
Because the quasipotential is defined via steady-state statistics, rather than ``local in time'' properties such as instantaneous fluxes or forces, its physical meaning is unclear in systems far from steady state~\citep{dechant2022geometric,dechant2022geometricCoupling,maes2014nonequilibrium}. 
Importantly, the situation is different in conservative systems. There, the quasipotential reduces to the nonequilibrium free energy, which is directly related to instantaneous forces by the conservative force expression~\eqref{eq:cons}.

In the following, we propose an alternative, local-in-time definition of the generalized free energy and excess/housekeeping decomposition. Our generalized free energy is defined as the conservative part of the thermodynamic forces, and the excess/housekeeping EPRs quantify the conservative/nonconservative contributions to dissipation. Using large deviations theory, we will show that our generalized free energy governs short-time current fluctuations in the transient regime. This may be contrasted to the quasipotential $\quasip$, which governs state fluctuations in the stationary regime. Our definition is well-suited to study dynamical properties of transient systems, but less useful for studying stability and fluctuations in stationarity. In this sense, our approach is different and complementary to the steady-state approach.
\end{newstuffenv}

\section{Variational approach: {conservative} systems}

\label{sec:variationaldb} 

We now introduce our first set of results. Specifically, we show that in conservative systems, the nonequilibrium free energy can be derived from a variational principle that makes no explicit
reference to equilibrium states. 
We extend our principle
to nonconservative systems in Section~\ref{sec:variational-ctive}.

Before proceeding, recall from Eq.~\eqref{eq:eprn} that the EPR can be expressed as the
relative entropy between forward and backward fluxes, $\epr=\DD(\jj\Vert\jjR)$.
This allows us to express the EPR in a variational way, 
\begin{align}
\epr=\max_{\ee\in\forcespace}\,\Big[\jj^{\top}\ee-\jjR^{\top}(e^{\ee}-\oneone)\Big]\,,\label{eq:legendreep}
\end{align}
\newstuff{where we maximize over all possible reaction observables.} 
To derive this expression, observe that $\min_{\ee}\DD(\jj\Vert\jjR\circ e^{\ee})=0$, since 
$\DD(\jj\Vert\jjR\circ e^{\ee})\vert_{\ee =\ff}=\DD(\jj\Vert \jj)=0$. 
Thus, we can write the EPR as
\begin{align*}
\epr & =\DD(\jj\Vert\jjR)-\min_{\ee}\DD(\jj\Vert\jjR\circ e^{\ee})\\
 & =\max_{\ee}\Big[\DD(\jj\Vert\jjR)-\DD(\jj\Vert\jjR\circ e^{\ee})\Big].
\end{align*}
Eq.~\eqref{eq:legendreep} follows by expanding the relative entropy
terms and simplifying. The optimal observable $\ee^{\star}=\ff$ is unique due to the strict concavity of the objective.

\newstuff{The variational expression~\eqref{eq:legendreep} is the Legendre transform of the function $\ee \mapsto \jjR^\top (e^{\ee} -\oneone)$ evaluated at $\jj$. As we will see below, this function is the cumulant generating function (CGF) of dynamical fluctuations. %
In large deviations, the variational expression~\eqref{eq:legendreep} is often called the ``Donsker-Varadhan'' form.}

Eq.~\eqref{eq:legendreep} provides a family of lower bounds on the EPR,
one for each reaction observable. The EPR is achieved by maximizing
over the choice of observable, and the thermodynamic forces are
recovered as the optimal observables.
In the setting of stochastic master equations, a related variational principle was recently proposed as a technique for thermodynamic inference~\citep{kim2020learning,otsubo2022estimating}.
Moreover, it %
applies to both conservative and
nonconservative systems. However, in conservative systems, the forces have the conservative 
form $\ff=\negBBgrad\poteq$. Therefore,
we may restrict the optimization to reaction observables having the form
$\ee=\negBBgrad\pot$ for some state observable $\pot\in\potspace$,
\begin{align}
\epr=\max_{\pot\in\potspace}\,\Big[-\jj^{\top}\BBgrad\pot-\jjR^{\top}(e^{-\BBgrad\pot}-\oneone)\Big].\label{eq:epr00} 
\end{align}
This expression may be written using only the forward
fluxes,
\begin{align}
\epr & =\max_{\pot\in\potspace}\,\Big[-\jj^{\top}\BBgrad\pot-\jj^{\top}(e^{\BBgrad\pot}-\oneone)\Big]\,,\label{eq:epr2}
\end{align}
where we used 
$\jjR^{\top}e^{-\BBgrad\pot}=\jj^{\top}e^{\BBgrad\pot}$ from Eq.~\eqref{eq:antisymmetry} and antisymmetry of the stoichiometric matrix.

Eq.~\eqref{eq:epr2} is a variational expression of the EPR
in conservative systems. Taking derivatives shows that the optimal $\potopt$
obeys
\begin{align}
\BBdiv(\jj\circ e^{\BBgrad\potopt})=-\newstuff{\BBdiv\jj}.\label{eq:poisson-2}
\end{align}
In conservative systems, the free energy potential satisfies this relation, as can be verified using Eq.~\eqref{eq:cons}, thus $\potopt=\poteq$. 
Moreover, due to strict concavity of the objective,$\poteq$ is the unique optimum, up to the choice of a nullvector of $\BBgrad$ that represents a conserved quantity. If desired, it is possible to identify the $\poteq$ that is consistent with the conserved quantities encoded in $\pp$ (see \smref{sm1}). 

We may consider our variational principle~\eqref{eq:epr2} from perspective of convex duality~\citep{boydConvexOptimization2004}. %
Observe that Eq.~\eqref{eq:epr2} involves the maximization of a
concave objective over state observables. %
By duality, it has an equivalent formulation as the minimization of a convex objective
over flux vectors:
\begin{align}
\epr=\min_{\jj^{\prime}\in\fluxspace}\DD(\jj^{\prime}\Vert\jjR)\suchthat\BBdiv\jj^{\prime}=\newstuff{\BBdiv\jj}.\label{eq:epr-var}
\end{align}
To show the equivalence, we write Eq.~\eqref{eq:epr-var} in its Lagrangian
dual form~\citep[Ch.~5]{boydConvexOptimization2004}, 
\begin{align}
\epr & =\max_{\potlag\in\potspace}\min_{\jj^{\prime}\in\fluxspace}\Big[\DD(\jj^{\prime}\Vert\jjR)+\potlag^{\top}\BBdiv(\jj^{\prime}-\newstuff{\jj})\Big],\label{eq:lagr0}
\end{align}
where $\potlag\in\potspace$ indicates the Lagrangian multipliers.
The inner optimization can be solved by taking derivatives. After a
bit of algebra, we find that the optimal fluxes have the form
$\jjR\circ e^{-\BBgrad\pot}$. Plugging back into~\eqref{eq:lagr0}
and simplifying shows equivalence to Eq.~\eqref{eq:epr2}. The optimal Lagrange multipliers are equal to $\poteq$, while the optimal
fluxes are equal to the actual forward fluxes $\jj^{\star}=\jjR\circ e^{-\BBgrad\poteq}=\jj$.

The variational expressions~\eqref{eq:epr2}-\eqref{eq:epr-var} comprise
our first set of results. They demonstrate that $\poteq$ may be derived in a simple way from the fluxes, without any explicit reference to equilibrium states.

\section{Variational approach: {nonconservative} systems}

\label{sec:variational-ctive}

This section contains most of our main results. We extend our variational principle to nonconservative systems, which leads to our definitions of the generalized free energy and the excess/housekeeping decomposition. We also show that our variational principle satisfies an important consistency condition under coarse-graining. We discuss connections to information geometry and discuss the linear-response regime of slow evolution.

\subsection{Overview}

\label{sec:variational-overview}

\newstuff{As we showed above, in conservative systems, the variational principle for EPR~\eqref{eq:legendreep} can be restricted to observables having the gradient form $\ee=\negBBgrad\pot$. 
This leads to a variational expression of EPR and $\poteq$, Eq.~\eqref{eq:epr2}.}

\newstuff{In nonconservative systems, restricting Eq.~\eqref{eq:legendreep} to gradient observables gives the following variational expression:} 
\begin{align}
\eprEX=\max_{\potlag\in\potspace}\Big[-\jj^{\top}\BBgrad\potlag-\jj^{\top}(e^{\BBgrad\potlag}-\oneone)\Big]\,.\label{eq:exdual}
\end{align}
\newstuff{This does not recover the full EPR unless all forces are conservative. Instead, the value of $\eprEX$ defines the excess EPR, 
the {conservative} contribution to dissipation.} 
For a temporally-extended process, the time-integrated excess EP is $\epEX=\int\eprEX(t)\,dt$. 

We have the lower bound $\eprEX\ge 0$, since the objective in Eq.~\eqref{eq:exdual} vanishes when $\potlag=\zz$. We also have the upper bound $\eprEX\le\epr$, since Eq.~\eqref{eq:exdual} is a restriction of the maximization~\eqref{eq:legendreep} to $\ee\in\mathrm{Im} \BBgrad$.

The optimal state observable $\potopt$ in Eq.~\eqref{eq:exdual} defines our generalized free energy potential, and $\BBgrad \potopt$ quantifies the conservative contribution to the thermodynamic forces. In the following, we  refer to $\potopt$ as the \emph{generalized potential}.  
The generalized potential satisfies the optimality condition,
\begin{align}
\BBdiv(\jj\circ e^{\BBgrad\potopt})=-\newstuff{\BBdiv\jj},\label{eq:poisson-1}
\end{align}
which is the analogue of Eq.~\eqref{eq:poisson-2}. \newstuff{Recall that $\newstuff{\BBdiv\jj}$ is the net production of species due to reaction fluxes $\jj$. Eq.~\eqref{eq:poisson-1} implies that $\potopt$ is defined by the property that, when the fluxes are exponentially tilted by $\potopt$, net production is reversed.}
The optimality condition~\eqref{eq:poisson-1} 
determines $\potopt$ up to the nullspace of $\BBgrad$, representing conserved quantities. Although not necessary for most of our results, $\potopt$ can always be chosen within this nullspace to satisfy the system's conservation laws (see \smref{sm1}).

Eq.~\eqref{eq:exdual} can also be rearranged as 
\begin{align}
\eprEX=\max_{\potlag\in\potspace}\big[-2\jj^{\top}\BBgrad\potlag-\jj^{\top}(e^{\BBgrad\potlag}-\BBgrad\pot-\oneone)\big].\label{eq:exdual2}
\end{align}
The first term is twice the average change of $\pot$ due to reaction fluxes. As we show below, the second term quantifies the fluctuations of $\pot$, and it is always nonnegative (since $e^x-x-1\ge 0$ for all $x$). 
Since excess EPR is nonnegative, the generalized potential obeys
\begin{align}
-\jj^{\top}\BBgrad\potopt \ge 0,
\label{eq:optdownhill}
\end{align}
\newstuff{which reduces to $-\dtpp^\top \potopt \ge 0$ in a closed system without external fluxes. This is analogous to Eq.~\eqref{eq:eprLossFE} in conservative systems, $-\dtpp^\top \poteq\ge 0$. 
Eq.~\eqref{eq:optdownhill} implies that the net production points downhill in the generalized potential.  
However, Eq.~\eqref{eq:optdownhill} does not imply Lyapunov stability in the same way as Eq.~\eqref{eq:eprLossFE}, because $\potopt$ need not be the gradient of any state function (like $\mathcal{F}$).}

The variational expression~\eqref{eq:exdual} is the central expression of this paper. 
Its physical meaning is explored further below in this section and in Section~\ref{sec:Physical-interpretations}, where we relate it to large deviations theory and thermodynamic speed limits. 
To our knowledge, this expression is largely unknown in stochastic thermodynamics, either in conservative or nonconservative systems. 
One exception is a paper by Shiraishi and Saito~\citep{shiraishi2019information},
who proposed a different variational expression for EPR in conservative
MJPs. In Section~\ref{subsec:shiraishi}, we show that the result 
of Ref.~\citep{shiraishi2019information} is a special
case of Eq.~\eqref{eq:epr2}, allowing us to extend the main result of Ref.~\citep{shiraishi2019information} to nonconservative MJPs.

The numerical values of excess EPR and the generalized potential 
can be found by solving the convex optimization~\eqref{eq:exdual} using standard numerical 
algorithms. Also, as shown in Section~\ref{subsec:cg}, these quantities can also
be found in closed form for a special class of systems (those with conservative forces after coarse-graining). Finally, for a time-extended process characterized by a flux trajectory $\jj(t)$, the generalized potential
can be found by solving an ordinary differential equation. The time derivative
of $\potopt(t)$ can be found by differentiating both sides of~\eqref{eq:poisson-1}
and rearranging. After simplifying, this gives
\begin{align}
\dot{\potopt}(t)=-\frac{1}{2}\HHH_{t}^{+}\BBdiv\big(\dot{\jj}(t)\circ(e^{\BBgrad\potopt(t)}+\oneone)\big),\label{eq:ode2}
\end{align}
where $\HHH_{t}^{+}$ is the pseudo-inverse of the positive-semidefinite matrix $\HHH_{t}:=\frac{1}{2}\BBdiv\diag(\jj(t)\circ e^{\BBgrad\potopt(t)})\BBgrad$. 
Given a known initial $\potopt(0)$, Eq.~\eqref{eq:ode2} can be solved to find $\potopt(t)$ at all $t$.

\subsection{Example: two-level MJP}

\label{subsec:2level}

To make things concrete, we illustrate our approach on a minimal example, the two-level MJP from Figure~\ref{fig:twolevel}. 
For this system, the dynamics obey 
\[
\dtpppx 2=j_{21}^{c}+j_{21}^{h}-j_{12}^{c}-j_{12}^{h},
\]
where $\dtpppx 1=-\dtpppx 2$ by conservation of probability. The optimality condition~\eqref{eq:poisson-1} reduces to
\begin{align*}
\dtpppx 2=-(j_{21}^{c}+j_{21}^{h})e^{\potoptxx 2-\potoptxx 1}+(j_{12}^{c}+j_{12}^{h})e^{\potoptxx 1-\potoptxx 2}.
\end{align*}
This condition is satisfied by the potential
\begin{align}
\potopt=\left(0,\ln\frac{j_{12}^{c}+j_{12}^{h}}{j_{21}^{c}+j_{21}^{h}}\right) \,,%
\label{eq:mjp2optpot}
\end{align}
which is unique up to a scalar constant (the nullspace of $\BBgrad$ is one dimensional, representing conservation of probability). 
Plugging $\potopt$ into Eq.~\eqref{eq:exdual} and simplifying gives
\begin{align}
\eprEX= (j_{21}^{c}+j_{21}^{h}-j_{12}^{c}-j_{12}^{h})\ln\frac{j_{21}^{c}+j_{21}^{h}}{j_{12}^{c}+j_{12}^{h}}.\label{eq:eprEX2leel}
\end{align}

In this example, the excess EPR is the same as the EPR incurred by
a system with two one-way fluxes: the forward flux $j_{21}^{c}+j_{21}^{h}$ and the reverse flux $j_{12}^{c}+j_{12}^{h}$. (We will see below that this exemplifies a more general ``coarse-graining'' principle.) Eq.~\eqref{eq:eprEX2leel} suggests an ``effective'' conservative force across the transition $1\to 2$,
\begin{align*}
\ln\frac{j_{21}^{c}+j_{21}^{h}}{j_{12}^{c}+j_{12}^{h}}=(\ln \pppx 1 + \beta^{\mathrm{eff}} E_1) - (\ln \pppx 2 +\beta^{\mathrm{eff}} E_2) 
\end{align*}
where 
$\beta^{\mathrm{eff}}=(\ln\frac{R_{21}^{h}+R_{21}^{c}}{R_{12}^{h}+R_{12}^{c}})/(E_{1}-E_{2})$ is an effective inverse temperature and $E_1,E_2$ refer to  microstate energies. 
Note that $\beta^{\mathrm{eff}}$ does not depend on the probability
distribution $\ppp$, only on the energy gap and the kinetics.

\subsection{Excess EPR and nonstationarity}

As mentioned, excess EPR captures the conservative contribution to EPR. Here we use duality to show that it can also be understood as quantifying the nonstationarity of the dynamics. 

The variational expression~\eqref{eq:exdual} has the dual form,
\begin{align}
\eprEX=\min_{\jj^{\prime}\in\fluxspace}\DD(\jj^{\prime}\Vert\jjR)\suchthat\BBdiv\jj^{\prime}=\newstuff{\BBdiv\jj},\label{eq:exdef}
\end{align}
derived in the same way as Eq.~\eqref{eq:epr-var}. 
The generalized potential $\potopt$ specifies the optimal Lagrange
multipliers for the constraints. The optimal fluxes, written as
\begin{align}
\jj^{\star}=\jjR\circ e^{-\BBgrad\potopt},\label{eq:optflux}
\end{align}
are generally not equal to the forward fluxes $\jj$, except in conservative systems. 
The excess EPR can be expressed using these optimal fluxes as
\begin{align}
\eprEX=\DD(\jj^{\star}\Vert\jjR).\label{eq:opt3}
\end{align}

Excess EPR vanishes when the reverse fluxes satisfy the constraints $\BBdiv\jjR=\BBdiv\jj$, which is equivalent to $\BBdiv\jj=\zz$ due to antisymmetry. The optimality condition~\eqref{eq:poisson-1} implies the generalized potential also vanishes when $\BBdiv\jj=\zz \implies \potopt=\zz$.  
\newstuff{In other words, the excess EPR and generalized potential vanish when net production vanishes, $\BBdiv\jj=\zz$. In a closed system without external fluxes, $\dtpp=\BBdiv\jj$, therefore they vanish precisely in steady state.}

Eq.~\eqref{eq:exdef} allows us to interpret our excess EPR as an information-theoretic {optimal transport} cost~\citep{gentilAnalogyOptimalTransport2017}. In general, ``optimal transport'' studies the minimal cost of transforming a system between two states, thus giving operational definitions of distance and speed~\citep{villaniOptimalTransportOld2009}. 
Eq.~\eqref{eq:exdef} implies that the excess EPR is the minimal cost necessary to achieve the same dynamical evolution as \newstuff{that induced by the 
true fluxes, $\BBdiv \jj$}. The information-theoretic cost function $\jj^{\prime}\mapsto \DD(\jj^{\prime}\Vert\jjR)$ quantifies the breaking of time-reversal symmetry relative to the reverse fluxes $\jjR$.  
We discuss the relation to another optimal-transport
distance (Wasserstein distance) in Section~\ref{sec:tsl}.

\subsection{Consistency under coarse-graining}

\label{subsec:cg} 
\global\long\def\eprEXcg{\bar{\sigma}_{\text{ex}}}%

Our generalized potential and excess EPR satisfy
an important consistency condition under coarse-graining. 
To introduce this idea, observe that a system may have multiple different
reactions with the same stoichiometry. In fact, this is a common
way to drive nonconservative systems. For example, in the two-level MJP discussed above (Figure~\ref{fig:twolevel}), the two reversible transitions between
levels 1 and 2 have identical stoichiometry. As a result, the incidence
matrix $\BBgrad$ has duplicate rows (1,3 and 2,4). As another example, in the Brusselator CRN (Figure~\ref{fig:crn}), the reversible reactions $X_{1}\rightleftharpoons X_{2}$
and $2X_{1}+X_{2}\rightleftharpoons3X_{1}$ have opposite stoichiometry.
As a result, $\BBgrad$ has duplicate rows (3,6 and 4,5).

We now introduce a coarse-graining procedure that combines reactions with
the same stoichiometry. Given a stoichiometric matrix $\BBbase$,
the coarse-grained stoichiometric matrix is defined as $\overline{\BBbase}\in\mathbb{Z}^{\numedge^{\prime}\times\numstate}$
($\numedge^{\prime}\le m$) where any duplicate rows (reactions with
the same stoichiometry) are merged. 
Coarse-grained forward fluxes $\overline{\jj}\in\mathbb{R}_{+}^{\numedge^{\prime}}$
are defined by summing fluxes $\jjrr$ from merged reactions, and similarly for the coarse-grained
reverse fluxes $\overline{\jjR}\in\mathbb{R}_{+}^{\numedge^{\prime}}$. 
In an MJP where the same transitions
are mediated by different reservoirs $\bath$, the coarse-graining
procedure will sum rate matrices induced by 
different reservoirs, $\bar{R}=\sum_{\bath}R^{\bath}$.

It is clear that this coarse-graining preserves net production, $\BBdiv\jj=\overline{\BBbase}^{\top}\overline{\jj}$. 
The quantity $\jj^{\top}(e^{\BBgrad\potlag}-\oneone)$, which appears in our variational principle~\eqref{eq:exdual}, is also invariant under coarse-graining:
\begin{align}
\jj{}^{\top}(e^{\BBgrad\potlag}-\oneone)=\overline{\jj}{}^{\top}(e^{\overline{\BBbase}\potlag}-\oneone).\label{eq:gd23}
\end{align}
In Section~\ref{sec:Physical-interpretations}, we show that
this is the cumulant generating function of dynamical fluctuations
of $\pot$, so Eq.~\eqref{eq:gd23} implies that these fluctuations 
are invariant under coarse-graining. Finally, our variational principle~\eqref{eq:exdual} is also invariant, therefore our excess EPR $\eprEX$ and generalized potential $\potopt$ are the same, regardless of whether they are defined using $(\jj,\BBbase)$ or $(\overline{\jj},\overline{\BBbase})$. 

This coarse-graining has practical implications, as 
it sometimes allows the generalized potential and excess EPR to
be found in closed form. Consider the class of systems
that have conservative coarse-grained forces: 
\begin{align}
\overline{\ff}=\ln(\overline{\jj}/\overline{\jjR})=-\overline{\BBgrad}\pot\label{eq:linf}
\end{align}
for some $\pot$. 
Then, the excess EPR is given by the regular EPR
at the level of the coarse-grained fluxes, 
\begin{align}
\eprEX(\jj)=\eprEXcg(\bar{\jj})=\epr(\bar{\jj}),\label{eq:linf2}
\end{align}
and $\potopt$ is equal to $\pot$ from Eq.~\eqref{eq:linf}. In such cases, $\potopt$ can be
found directly by inspecting the linear system of equations~\eqref{eq:linf},
eliminating the need for numerical optimization. This technique was implicitly used to find $\potopt$ for the two-level MJP in the previous
example, as in Eq.~\eqref{eq:mjp2optpot}. The same technique will be used in 
the Brusselator example in Section~\ref{sec:brusselator}.

\subsection{Housekeeping entropy production and the excess/housekeeping decomposition}

\label{sec:ex-vs-hk}

We now consider the \emph{housekeeping EPR}, the nonconservative part of the dissipation, defined as
\begin{align}
\eprHK:=\epr-\eprEX.
\end{align}
Since the excess EPR satisfies $0\le\eprEX\le\epr$, the housekeeping
EPR also obeys the bounds $0\le\eprHK\le\epr$. 
For a temporally-extended process, the time-integrated housekeeping
EP is $\epHK=\int\eprHK(t)\,dt$.

The housekeeping EPR has a variational expression:
\begin{align}
\eprHK=\min_{\potlag\in\potspace}\DD(\jj\Vert\jjR\circ e^{\negBBgrad\potlag})=\DD(\jj\Vert\jj^{\star}),\label{eq:hkdef}
\end{align}
where the optimal fluxes are $\jj^{\star}=\jjR\circ e^{-\BBgrad\potopt}$
as in Eq.~\eqref{eq:optflux}. This result is derived by plugging
Eq.~\eqref{eq:exdual} into $\eprHK=\DD(\jj\Vert \jjR)-\eprEX$, expanding the definition of $\DD$, and rearranging. 

To clarify the meaning of Eq.~\eqref{eq:hkdef}, we introduce the notation
\begin{align}
\DDexpFam(\ee\Vert\ee^{\prime}):=\DD\big(\jjR\circ e^{\ee}\Vert\jjR\circ e^{\ee^{\prime}}\big)\label{eq:ddd3}
\end{align}
to indicate the relative entropy between pairs of flux vectors in the exponentially-tilted parametric family
\begin{align}
\ee\mapsto\jjR\circ e^{\ee}.\label{eq:expfam}
\end{align}
Using this notation, the housekeeping EPR may be written as
\begin{align}
\eprHK=\min_{\potlag\in\potspace}\DDexpFam(\ff\Vert\negBBgrad\potlag)=\DDexpFam(\ff\Vert\negBBgrad\potopt).\label{eq:hkdef-1}
\end{align}
In this form, we see that the housekeeping EPR quantifies the information-geometric distance between the actual forces $\ff$ and the closest conservative forces $\negBBgrad\potlag$ (for some $\potlag$). It vanishes when the forces are conservative, and otherwise quantifies their ``nonconservative-ness''. 
From this perspective, the generalized potential $\potopt$, which achieves the minimum in Eq.~\eqref{eq:hkdef-1}, provides the ``best conservative approximation'' of the forces.

\begin{newstuffenv}
Eq.~\eqref{eq:hkdef-1} implies that housekeeping EPR vanishes whenever $\ff$ belongs to the image of the stoichiometric matrix $\BBgrad$. This leads to a simple necessary condition for non-vanishing housekeeping EPR, expressed as an algebraic property of $\BBbase$:
\begin{align} 
\eprHK>0\quad\text{only if}\quad\operatorname{rank}\BBbase < \numedge/2\,, \label{eq:rankprop}
\end{align}
since otherwise it is always possible to express $\ff=\BBgrad \pot$ for some $\pot$.  
The relevant dimensionality is $\numedge/2$ (rather than $\numedge$) due to the antisymmetry of $\ff$ and $\BBbase$~\footnote{To derive Eq.~\eqref{eq:rankprop} formally, recall that $\ffrr=-\ffrrR$ by antisymmetry, hence $\ff$ belongs to the antisymmetric subspace $V=\{\ee \in \forcespace : \eerr=-\eerrR\}$ of dimension $\numedge/2$. The stoichiometric matrix also obeys the antisymmetry $\BBgrad_{\rr \cdot}=-\BBgrad_{\negedge\cdot}$, therefore $\operatorname{Im}\BBgrad\subseteq V$. When $\operatorname{rank}\BBbase = \numedge/2$, we must have $\operatorname{Im}\BBgrad= V\ni \ff$.}.
\end{newstuffenv}

\begin{newstuffenv}Interestingly, the excess and housekeeping decomposition can be defined at the level of the forces, fluxes, and dissipation of individual reactions. This fine-grained decomposition allows us to classify the contribution of individual reactions to state evolution versus cyclic fluxes. 
The forces are decomposed into conservative and nonconservative parts as $\ff= \ff_\ex + \ff_\hk$, where 
\begin{align}
\ff_\ex :=-\BBgrad \potopt=\ln \frac{\jj^{\star}}{\jjR} \quad\;\; \ff_\hk :=\ff+\BBgrad \potopt=\ln\frac{\jj}{\jj^{\star}}\,.
\label{eq:ffex}
\end{align}

Similarly, the fluxes are decomposed as $\jj= \jj_\ex + \jj_\hk$, where
\begin{align}
\jj_\ex &:=\jj^{\star} &\qquad &&\jj_\hk &:=\jj-\jj^{\star}\,.
\label{eq:jjex}
\end{align}
We draw attention to two subtle points. First, if one defines the ``reverse'' excess fluxes as $[\tilde{\jj}_\ex]_{\rr}=[{\jj}_\ex]_{\negedge}$, then the excess fluxes have thermodynamic forces $\ln(\jj_\ex/\jjR_\ex)=\ff_\ex-\ff_\hk=-2\BBgrad\potopt-\ff$; thus, one should not think of the excess fluxes as being conservative for the potential $\potopt$. Second, the housekeeping fluxes $\jj_\hk$ should be interpreted as net fluxes (rather than one-way), because their entries may be positive or negative.  
These are cyclic fluxes ($\BBdiv \jj_\hk=\zz$) that contribute to dissipation without
changing the state, and they vanish if and only if the forces are conservative.

Finally, we decompose the dissipation incurred by individual reactions into excess and housekeeping contributions. In the same way that we decomposed total EPR in Eq.~\eqref{eq:epr-rr}, we write $\eprEX = \sum_\rr \sigma_\ex^{(\rr)}$ and $\eprHK = \sum_\rr \sigma_\hk^{(\rr)}$, where the nonnegative contributions from reaction $\rr$ are
\begin{align}
\begin{aligned}
\sigma_\ex^{(\rr)} &:= \jjrr^\star(f_{\ex,\rr} -1+e^{-f_{\ex,\rr}})\ge 0 \\
\sigma_\hk^{(\rr)} &:= \jjrr(f_{\hk,\rr} -1+e^{-f_{\hk,\rr}})\ge 0
\end{aligned}
\label{eq:eprexhk-rr}
\end{align}
which may be derived from Eqs.~\eqref{eq:opt3} and \eqref{eq:hkdef} respectively. %
$\sigma_\ex^{(\rr)}$ and $\sigma_\hk^{(\rr)}$ vanish only when $f_{\ex,\rr}=0$ and $f_{\hk,\rr}=0$, respectively. 
We note that the excess/housekeeping decomposition does not always commute with the reaction decomposition, so in general $\sigma^{(\rr)} \ne \sigma_\ex^{(\rr)}+ \sigma_\hk^{(\rr)}$.
\end{newstuffenv}

\begin{figure}
\includegraphics[width=1\columnwidth]{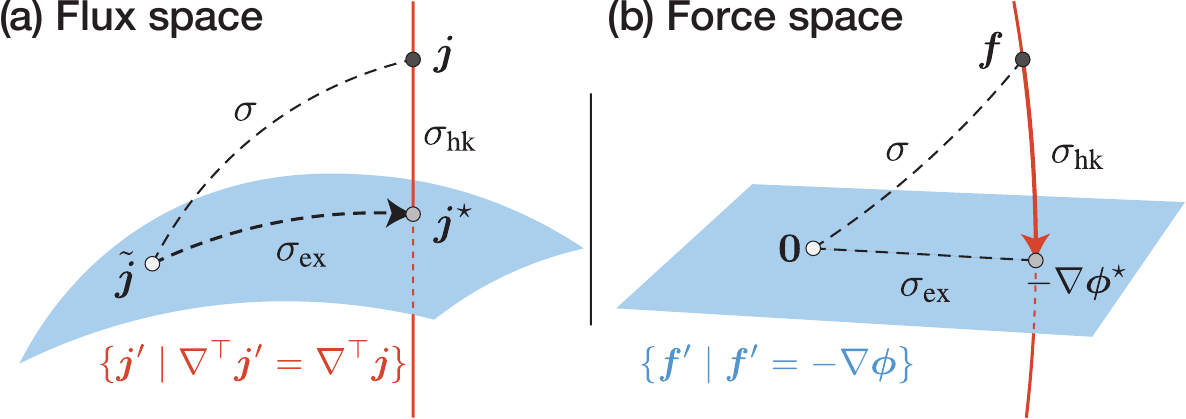}

\caption{\textbf{Information-geometric interpretation of excess/housekeeping
decomposition}. 
\textbf{(a)} The EPR $\protect\epr=\protect\DD(\protect\jj\Vert\protect\jjR)$
is the relative entropy from the forward fluxes $\protect\jj$ to
the reverse fluxes $\protect\jjR$. Excess EPR $\protect\eprEX$ is
defined by the projection of $\protect\jjR$ onto the set of
fluxes that give the actual time evolution (orange line), Eq.~\eqref{eq:exdef}.
Excess/housekeeping provide an orthogonal decomposition of the
EPR in flux space, Eq.~\eqref{eq:pyth-1}. \textbf{(b)} In the space
of forces, the EPR $\protect\epr=\protect\DDexpFam(\protect\ff\Vert\protect\zz)$
is the relative entropy from the actual forces $\protect\ff$ to the
origin $\protect\zz$. Housekeeping EPR is defined by the (dual) projection
of the forces onto the set of conservative forces (blue plane),
Eq.~\eqref{eq:hkdef-1}. Excess/housekeeping provide an orthogonal
decomposition of the EPR in force space, Eq.~\eqref{eq:pyth}. The
two projections meet at the point corresponding to the generalized
potential $\protect\potopt$.}\label{fig:pyth}
\end{figure}

\subsection{Information-geometric interpretation}
Our excess/housekeeping decomposition can be interpreted
in terms of information-geometry, as shown in Figure~\ref{fig:pyth}.
In the space of fluxes, we write our decomposition as
\begin{align}
\underbrace{\DD(\jj\Vert\jjR)}_{\begin{matrix}\epr\end{matrix}}=\underbrace{\DD(\jj^{\star}\Vert\jjR)}_{\begin{matrix}\eprEX\end{matrix}}+\underbrace{\DD(\jj\Vert\jj^{\star})}_{\begin{matrix}\eprHK\end{matrix}}.\label{eq:pyth-1}
\end{align}
This expression is an instance of the Pythagorean relation from information
geometry~\citep{amari2016information,collins2002logistic}. It is analogous to the Pythagorean
relation from Euclidean geometry, except that squared Euclidean distance
is replaced by relative entropy. Excess and housekeeping 
provide an orthogonal decomposition of EPR, as illustrated
in Figure~\ref{fig:pyth}(a). 
Orthogonality reflects the fact that the excess EPR~\eqref{eq:exdef} is defined by an (information-geometric) projection of $\jjR$ onto a linear subspace~\cite{collins2002logistic}: the subspace of fluxes that induce the same time evolution as the actual fluxes, $\BBdiv\jj^{\prime}=\newstuff{\BBdiv\jj}$. 

The same decomposition can be expressed in the dual space of forces,
 see Figure~\ref{fig:pyth}(b). Using notation~\eqref{eq:ddd3},
we write 
\begin{align}
\underbrace{\DDexpFam(\ff\Vert\zz)}_{\begin{matrix}\epr\end{matrix}}=\underbrace{\DDexpFam(\negBBgrad\potopt\Vert\zz)}_{\begin{matrix}\eprEX\end{matrix}}+\underbrace{\DDexpFam(\ff\Vert\negBBgrad\potopt)}_{\begin{matrix}\eprHK\end{matrix}}\label{eq:pyth}
\end{align}
Here, orthogonality arises because the housekeeping EPR~\eqref{eq:hkdef-1}
is defined by the dual projection of the forces $\ff$ onto the flat
manifold~\cite{amari2016information}: the manifold of conservative forces, $\ff^{\prime}=\negBBgrad\pot$ for some $\pot$. 
These two projections recover the same point, corresponding to the generalized potential $\potopt$~\citep{csiszarIdivergenceGeometryProbability1975}.

Surprisingly, there has been almost no work on decomposing entropy production using the information-geometric Pythagorean
relation at the level of fluxes or trajectories. One exception is Ref.~\citep{ito2020unified}, where entropy production was decomposed into partial contributions from individual subsystems. Also, Ref.~\cite{Kobayashi2022} explored information-geometric decompositions using a different divergence (not relative entropy).

\subsection{Linear response}
\label{subsec:lr}

\newcommand{\HHhs}{G}

Here we consider our variational principle in the linear-response regime 
$\BBdiv\jj \approx \zz$ around vanishing net production. For closed systems, $\dtpp = \BBdiv\jj$, so this is the regime of slow evolution.  Below, we show that  our generalized potential obeys Onsager relations, with transport coefficients given by short-time diffusion coefficients. In this sense, our linear-response regime generalizes Onsager theory to far-from-equilibrium and nonconservative systems.

Recall that the $\potopt=\zz$ when $\BBdiv\jj=\zz$. To consider the linear-response regime, we expand the variational principle~\eqref{eq:exdual2} to second order around $\pot\approx \zz$, giving
\begin{align}
\eprEX\approx\max_{\potlag\in\potspace}\big[-2\pot^\top \BBdiv \jj -\pot^{\top}\HH\pot\big],\label{eq:ovv}
\end{align}
where we introduced the matrix
\begin{align}
 \HH := \frac{1}{2}\BBdiv\diag(\jj)\BBgrad\,.
 \label{eq:hsss}
\end{align}
For the special case of an MJP with rate matrix $R$, it has a more explicit expression as
\begin{align}
\HH_{ji} = \frac{1}{2}\begin{cases} 
-\pppx{i} R_{ji}- \pppx{j} R_{ij} & i \ne j\\
\sum_{k (\ne i)} (\pppx{i} R_{ki}+\pppx{k} R_{ik}) & i = j
\end{cases}
\label{eq:mjpH}
\end{align}
Eq.~\eqref{eq:ovv} can be solved to write the excess EPR as
\begin{align}
\eprEX\approx 
\jj^\top \BBgrad\HH^{+}\BBdiv \jj. %
\label{eq:eprLR}
\end{align}
where $\HH^+$ is the pseudo-inverse.

The optimal observable in Eq.~\eqref{eq:ovv} satisfies the linear-response version of~\eqref{eq:poisson-1}~\footnote{The equation $\BBdiv\jj=-\HH\potopt$ only determines $\potopt$ up to the nullspace of $\BBgrad$. In the linear-response regime, the choice $\potopt = -\HH^+(\jj)\BBdiv\jj$ satisfies the correct conservation laws, as described in \smref{sm1}.},
\begin{align}
\BBdiv\jj\approx-\HH\potopt \qquad\qquad \potopt \approx -\HH^+ \BBdiv\jj\,,\label{eq:can3-2}
\end{align}
This is a linear Onsager
phenomenological relation~\citep[Ch.~7]{doi2021onsager} that connects the generalized potential $\potopt$ to dynamics. The matrices $\HH,\HH^+$ are positive-semidefinite and symmetric, 
thus Onsager's reciprocal relations are satisfied. The matrix $\HH$ specifies the mobility coefficients, 
while $\HH^+$ specifies the friction coefficients. 
\newstuff{In the next section, we will see that $\HH$ governs the variance of short-time dynamical fluctuations. 
The linear-response approximation is valid when  net production ($\BBdiv\jj$) is small, relative to the scale of dynamical fluctuations ($\HH$).}

\newstuff{Next, we consider a closed system, where $\dtpp = \BBdiv  \jj$. Then, we can express the excess EPR as}
\begin{align}
\eprEX\approx 
\dtpp^\top \HH^{+}\dtpp\,. %
\label{eq:eprLRreim}
\end{align}
This defines a Riemannian geometry over thermodynamic states: $\eprEX$ acts as the square of the line element and the friction tensor $ \HH^{+}$ acts as the Riemannian metric (note that it implicitly depends on the state and time via the fluxes $\jj(\pp(t),t)$). 
This can be used to define thermodynamic length for slow-evolving nonconservative systems, thereby generalizing the thermodynamic geometry originally developed for conservative systems~\citep{sivak_thermodynamic_2012}.

Eq.~\eqref{eq:eprLRreim} does not assume proximity or existence of the steady-state.  Nonetheless, we may also consider a closed system that remains close to steady state, $\pp\approx \ppss$. We can then further approximate the excess EPR as 
\begin{align}
\eprEX\approx 
\dot{\pp}^{\sss\top}\HH^{+}_{\sss}\dot{\pp}^{\sss}\,,\label{eq:eprLRss}
\end{align}
where $\HH^{+}_{\sss}$ is the friction matrix evaluated at steady-state fluxes $\jj(\ppss(t),t)$ and $\dot{\pp}^{\sss}$ is the change of the steady state due to driving. 
Eq.~\eqref{eq:eprLRss} is quadratic in the speed of driving $\dot{\pp}^{\sss}$, so the total excess EP incurred over the course of a process, $\epEX(\ft)=\int_{0}^{\ft}\eprEX(t)\,dt$, 
vanishes in the limit of slow driving ($T\to\infty$ and $\dot{\pp}^{\sss}=O(1/T)$). This can be interpreted as a generalized Clausius equality~\cite{komatsu2008steady} for our excess EP.

For closed systems, the linear-response analysis highlights the similarity to Onsager's variational principle (OVP)~\citep{onsager1931reciprocal1,gyarmati1970non,doi2021onsager}, also called the ``least dissipation principle''. In its usual form, OVP says that the evolution $\dtpp$ is determined by a balance free energy input and dissipation due to friction. In our notation, it can be expressed as a variational principle for the state evolution, given a fixed generalized potential~$\potopt$: %
\begin{align}
\eprEX \approx \max_{\dot{\bm{y}}\in \mathrm{im} \BBdiv} \big[-2\dot{\bm{y}}^\top \potopt-\dot{\bm{y}}^\top \HH^+_{\sss}\dot{\bm{y}} \big]\,.\label{eq:ovp}
\end{align}
where the optimal $\dot{\bm{y}}$ recovers the system's actual evolution,
\begin{align}\dtpp\approx -\HH_{\sss}\potopt\,.\label{eq:onsevol0}
\end{align}
In our setting, however, we are interested in the ``inverse'' problem of inferring the generalized potential $\potopt$
from a given state evolution $\dtpp$, giving Eq.~\eqref{eq:ovv}. For a closed system near steady-state, it gives 
\begin{align}\potopt \approx -\HH^+_{\sss} \dtpp\,.\end{align}
This inverse form
of the OVP is sometimes called ``Gyarmati's
variational principle''~\citep{gyarmati1970non,martyushev2006maximum,verhas2014gyarmati}.

\global\long\def\CGFee{\Lambda}%
\global\long\def\CGFpot{\Lambda_{\pot}}%

\section{Large deviations and thermodynamic uncertainty
relations}

\label{sec:Physical-interpretations}

Above, we introduced the variational principle~\eqref{eq:exdual} and used it to define our generalized potential and excess/housekeeping decomposition.  
Here, we show that this variational principle has an intuitive physical interpretation
in terms of dynamical large deviations. In particular, we show that our excess EPR governs the irreversibility of state dynamics and that the generalized potential can be understood as the ``most irreversible'' state observable. 
We also derive a thermodynamic uncertainty relation that allows for practical thermodynamic inference.

\global\long\def\VV{n}%
\global\long\def\NNrr{J_{\rr}^{(\VV)}}%
\global\long\def\NN{\bm{J}^{(\VV)}}%

\global\long\def\CGFee{\Lambda}%
\global\long\def\CGFpot{\Lambda_{\pot}}%
\global\long\def\timeinterval{dt}%

\begin{newstuffenv}
\subsection{Dynamical large deviations: excess EPR}
\label{sec:ldpexcessepr}

Imagine that one makes $\nnn$ independent measurements of a Markovian stochastic system over a short time interval $[t,t+dt]$. These copies may represent different particles, unit volumes in a chemical reactor, or trial runs of the same experiment~\cite{notehydro}. In these measurements, the number of times each reaction occurs will exhibit stochastic fluctuations. For each reaction $\rr$, let the random variable $\NNrr$ indicate the \emph{empirical flux}, the mean number of reaction events per copy and per unit time. 
The reactions may correspond  to transitions between microstates, as in a general MJP, or to macroscopic chemical reactions, as in a chemical master equation. 
We write $\NN=(J_{1}^{(\VV)},\dots,J_{\numedge}^{(\VV)})$ to indicate the vector of empirical fluxes across all reactions.

The empirical fluxes are said to obey a \emph{large deviations principle (LDP)} if, for large $\nnn$, their probability distribution scales as
\[
\ldpProb{\NN\approx\bm{g}}\asymp e^{-\nnn\,\Psi(\bm{g})\,dt}\,,
\]
where $\Psi(\bm{g})$ is the rate function. This may be termed a many-particle dynamical LDP, since it concerns fluctuations of a dynamical quantity (the fluxes) across independent copies.

In MJPs~\cite{maes_and_2008} and ideal CRNs~\cite{lazarescu_large_2019}, the rate function is known to be the generalized 
relative entropy $\DD$~\eqref{eq:klflux}. In particular,  the empirical flux LDP is
\begin{align}
\ldpProb{\NN\approx\bm{g}}\asymp e^{-\nnn \,\DD(\bm{g}\Vert\jj)\,dt}\,,
\label{eq:ddldp}
\end{align}
where $\jj=\mathbb{E}[\NN]$ are the mean fluxes. 
(See Lazarescu et al.~\cite{lazarescu_large_2019} for the derivation for well-mixed CRNs with elementary reactions and mass-action kinetics.) In the rest of this section, we assume that the rate function has this form. 

EPR has a simple interpretation in terms of statistical fluctuations. 
In limit of many copies, the empirical fluxes converge
to their expectation values, $\jj$. At a finite $\nnn$, however, there is a finite probability that the empirical fluxes move ``backward in time'' due to a statistical fluctuation, taking the value of the expected \emph{reverse} fluxes $\jjR$. 
Given  Eq.~\eqref{eq:ddldp}, the probability of this time-reversal scales as the EPR,
\begin{align}
\ldpProb{\NN\approx\jjR}\asymp e^{-\nnn\,\epr(\jj)\,dt}\,,
\label{eq:vv87}
\end{align}
which follows from $\DD(\jjR\Vert\jj)=\DD(
\jj\Vert\jjR)=\epr(\jj)$ (assuming no odd variables).  
Time-reversal is exponentially rare, except in equilibrium ($\epr=0$) when the system has no  direction of time. 
Eq.~\eqref{eq:vv87} gives a statistical interpretation of EPR, independent of thermodynamic quantities like heat and work. 

Instead of considering time-reversal of all fluxes, as in Eq.~\eqref{eq:vv87}, one may consider the time-reversal of net production, i.e., the state change due to reactions. The empirical net production is given by the random variable $\BBdiv\NN$, and the probability of its time-reversal is $\ldpProb{\BBdiv\NN\approx-\BBdiv\jj}\asymp e^{-\nnn \, \Psi^{\prime} dt}$.  
The rate function $\Psi^{\prime} $ may be computed using the ``contraction principle''~\cite[Sec.~2.3]{varadhan_large_2016}, which says that LDP governing a set of events is determined by the most likely event. In this case, the rate function is expressed via the optimization problem,
\begin{align*}
\Psi^{\prime}=\min_{\bm{g}\in\fluxspace}\DD(\bm{g}\Vert \jj)\suchthat\BBdiv\bm{g}=-\BBdiv\jj \,.
\end{align*}
Using a change of variables and antisymmetry~\eqref{eq:antisymmetry}, we see that this is equivalent to the variational expression~\eqref{eq:exdef} for excess EPR. Thus, for the empirical net production, the probability of time-reversal scales as the excess EPR,
\begin{align}
\ldpProb{\BBdiv\NN\approx-\BBdiv\jj}\asymp e^{-\nnn\,\eprEX(\jj)\,dt}\,.
\label{eq:ldpexcessEPR}
\end{align}
This provides a physically meaningful and operational 
interpretation of excess EPR in terms of statistical irreversibility.

It is illuminating to compare Eq.~\eqref{eq:ldpexcessEPR}, the large deviations expression for excess EPR, and Eq.~\eqref{eq:ldpcccnc}, the large deviations expression for the steady-state quasipotential (described in Section~\ref{subsec:ss}).  
The quasipotential governs occupation fluctuations in stationarity. It is useful for studying steady-state fluctuations and stability, including classic problems such as escape from a stochastic attractor. Our excess EPR, on the other hand, governs the dynamical fluctuations of local-in-time fluxes. As we will see, it is useful for studying the dynamical properties, including thermodynamic uncertainty relations and thermodynamic speed limits. 
As we will see in the next section, it is also possible to interpret our generalized potential in terms of dynamical fluctuations.
\end{newstuffenv}

\subsection{Dynamical large deviations: generalized potential}

In this section, we interpret our generalized potential $\potopt$ in terms of large deviations.

\global\long\def\potRVn{\overline{\Delta\phi}_{\nnn}}%
\global\long\def\potRV{\Delta\phi}%

Suppose that one is interested in a fluctuating state observable $\pot$ (e.g., energy, position, etc.) across $\nnn$ independent measurements over time $[t,t+dt]$. The empirical mean change of the observable is captured by the random variable
\begin{align}
\potRVn=\pot^\top \BBdiv \NN \,dt\,. %
\label{eq:rvphi}
\end{align}
Specifically, this is the empirical change due to internal reactions. In a closed system without external fluxes, $\potRVn$ is equal to the change over time.

\begin{figure}
\begin{centering}
\includegraphics[width=0.75\columnwidth]{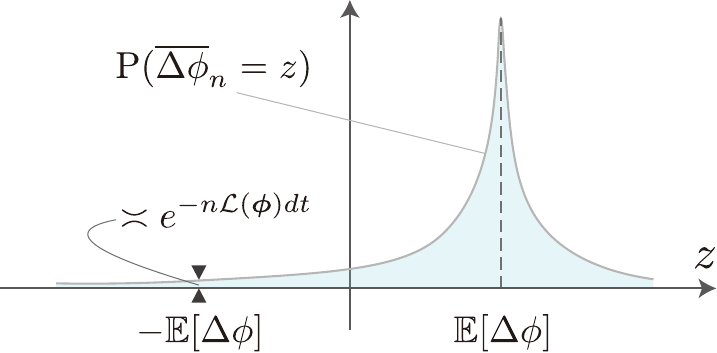}
\par\end{centering}
\caption{\textbf{Large deviations interpretation of the irreversibility measure
$\protect\cgfLeg(\protect\pot)$.} 
The change of state observable
$\protect\pot$ due to reactions over time $[t,t+dt]$ is measured in $\protect\nnn$ independent copies, and the empirical mean change 
is captured by the random variable $\potRVn$. Here we show schematically the probability of different outcomes %
$\potRVn=z$. 
For large $\protect\nnn$, the probability distribution
is peaked at its expectation value $z=\mathbb{E}[\protect\potRV]$. The probability
that the empirical mean moves in reverse, $z=-\mathbb{E}[\protect\potRV]$, decays exponentially in $\nnn$ as $\asymp e^{-n\protect\cgfLeg(\protect\pot)\,dt}$.
The generalized potential $\protect\potopt$ is the ``most irreversible'' observable, having the largest $\protect\cgfLeg(\protect\pot)$, and the excess EPR $\eprEX$ is its degree of irreversibility, see Eq.~\eqref{eq:eprLDP}.}\label{fig:ldp}
\end{figure}

In the limit of many copies, $\potRVn$ converges to $\mathbb{E}[\potRV]=\pot^\top \BBdiv\jj\, dt$.  
As above, we may consider the irreversibility of $\pot$ in terms of the probability that, due to a statistical fluctuation, $\potRVn$ moves ``backwards in time'' relative to the expectation. This probability can be expressed as
\begin{align}
\ldpProb{\potRVn \approx-\mathbb{E}[\potRV]}\asymp e^{-\nnn\cgfLeg(\pot)\,dt}\,.\label{eq:m2}
\end{align}
In Figure~\ref{fig:ldp}, we illustrate the meaning of the rate function $\cgfLeg(\pot)\,dt$, which is a fundamental measure of the irreversibility of $\pot$.  It is large for ``fast'' observables that change quickly relative to the scale of their dynamical fluctuations. 

$\cgfLeg(\pot)$ can be computed using the contraction principle, as 
$\min_{\bm{g}}\DD(\bm{g}\Vert \jj)$ subject to $\pot^\top\BBdiv\bm{g}=-\pot^\top \BBdiv\jj$. This optimization can be expressed in its dual form as
\begin{align}
\cgfLeg(\pot)=\max_{\lambda\in\mathbb{R}}\big[\lambda(-\jj^{\top}\BBgrad\pot)-\jj^{\top}(e^{\lambda\BBgrad\pot}-\oneone)\big].\label{eq:lcgf}
\end{align}
By comparing with our variational principle~\eqref{eq:exdual} and with a bit of rearranging, we may write the excess EPR as
\begin{align}
\eprEX=\max_{\potlag\in\potspace}\cgfLeg(\pot)=\cgfLeg(\potopt),\label{eq:eprLDP}
\end{align}
This shows that the generalized potential $\potopt$
is the ``most irreversible'' observable, and that the excess EPR
$\eprEX$ quantifies its degree of irreversibility.

\subsection{TUR and thermodynamic inference}

\label{subsec:TI}

Here we discuss our variational principle as a thermodynamic uncertainty relation (TUR), which relates fluctuation statistics of observables to excess EPR.  This TUR sets a fundamental bound on allowed statistics, limiting the ``speed'' of state observables  relative to the scale of their fluctuations. The TUR also provides a way to perform thermodynamic inference of excess EPR and the generalized potential from empirical measurements. 

Before proceeding, we remark that our large-deviation results, such as Eqs.~(\ref{eq:ldpexcessEPR},\ref{eq:m2},\ref{eq:eprLDP}), in theory provide a relationship between excess EPR, the generalized potential, and measurement statistics. In practice, however, such results are often not practical because they involve the statistics of exponentially rare events. Our TUR, on the other hand, does not depend on exponentially rare events.

To introduce the TUR, consider again the change of observable $\pot$ due to reactions over time interval $[t,t+dt]$. We use the random variable $\potRV$ to indicate this change \emph{in a single system} (it should be contrasted with the random variable $\potRVn$~\eqref{eq:rvphi}, the empirical mean change across $n$ system copies). 

The statistics of $\potRV$ are encoded in the cumulant generating function (CGF), $\CGFee_{\pot}(\lambda)=\ln\mathbb{E}e^{\lambda\potRV}$. 
As an example, in an MJP where $\potx$ is the position
of site $i$ on a one-dimensional lattice, $\potRV$ is the short-time displacement. The first derivative of the CGF gives the mean displacement, the second derivative gives the mean square displacement, etc. 

A key result in large deviations theory, called Cramér's theorem~\cite{varadhan_large_2016}, states that the large-deviations rate function of $\potRVn$ is the Legendre transform of the CGF of $\potRV$. Given Eq.~\eqref{eq:lcgf}, the rate function  $\cgfLeg(\pot) \,dt$   is the Legendre transform of  $\CGFee_{\pot}(\lambda)=\jj^{\top}(e^{\lambda\BBgrad\pot}-\oneone)\,dt$, where we assume Poissonian fluctuations and work to first order in $dt$.   
The cumulants are explicitly given by $K_{\pot}^{(k)}=\partial_{\lambda}^{(k)}\CGFee_{\pot}(0)\approx\jj^\top [\BBgrad\pot]^{k}\,dt$. 
For example, the mean and variance are 
\begin{align*}
 K_{\pot}^{(1)}&=\mathbb{E}[\potRV] = \pot^\top \BBdiv \jj \, dt
 \\
 K_{\pot}^{(2)}&=\mathrm{Var}(\potRV) = \pot^\top \BBdiv \diag(\jj) \BBgrad\pot \,dt  
\end{align*}
We note that the variance is equal to the mean-squared displacement (MSD) $\mathbb{E}[(\potRV)^2]$ to first order in $dt$, and that it can be expressed as $K_{\pot}^{(2)}=2\pot^\top \HH \pot\,dt$ in terms of our mobility matrix $\HH$~\eqref{eq:hsss}.  
In this sense, $\HH$ is a diffusion matrix that governs the short-time MSD of all state observables.

We can also express the variational principle for excess EPR in terms of cumulants. Expanding the exponentials in Eq.~\eqref{eq:exdual2} gives
\begin{align}
\eprEX=\max_{\pot\in\potspace}\,\frac{1}{dt}\Big[-2K_{\pot}^{(1)}\,-\sum_{k=2}\frac{1}{k!}K_{\pot}^{(k)}\Big].\label{eq:estEPR}
\end{align}
This provides a set of bounds on the excess EPR, one for each state observable $\pot$.  Each bound involves two terms: the first is proportional to the observable's ``speed'' (the mean displacement $-2K_{\pot}^{(1)}$), the second quantifies the size of its fluctuations (via a sum of higher cumulants). This fluctuation term is nonnegative, as discussed near Eq.~\eqref{eq:exdual2}. Note that the sum of higher cumulants converges, so in practice it may be truncated at some finite order.

Eq.~\eqref{eq:estEPR} can be interpreted as a short-time TUR~\cite{manikandan2020inferring} for excess EPR that specifies a tradeoff between speed and fluctuations. It is also an example of a ``higher-order'' TUR~\cite{dechant2020fluctuation},
since it involves not only the mean and variance but also higher cumulants.
Higher-order TURs can be made tight by appropriate
choice of observable, unlike traditional TURs (involving only the mean and variance) that cannot always be made tight in far-from-equilibrium discrete systems~\cite{otsubo2020estimating}. 
In \smref{tur}, we derive a weaker but simpler bound that involves simpler statistics.

\newstuff{The TUR~\eqref{eq:estEPR} permits thermodynamic inference of $\eprEX$ and $\potopt$, using similar techniques as for inference of EPR and thermodynamic forces from short-time trajectories~\cite{kim2020learning,otsubo2020estimating,otsubo2022estimating,aguilera2025inferring}.  Suppose that, in a closed system, one makes $\nnn$ two-point measurements of the change of a state observable during a short time interval, $\phi(t)\mapsto \phi(t+dt)$. These measurements can be used to compute the cumulants $K_{\pot}^{(k)}$, which give a lower bound on the excess EPR when plugged into the TUR~\eqref{eq:estEPR}. 
The TUR can be further tightened by scaling the observable as $\lambda\pot$ by $\lambda\in \mathbb{R}$ and optimizing over $\lambda$, in this way recovering the overall irreversibility $\cgfLeg(\pot)$~\eqref{eq:lcgf}.}

\newstuff{If it is possible to measure more than one observable ($\pot_{(1)}, \pot_{(2)}, ...$), tighter bounds can be found by numerical optimization of the TUR~\eqref{eq:estEPR} over linear combinations of these observables.  In the fine-grained regime where $\numstate$ linearly-independent observables can be acquired (e.g., such as identity of starting and ending microstates), then the measurements span the entire space of observables $\pot\in\potspace$ and numerical optimization should recover the full excess EPR and the generalized potential $\potopt$ at optimality. We note that the objective in TUR~\eqref{eq:estEPR} is a concave function of $\pot$, thus optimization can be carried out using very efficient numerical algorithms. We also note that, unlike some EPR estimation techniques, our method never needs to distinguish which reaction or reservoir mediated a given transition.}

\section{Thermodynamic speed limits}

\label{sec:tsl}

\global\long\def\taumin{T_{\min}}%
\global\long\def\atanh{\tanh^{-1}} 
\global\long\def\Act{{A}}
\global\long\def\chngObs{K_{\pot}^{(1)}} 
\global\long\def\actOBS{A_{\pot}}%
\global\long\def\Lobs{L_{\obs}}%
\global\long\def\avgAct{\langle{\Act}\rangle }%
\global\long\def\totAct{\mathcal{A}}%
\global\long\def\totEvol{\bm{V}} 
\global\long\def\actOBSpos{A^{+}_{\pot}}%
\global\long\def\actOBSneg{A^{-}_{\pot}}%
\global\long\def\wassLen{\mathscr{L}}
\global\long\def\wassTot{\mathscr{W}}

In this section, we derive thermodynamic speed limits (TSLs). We first introduce a short-time TSL, which relates excess EPR to the instantaneous
speed of evolution at a given time. We then introduce a finite-time TSL, which relates 
integrated excess EP to time and trajectory length. Because excess
EPR is always smaller than EPR, our TSLs also imply bounds on
the total entropy production. 

\subsection{Short-time speed limits}

\global\long\def\eprTSL{\sigma_{\text{TSL}}}%
\global\long\def\epTSL{\Sigma_{\text{TSL}}}%
\global\long\def\evoll{\BBdiv \jj}

We first derive a ``short-time'' TSL that relates excess EPR to two simple quantities. The first quantity is the \emph{activity}, the total number of reactions per unit time,
\begin{align}\Act:=\sum_{\rr}\jjrr=\left\Vert \jj\right\Vert _{1}\,.
\end{align}
\newstuff{The second quantity is the {net production} due to reactions, $\BBdiv \jj$. 
In a closed system without external fluxes, this is equal to the  state evolution, $\BBdiv \jj=\dtpp$. For an open system with external fluxes, it includes both the state evolution and the external fluxes, $\BBdiv \jj=\dtpp+\II$. Finally, for an open system in steady state, it is equal to the external fluxes, $\BBdiv \jj=\II$.} 

\begin{newstuffenv}
Our TSL quantifies the minimal excess EPR compatible with a given net production $\evoll$ and activity $\Act$, 
\begin{equation}
\eprTSL:=\min_{\bm{g}\in\fluxspace}\eprEX(\bm{g})\!\suchthat\! \BBdiv\bm{g}=\evoll,\left\Vert \bm{g}\right\Vert _{1}=\Act\label{eq:tsldef}
\end{equation}
$\eprTSL$ is a function of the $\evoll$ and $\Act$, though we usually leave this implicit in our notation. 
Also note that without the activity constraint, the minimum would be zero, since one can increase forward and reverse fluxes while keeping their difference fixed. 
In \smref{tsl}, we derive the form of the optimal fluxes in Eq.~\eqref{eq:tsldef} and show that they are conservative.

In addition, in that SM we show that our TSL is closely related to the \emph{(1-)Wasserstein speed} from optimal transport~\cite{dechant2022minimum,van2022thermodynamic,nagayama2025infinite}. 
The Wasserstein
speed quantifies the minimum activity required to achieve net production $\evoll$ in a system with stoichiometric
matrix $\BBbase$,
\begin{align}
\dot{W}:=\min_{\bm{a}\in\fluxspace}\left\Vert \bm{a}\right\Vert _{1}\suchthat\BBdiv\bm{a}=\evoll\,.\label{eq:wassdef}
\end{align}
Importantly, $\dot W$ is sensitive
to the system's network topology, as encoded in the matrix $\BBbase$. For example, the same net production requires a faster Wasserstein speed
on a 1-dimensional chain
than on a fully connected network. In the SM, we show that $\dot{W}$ may be efficiently computed using its dual form.

Our TSL can be expressed in terms of $\dot{W}$ as
\begin{align}
 \eprTSL = 2\dot{W} \,\atanh\frac{\dot{W}}{\Act}\,.\label{eq:tslW0} 
\end{align}
The right side depends on the ratio between the minimum required activity ($\dot{W}$) and the actual activity ($\Act$). The bound diverges as $\dot{W}\to \Act$, which corresponds to the limit where all activity is channeled into production. We note that this limit can only be achieved by making reactions absolutely irreversible ($\jjrr>0,\jjRrr=0$), since reversibility increases activity without contributing to net production. 

To summarize, we have the following hierarchy of bounds:
\begin{align}
\epr\ge\eprEX\ge\eprTSL\ge\frac{2\dot{W}^{2}}{\Act}\,.\label{eq:hierbounds}
\end{align}
The last quadratic bound follows from Eq.~\eqref{eq:tslW0} and $x\atanh x\ge x^2$. It is tight in the slow-production limit ($\dot{W}\ll\Act$), but it is much weaker than Eq.~\eqref{eq:tslW0} for fast production, since it does not diverge when $\dot{W}\to  \Act$. 
In principle, all of these bounds in Eq.~\eqref{eq:hierbounds} are achievable.

By inverting Eq.~\eqref{eq:tslW0}, we may write our TSL in a more familiar ``speed limit'' form, as an upper bound on the speed as a function of excess EPR and activity:
\begin{align}
 \dot{W} = \Act\,\Phi^{-1}({\eprTSL}/{2 \Act}) \le \Act\,\Phi^{-1}({\eprEX}/{2 \Act})
\end{align}
where $\Phi^{-1}$ is the inverse function of $\Phi(x):=x\atanh x$.

Finally, we remark on the case of steady-state open systems, since these play an important role in many biological and chemical studies. For an open system in steady state, net production balances the external fluxes, $\evoll=\II$, and our TSL~\eqref{eq:tsldef} identifies the steady-state fluxes that minimize excess EPR under an activity constraint. This has some similarity to ``flux-balance analysis'' (FBA)~\cite{orth2010flux}, a biological modeling technique which identifies steady-state metabolic fluxes that optimize some (typically linear) objective under constraints. (See Refs.~\cite{fleming2012variational,kschischo2010gentle,pinero2024optimization} for connections between FBA and nonequilibrium thermodynamics). From a practical perspective, our TSL bounds the dissipation using three accessible pieces of information: the stoichiometry $\BBbase$, the external fluxes $\II$, and the overall activity $\Act$. We demonstrate the utility of this bound when we analyze metabolic networks in 
Section~\ref{sec:metabolic-example}.

\subsection{Relation to other TSLs}

Previous work~\cite{dechant2019thermodynamic,van2022thermodynamic,nagayama2025infinite} has considered the analogue of $\eprTSL$ but for total EPR,
\begin{equation}
\eprTSL^\prime :=\min_{\bm{g}\in\fluxspace}\epr(\bm{g})\suchthat \BBdiv\bm{g}=\evoll,\left\Vert \bm{g}\right\Vert _{1}=\Act\,.\label{eq:tsldef-ep}
\end{equation}
However, as mentioned above, the optimal fluxes in Eq.~\eqref{eq:tsldef} are conservative, which implies that the two TSLs are equivalent: $\eprTSL^\prime=\eprTSL$. This also shows that the activity TSL~\eqref{eq:tsldef-ep} only recovers the excess component of total EPR. We note that the Wasserstein expression~\eqref{eq:tslW0} for $\eprTSL^\prime$ was derived for MJPs in Ref.~\cite{dechant2019thermodynamic,van2022thermodynamic} and CRNs in Ref.~\cite{nagayama2025infinite}.

For MJPs, the Wasserstein speed can be bounded as $\dot{W}\ge \Vert \dtppp\Vert_{1}/2$. The quantity $\Vert \dtppp\Vert _{1}/2$, called the \emph{total variation (TV) speed}, is a simple and optimization-free quantity which is not sensitive to system topology~\citep{dechant2022minimum}. Our results imply the bounds
\begin{align}
\eprEX\ge\Vert \dtppp\Vert _{1}\,\atanh\frac{\Vert \dtppp\Vert _{1}}{2\Act} \ge\frac{\Vert \dtppp\Vert _{1}^{2}}{2\Act}\,. \label{eq:tvbounds}
\end{align}
The last bound recalls a quadratic TSL for HS excess EPR~\citep{van2020unifiedCSLTUR},
\begin{align}
\eprEXhs\ge\frac{\Vert \dtppp\Vert _{1}^{2}}{2\Act}\,.
\end{align}

HS excess EPR does not obey the stronger TV bound, $\eprEXhs\not\ge\Vert \dtppp\Vert _{1}\,\atanh({\Vert \dtppp\Vert _{1}}/{2\Act})$, although some intermediate inequalities have been shown~\citep{delvenne2021thermo,lee2022speed}.  It also does not obey the analogue of the Wasserstein TSL
\eqref{eq:tslW0}, as will be seen in the unicyclic system considered
in Section~\ref{sec:Examples}~\footnote{A slightly weaker version of the Wasserstein TSL~\eqref{eq:tslW0} was recently
proposed for HS excess EPR in Ref.~\citep{delvenne2021thermo}. Unfortunately,
the bound proposed in that work is not always valid; a counter-example may be found using unicyclic
system from Section~\ref{subsec:Unicyclic-MJP}.}.
\end{newstuffenv}

\subsection{Finite-time speed limits}

\begin{newstuffenv}
In this section, we derive ``finite-time'' TSLs for the excess EP incurred by a time-extended process. 

Recall that the excess EP incurred during time $t\in [0,\ft]$ is given by $\epEX=\int_{0}^{\ft}\eprEX(t)\,dt$. We define our finite-time TSL as the minimal excess EP required by any trajectory with time-dependent net production $\evoll(t)$ and activity $\Act(t)=\Vert \jj(t)\Vert_1$. Formally, we optimize over all time-dependent fluxes compatible with these constraints:
\begin{align}
\epTSL &:= \min_{\bm{g}(t)%
} \int_0^\ft \eprEX(\bm{g}(t)) \,dt\label{eq:tottsl}\\
&\quad\suchthat \BBdiv \bm{g}(t) = \evoll(t), \Vert \bm{g}(t)\Vert_1=\Act(t) \;\forall t \nonumber
\end{align}
The minimization decouples over different time points, allowing us to express it as an integral over the short-time TSL:
\begin{align}
 \epTSL = \int_0^\ft \eprTSL(\evoll(t),\Act(t)) \,dt\,.
\end{align}
As mentioned, $\eprTSL$ can be equivalently defined as a minimization of total EPR (rather than excess EPR), thus Eq.~\eqref{eq:tottsl} can also be equivalently defined as a minimization of total EP.

Using Eq.~\eqref{eq:tslW0}, we may relate our TSL to the Wasserstein distance from optimal transport: 
\begin{align}
\epTSL = \int_0^\ft 2\dot{W}(t) \atanh \frac{\dot{W}(t)}{\Act(t)} \,dt\,,
\label{eq:tslEPtot}
\end{align}
where $\dot{W}(t)$ is the Wasserstein speed~\eqref{eq:wassdef} corresponding to net production $\BBdiv \jj(t)$. The bound $\epEX\ge \epTSL$ is achieved when the fluxes are chosen to be optimal at each time point $t\in[0,\ft]$, as described in \smref{tsl}. 

We can derive another TSL for the time-integrated activity and Wasserstein length:
\begin{align} 
 \totAct:=\int_{0}^{\ft}\Act(t)\,dt \qquad \wassLen :=\int_{0}^{\ft}\dot{W}(t)\,dt
\end{align}
Using this definition, we may lower bound $\epTSL$ as 
\begin{align}
\epTSL \ge \epTSL^\wassLen := 2 \wassLen\, \atanh \frac{\wassLen}{\totAct} \,.
\label{eq:tslwassLenFT}
\end{align}
To derive this, we multiply and divide the term inside the integral in Eq.~\eqref{eq:tslEPtot} by $\Act(t)/\totAct$. We then apply Jensen's inequality to the convex function $x\atanh x$ with $x=\dot{W}/\Act$. 

From a geometric perspective, the Wasserstein length $\wassLen$ is the length of a given trajectory through concentration space. The Wasserstein length also has an operational interpretation: it is the minimal integrated activity (total number of reaction events) needed to implement the trajectory with a given stoichiometric matrix.

The above inequalities require detailed information about the system's trajectory. Our final result is another TSL that does not depend on such detailed information. To derive it, we introduce the integrated net production:
\begin{align}
 \bm{V}:=\int_{0}^{\ft}\evoll(t)\,dt%
 \,.
\end{align}
For a closed system, the integrated net production is the change of the state, $\bm{V}=\pp(T)-\pp(0)$. For an open system, it accounts for the integrated outflow as well as the change of state, $\bm{V}=\pp(T)-\pp(0)+\int_0^T \II(t) \,dt$. Finally, for an open system in steady state, or (more generally) a cyclic process $\pp(T)=\pp(0)$, the integrated net production is the integrated outflow: $\bm{V}=\int_0^T \II(t) \,dt$. In all cases, $\bm{V}$ depends only on the initial and final state and the integrated flows. 

The Wasserstein distance associated with integrated net production is the integrated-time analogue of Eq.~\eqref{eq:wassdef}:
\begin{align}
\wassTot:=\min_{\bm{a}\in\fluxspace}\left\Vert \bm{a}\right\Vert _{1}\suchthat\BBdiv\bm{a}=\bm{V}\,.\label{eq:wassdefInt}
\end{align}
For a closed system, $\wassTot$ is the geodesic distance in Wasserstein space between $\pp(0)$ and $\pp(T)$. More generally, using the triangle inequality, we may show $\wassTot \le \wassLen$, with equality for constant-speed trajectories. This leads to our final TSL:
\begin{align}
\epTSL^\wassLen \ge \epTSL^\wassTot := 2 \wassTot\, \atanh \frac{\wassTot}{\totAct} \,.
\label{eq:finiteTSLW0}
\end{align}
Summarizing, we have the sequence of bounds
\begin{align}
\ep \ge \epEX \ge \epTSL \ge \epTSL^\wassLen \ge \epTSL^\wassTot\,. 
\end{align}\end{newstuffenv}

We may also derive a more familiar form of the TSL, as a bound on
the minimal time required to traverse a given distance. Consider the time-averaged dynamical activity, defined as the number of reactions
per time, $\avgAct:=\totAct/T$. Rearranging the results above gives
\begin{align} 
\ft & \ge\frac{\wassLen}{\avgAct}\coth\frac{\epEX}{2\wassLen}\ge\frac{\wassTot}{\avgAct}\coth\frac{\epEX}{2 \wassTot}.\label{eq:timelimit}
\end{align}
Since $\coth(x)\ge1$, there is a finite minimal time needed to undergo a trajectory
of a given length,
\[
T\ge\wassLen/\avgAct\ge \wassTot/\avgAct.
\]
This latter bound is relevant in the highly irreversible limit where $\epEX$ diverges. Conversely, these TSLs imply that $\epEX$ diverges as $-\ln(\ft-\wassLen/\avgAct)$
as $\ft\to\wassLen/\avgAct$. This is stronger than the
$1/\ft$ scaling reported in conventional TSLs~\citep{aurell2011optimal,shiraishi_speed_2018,van2021geometrical,nakazato2021geometrical,yoshimura2021thermodynamic,hamazaki2022speed,kohei2022},
which is only tight in the limit of slow production~\citep{berut2012experimental,zhen2021universal,zhen2022inverse}.

\subsection{Information-theoretic speed limit for relaxing MJPs}

\label{subsec:shiraishi}

\global\long\def\qqx#1{q_{#1}}
\global\long\def\qq{\bm{q}}

Our final result is a finite-time TSL that relates excess EP and the
speed of relaxation. This result is different from the TSLs
derived above: it only applies to MJPs, it only applies to time-symmetric processes, and it does not reference Wasserstein distance. Nonetheless, it shows the connection between our approach and previous literature~\cite{shiraishi2019information}, and 
it provides a useful bound based on an interpretable
and empirically-accessible information-theoretic notion of distance.

Consider an MJP that undergoes a process over $t\in[0,\ft]$, during
which the system's probability distribution goes from $\ppp(0)$ to 
$\ppp(\ft)$. Suppose that the process involves an autonomous relaxation
(time-independent control parameters), or more generally that the driving
is time-symmetric, such that the control parameters at time $t$ are the
same as at time $\ft-t$. Given these assumptions, we show in \smref{sm3}
that the variational principle for excess EPR~\eqref{eq:exdual} can
be expressed as
\begin{align}
\eprEX=\max_{\qq}\,\Big[-\frac{d}{dt} D(\ppp(t)\Vert\qq(-t)) \Big].\label{eq:shi-gen-1}
\end{align}
where the maximization is over all probability distributions over
the microstates. The notation $\qq(-t)$ indicates that $\qq$ evolves
``backwards in time'', 
\begin{align}
-\frac{d}{dt}\qqx{\xx}(-t)=\sum_{\yy(\ne\xx),\bath}(\qqx{\yy}(-t)\Rija-\qqx{\xx}(-t)\Rjia).\label{eq:gdd2-1}
\end{align}
Thus, $\eprEX$ is the fastest rate
of contraction of relative entropy between the actual distribution 
$\ppp$ evolving forward-in-time and any other distribution evolving
backward-in-time. The optimum in Eq.~\eqref{eq:shi-gen-1} is achieved
by the pseudo-canonical distribution $\ppp^{\star}\propto\ppp\circ e^{-\potopt}$ 
defined by the generalized potential $\potopt$. 

Eq.~\eqref{eq:shi-gen-1} gives the following information-theoretic
bound, 
\begin{align}
\epEX(\ft)\ge D(\ppp(0)\Vert\ppp(\ft)),\label{eq:info-relax-bound}
\end{align}
The relative entropy $D(\ppp(0)\Vert\ppp(\ft))$ is an information-theoretic
measure of the distance traversed by the system's state over time $t\in[0,\ft]$.
It is interpretable and practically accessible by measuring the state
at two timepoints.

Eq.~\eqref{eq:shi-gen-1} and~\eqref{eq:info-relax-bound} generalize
the main results of Ref.~\citep{shiraishi2019information}, which
derived the same expressions for EPR in conservative MJPs. In particular, the derivation
of Eq.~\eqref{eq:info-relax-bound} proceeds in the same way as~\citep[Eq.~(3),][]{shiraishi2019information}.
Specifically, we choose $\qq(0)=\ppp(\ft)$ in Eq.~\eqref{eq:shi-gen-1} and then
integrate over $t\in[0,\ft/2]$. Under the assumption of time-symmetric
driving, $\qq(t)=\ppp(\ft-t)$ is a solution to Eq.~\eqref{eq:gdd2-1},
so the time integral gives
\begin{align*}
\epEX(\ft/2) & \ge-\int_{0}^{\ft/2} \frac{d}{dt} D[\ppp(t)\Vert\ppp(\ft-t)]\,dt\\
 & =D[\ppp(0)\Vert\ppp(\ft)]-
D[\ppp(\ft/2)\Vert\ppp(\ft/2)] \\
&=D[\ppp(0)\Vert\ppp(\ft)].
\end{align*}
Since %
$\eprEX(t)\ge0$ for $t\in [\ft/2,\ft]$, we have the sequence of bounds:
\[
\epEX(\ft)\ge\epEX(\ft/2)\ge D[\ppp(0)\Vert\ppp(\ft)].
\]

A bound like Eq.~\eqref{eq:info-relax-bound} was conjectured for HS excess EP in Ref.~\citep{gu2023speed}. However, that conjecture does not hold, as we will see in the unicyclic system considered in the next section.

\global\long\def\driving{\gamma}%
\begin{figure*}[!t]
\includegraphics[width=1\textwidth]{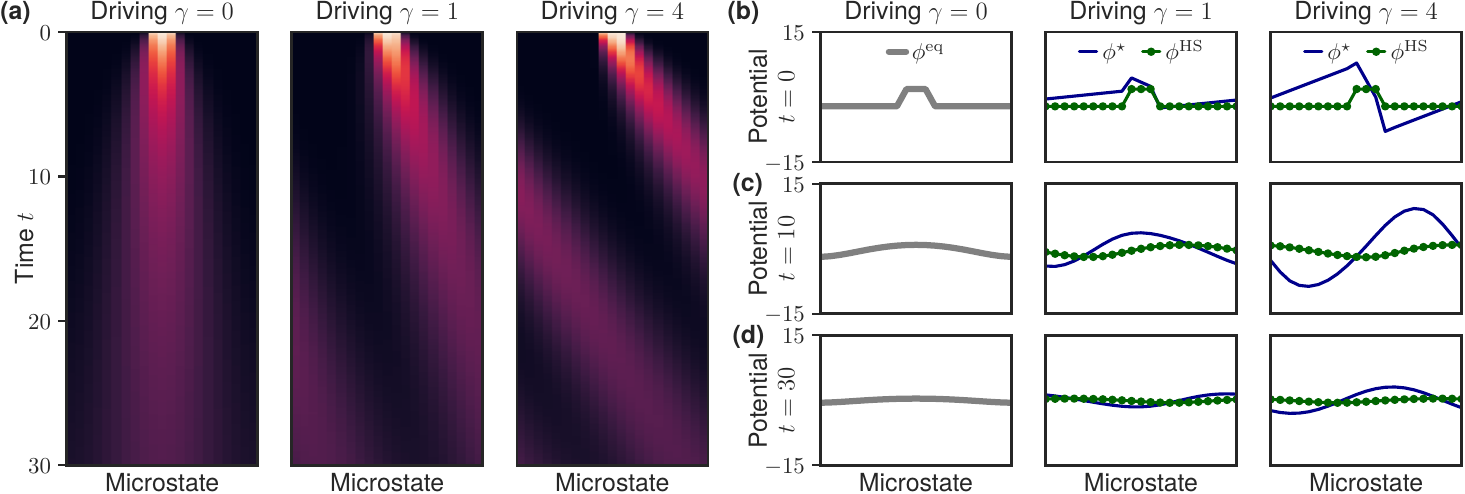}\caption{
\newstuff{
\textbf{Time evolution and generalized potentials for unicyclic MJP}. \textbf{(a)}: Time evolution of the probability distribution
for three driving strengths ($\protect\driving=0,1,4$). The system has 21 microstates, and the initial
distribution is concentrated on three microstates $i\in\{10,11,12\}$.
\textbf{(b)} Our generalized potential $\protect\potopt$ (blue) and steady-state
potential $\protect\potss$ (dotted green) for different driving strengths, evaluated
at $t=0$. 
\textbf{(c)}-\textbf{(d)} Generalized free energy $\protect\potopt$ and steady-state
potential $\protect\potss$ for different driving strengths, now evaluated at $t=10$ and $t=30$. Both potentials vanish as the system approaches the uniform steady state.
}}\label{fig:cyclicA}
\end{figure*}

\section{Examples}

\label{sec:Examples}

We now illustrate our approach on two examples: a unicyclic MJP and a nonlinear CRN (Brusselator).

\subsection{Unicyclic MJP}

\label{subsec:Unicyclic-MJP}

In our first example, we consider a uniform cyclic MJP. This example
will be useful to illustrate the difference between our
decomposition and the HS one.

Our MJP has $d=21$ microstates, and its transition rates are parameterized as 
\begin{align}
R_{i+1,i}=\frac{1}{1+e^{-\driving}},\quad R_{i-1,i}=\frac{e^{-\driving}}{1+e^{-\driving}},\label{eq:ratematrix}
\end{align}
where the indexing of microstates $i$ is taken mod $\numstate$.
The parameter $\driving$ determines the strength of nonconservative
driving around the cycle. The overall timescale is normalized, so
that the escape rates do not depend on $\gamma$ ($-R_{ii}=1$ for
all $i$). Therefore, the dynamical activity is $\Act=1$, regardless of the
$\gamma$ or the distribution $\ppp$. 
As the initial probability distribution $\ppp(0)$, we assign elevated probability to three
microstates, $\pppx{10}(0)=\pppx{11}(0)=\pppx{12}(0)=0.3$,
with the remaining probability split among the other 18 microstates, $\pppx{i}(0)=0.1/18$.

Figure~\ref{fig:cyclicA}(a) shows the time evolution of the
system's state for three values of the driving strength: $\driving=0$, $\driving=1$, and $\driving=4$. For $\driving=0$,
the system is conservative and it relaxes to equilibrium by diffusing symmetrically.
For $\driving=1$ and $\driving=4$, clockwise/counter-clockwise symmetry is broken and the system exhibits a decaying
oscillation around the cycle in the direction $i\to i+1$.

Figure~\ref{fig:cyclicA}(b) shows the steady-state potential
$\potss=\ln(\ppp/\pppss)$ and our generalized potential
potential $\potopt$ (found by numerical optimization) at the initial state $\ppp(0)$. For the conservative system with
$\driving=0$, the potentials are equal to each other and to the standard nonequilibrium free energy, $\potopt=\potss=\poteq=\ln(\ppp/\pppeq)$.
For stronger driving, $\potopt$ becomes larger in magnitude, more asymmetric, and increasingly 
``cliff-like'', reflecting the system's increasingly fast and asymmetric relaxation dynamics. On the other hand, because the initial state is symmetric and the steady state is uniform for all $\driving$, the steady-state potential 
$\potss$ is always symmetric and does not vary with $\driving$. 

\newstuff{Figure~\ref{fig:cyclicA}(c)-(d) shows the steady-state potential $\potss$
 and our generalized potential $\potopt$ for the same three driving strengths, but now evaluated on the time-evolved states $\ppp(t)$ at $t=10$ and $t=30$. Both potentials become smoother over time, gradually approaching the all-zero vector as the system approaches the uniform steady-state distribution. 
Our generalized potential maintains the asymmetry that reflects the asymmetry of relaxation dynamics, and its magnitude reflects the strength of driving. 
On the other hand, the steady-state potential remains essentially symmetric about its maximum, similar to the time-evolved distribution $\ppp(t)$. It has some dependence on $\driving$ due to the fact that different $\driving$ leads to different time-evolved states $\ppp(t)$, but this dependence is weak. 
}

\begin{figure*}[!t]
\includegraphics[width=1\textwidth]{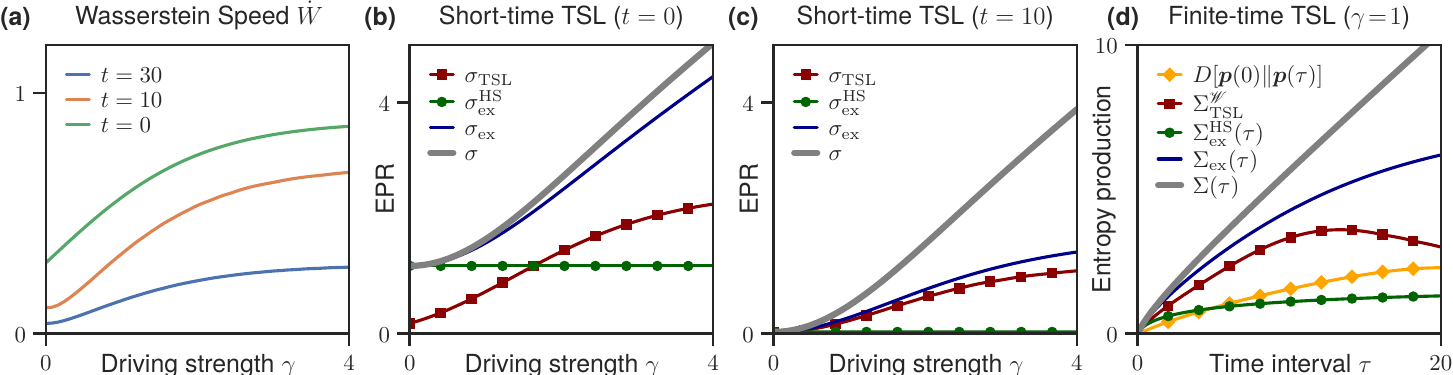}\caption{
\newstuff{
\textbf{Wasserstein TSL on unicyclic MJP.} \textbf{(a)} Wasserstein speed increases
with driving strength when evaluated on state $\ppp(t)$ at $t\in \{0,10,30\}$. \textbf{(b)} EPR, our excess EPR and HS excess EPR for the initial state $\ppp(0)$, as a function of the driving strength. The Wasserstein short-time
TSL $\eprTSL$~\eqref{eq:tslW0} provides a lower bound on our excess EPR. 
The HS excess EPR and steady-state
potential are not sensitive to the driving strength $\protect\driving$. 
\textbf{(c)} Same as in (b), but now shown for the time-evolved state $\ppp(t)$ at $t=10$. 
\textbf{(d)} Time-integrated EP, excess
EP, and HS excess EP as a function of the time interval, considering the spatiotemporal trajectory shown in Figure~\ref{fig:cyclicA}(a) for $\driving=1$. 
We also show the finite-time Wasserstein
TSL $\epTSL^{\mathscr{W}}$~\eqref{eq:finiteTSLW0} and information-theoretic bound~\eqref{eq:info-relax-bound}. Observe that the Wasserstein and information-theoretic
bounds only hold for our excess EP, not the HS excess EP.
}}\label{fig:cyclicB}
\end{figure*}

\newstuff{Figure~\ref{fig:cyclicB} illustrates our thermodynamic speed limit (TSL). Figure~\ref{fig:cyclicB}(a) shows the Wasserstein speed $\dot{W}$~\eqref{eq:wassdef} as a function of the driving strength, evaluated
at the initial state $\ppp(0)$ and the time-evolved state $\ppp(t)$ at $t=10$ and $t=30$. At all times, the Wasserstein speed increases with driving
strength. Thus, the state evolves faster when it undergoes
a directed force around the cycle, compared to when it undergoes
an undriven symmetric diffusion.
}

\newstuff{Figure~\ref{fig:cyclicB}(b) illustrates our short-time Wasserstein TSL on the initial state $\ppp(0)$. This shows the EPR, our excess EPR, the HS excess EPR, and the TSL lower bound $\eprTSL=2\dot{W}\atanh({\dot{W}}/{\Act})$~\eqref{eq:tslW0} as a function of the driving strength $\driving$. 
The EPR, 
our excess EPR, and the Wasserstein TSL increase with stronger driving. However, the HS excess
EPR remains constant, being insensitive to the driving strength. We verify the inequality $\eprEX\ge\eprEXhs$ from Eq.~\eqref{eq:hscomparison}. We also verify that the Wasserstein TSL holds for our excess EPR, but not the HS excess EPR.
 Figure~\ref{fig:cyclicB}(c) shows that same quantities, but now evaluated on the time-evolved state $\ppp(t)$ at $t=10$. Observe that by time $t=10$, HS excess EPR has almost entirely vanished. Our excess EPR and TSL bound also decrease over time, but not as quickly. The short-time Wasserstein TSL is nearly tight at $t=10$ across a range of driving strengths.

To summarize, Figure~\ref{fig:cyclicA}(b) and Figure~\ref{fig:cyclicB}(b) show that our generalized potential and excess
EPR are sensitive to the direction and speed of evolution of the system's
state. On the other hand, in this system, the steady-state potential and HS excess EPR are not sensitive to the strength of driving or speed of evolution.

Finally, Figure~\ref{fig:cyclicB}(d) illustrates our finite-time TSLs. Given the spatiotemporal trajectory
 from Figure~\ref{fig:cyclicA}(a) for $\driving=1$, we plot the time-integrated EP, $\ep(\tau)=\int_{0}^{\tau}\epr(t)\,dt$, as a function of the
time interval $\tau$, and similarly for $\epEX(\tau)$ and $\epEXHS(\tau)$. 
 We verify 
the bounds $\ep(\tau)\ge\epEX(\tau)\ge\epEXHS(\tau)$, the information-theoretic bound $\epEX(\tau)\ge D[\ppp(0)\Vert\ppp(\tau)]$, and the Wasserstein TSL $\epTSL^\wassTot$~\eqref{eq:finiteTSLW0}. The Wasserstein TSL captures most of the excess EPR until $\tau \approx 10$, at which point it begins to decrease. 
We observe that the Wasserstein and information-theoretic bounds do not hold for HS
excess EPR.} 

\subsection{Brusselator CRN}

\label{sec:brusselator}

\global\long\def\Tcyc{T_{\text{cyc}}}%
\global\long\def\kTwoVal{17}%
\global\long\def\kThreeMin{6}%
\global\long\def\kThreeMax{11}%
\global\long\def\kThreeChoices{\{\kThreeMin,9,\kThreeMax\}}%
\global\long\def\kThreeDecomp{8}%
\global\long\def\kThree{k_{3}^{+}}%

\begin{figure*}[t]
\includegraphics[width=1\textwidth]{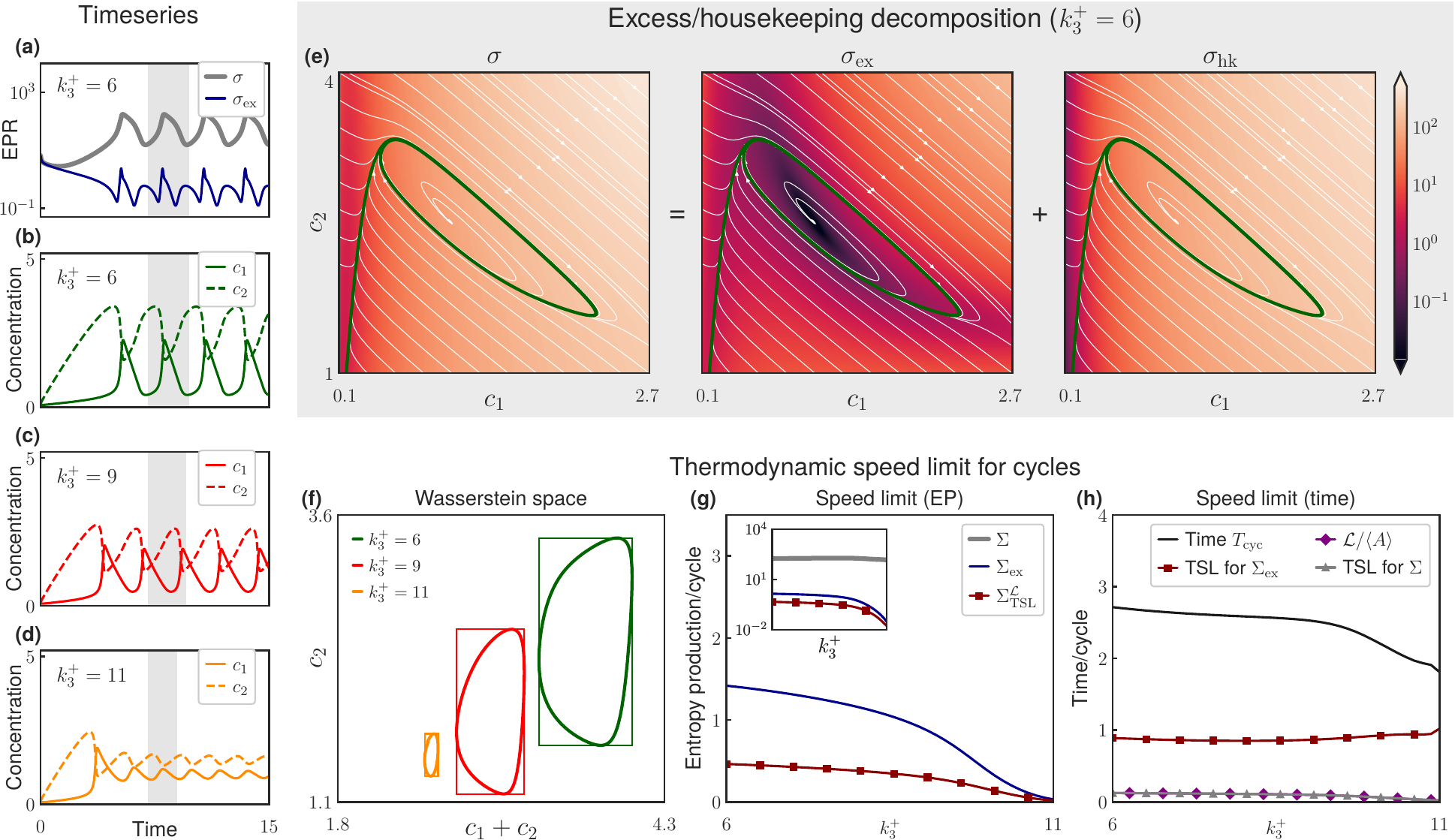}\caption{\textbf{Brusselator example} (see also Figure~\ref{fig:crn}). \textbf{(a)} Timeseries of $\protect\epr$
and $\protect\eprEX$ during a trajectory ($k_3^+=6$, $k_{1}^{+}=k_{1}^{-}=k_{2}^{-}=k_{3}^{-}=1,k_{2}^{+}=\protect\kTwoVal,\protect\kThree=\protect\kThreeMin$). \textbf{(b)-(d)} Evolution of concentrations $\cccx{1}(t),\cccx{2}(t)$ for $k_{3}^{+}\in\protect\kThreeChoices$
(other parameters as above). The length of a single oscillator cycle is marked in gray. Excess EPR tends to be larger during faster evolution. \textbf{(e)}
 $\protect\epr$ decomposed into $\protect\eprEX$ and
 $\protect\eprHK$ as a function of concentrations, parameters as in \emph{(a)}. Streamlines (white) and actual
trajectory (green) are overlaid. The excess EPR is sensitive to dynamical
features, such as the limit cycles and fixed points. \textbf{(f)} %
The cycles from
\emph{(b)-(d)} are shown in coordinates $(\cccx{1}+\cccx{2},\cccx{2})$ in
Wasserstein space. %
The cycle Wasserstein length
$\wassLen$ is the perimeter length
of the bounding boxes (see text for details). \textbf{(g)} The Wasserstein TSL $\epTSL^{\wassLen}$~\eqref{eq:tslwassLenFT} bounds EP and excess EP. Inset shows semi-logarithmic scale. \textbf{(h)} Cycle time is bound by the time TSLs~\eqref{eq:bruss-tsl2}.}\label{fig:bruss-new}
\end{figure*}

In our second example, we consider the Brusselator~\citep{prigogine1968symmetry},
a well-known CRN model of a chemical oscillator, shown above in Figure~\ref{fig:crn}.

To analyze this system, we use the fact that two of the reactions,
$X_{1}\rightleftharpoons X_{2}$ and $3X_{1}\rightleftharpoons2X_{1}+X_{2}$,
have the same stoichiometry. We then introduce the reaction-level
coarse-graining described in Section~\ref{subsec:cg}, giving 
the following coarse-grained stoichiometric matrix and forward/reverse
fluxes:
\[
\overline{\BBbase}\!=\!\!\left[\begin{array}{cc}
\vspace{-10pt}\\
\vspace{10pt}1 & 0\\
\vspace{10pt}\!{-1} & 0\\
\vspace{10pt}\!{-1} & 1\\
\vspace{5pt}1 & \!\!{-1}\!
\end{array}\right]\;\,\bar{\jj}\!=\!\!\left[\begin{array}{c}
\vspace{-10pt}\\
\vspace{10pt}{\scriptstyle k_{1}^{+}}\\
\vspace{10pt}{\scriptstyle \cccx {1}k_{1}^{-}}\\
\vspace{10pt}{\scriptstyle \cccx 1(k_{2}^{+}+\cccx 1^{2}k_{3}^{-})}\\
\vspace{5pt}{\scriptstyle \cccx 2(k_{2}^{-}+\cccx 1^{2}k_{3}^{+})}
\end{array}\right]\;\,\bar{\jjR}\!=\!\!\left[\begin{array}{c}
\vspace{-10pt}\\
\vspace{10pt}{\scriptstyle \cccx {1}k_{1}^{-}}\\
\vspace{10pt}{\scriptstyle k_{1}^{+}}\\
\vspace{10pt}{\scriptstyle \cccx 2(k_{2}^{-}+\cccx 1^{2}k_{3}^{+})}\\
\vspace{5pt}{\scriptstyle \cccx 1(k_{2}^{+}+\cccx 1^{2}k_{3}^{-})}
\end{array}\right]
\]
The resulting coarse-grained forces are conservative, 
\[
\bar{\ff}=\left(\;
\ln\frac{\bar{j}_{1}}{\bar{j}_{2}},\;\; \ln\frac{\bar{j}_{2}}{\bar{j}_{1}}, \;\; \ln\frac{\bar{j}_{3}}{\bar{j}_{4}}, \;\; \ln\frac{\bar{j}_{4}}{\bar{j}_{3}}\;\right)=-\overline{\BBgrad}\potopt,
\]
where generalized potential is
\begin{align}
\potopt & =\left(\ln\frac{\bar{j}_{2}}{\bar{j}_{1}},\;\;\ln\frac{\bar{j}_{2}}{\bar{j}_{1}}+\ln\frac{\bar{j}_{4}}{\bar{j}_{3}}\right)\\
 & =\left(\ln\frac{\cccx 1k_{1}^{-}}{k_{1}^{+}},\;\;\ln\frac{\cccx 1k_{1}^{-}}{k_{1}^{+}}+\ln\frac{\cccx 2(k_{2}^{-}+\cccx 1^{2}k_{3}^{+})}{\cccx 1(k_{2}^{+}+\cccx 1^{2}k_{3}^{-})}\right).
\end{align}
Since the coarse-fluxes are conservative, the excess EPR is 
 the EPR of the coarse-grained fluxes (see Section~\ref{subsec:cg}):
\begin{align}
\eprEX & =\epr(\bar{\jj})\,.%
\label{eq:brussdecomp}
\end{align}
The housekeeping EPR is the remainder, $\eprHK =\epr(\jj)-\epr(\bar{\jj})$.

\begin{newstuffenv}
The excess/housekeeping decomposition can also be performed at the level of individual reactions, as described above in Section~\ref{sec:ex-vs-hk}. For the Brusselator, the reactions $\varnothing \rightleftharpoons X_1$ are not coarse-grained, and one can verify that their forces obey $\ffrr = -[\BBgrad \potopt]_\rr$. Thus, according to Eqs.~\eqref{eq:ffex}-\eqref{eq:jjex}, their fluxes and forces are purely excess, with no housekeeping contribution. 
On the other hand, the reactions $X_{1}\rightleftharpoons X_{2}$ and $3X_{1}\rightleftharpoons2X_{1}+X_{2}$ are coarse-grained, so they have $\ffrr \ne -[\BBgrad \potopt]_\rr$ and exhibit nonvanishing housekeeping forces and fluxes.  
Their housekeeping fluxes quantify the flow around the futile cycle $X_{1}\rightharpoonup X_{2}$ followed by $2X_{1}+X_{2}\rightharpoonup 3X_1$, which dissipates EP but has no net effect on species counts.
\end{newstuffenv}

We now provide a concrete numerical example. We consider rate constants $k_{1}^{+}=k_{1}^{-}=k_{2}^{-}=k_{3}^{-}=1,k_{2}^{+}=\kTwoVal$, 
while varying $\kThree$ in the range of $\kThree\in[\kThreeMin,\kThreeMax]$.
For these parameter values, the system exhibits limit-cycle oscillations.
Time-dependent concentrations $\cccx {1}(t),\cccx {2}(t)$ for three different
choices of $k_{3}^{+}$ are shown in Figure~\ref{fig:bruss-new}(b)-(d).
In addition, Figure~\ref{fig:bruss-new}(a) shows the EPR $\epr(t)$
and excess EPR $\eprEX(t)$ for $\kThree=\kThreeMin$. Excess EPR
tends to be large when the concentrations are changing rapidly. %

Figure~\ref{fig:bruss-new}(e) shows the decomposition of EPR into
excess and housekeeping components, as a function of the two concentrations
$\ccc=(\cccx 1, \cccx 2)$. We focus on the $\kThree=\kThreeMin$ system,
the same one shown in Figure~\ref{fig:bruss-new}(a)-(b). Streamlines show the dynamical evolution
$\dtpp$ at each point in concentration space, and the actual trajectory
from Figure~\ref{fig:bruss-new}(b) shown in green. 
Excess EPR highlights dynamical structure: bright regions correspond to fast evolution; dark regions to slow evolution (the limit cycle and unstable fixed point). 
This is in contrast to the EPR and the housekeeping EPR, which do not appear to have any clear relationship to state dynamics.

We do not compare our analysis to the HS decomposition because,
for the parameters considered here, the Brusselator does not have a steady
state (no stable fixed point). Formally, the HS decomposition could be
defined by taking $\cccss$ as the unstable fixed point, however this
is not physically meaningful and it leads to negative values of
$\eprEXhs$ (see~\citep[Fig.~7,][]{kohei2022}).

Next, we illustrate our finite-time TSL by deriving a bound on the
dissipation and time needed to complete one oscillator cycle. For
a trajectory that enters a limit cycle, we measure the cycle time $\Tcyc$ and the integrated activity $\totAct$, excess EP $\epEX$, and EP $\ep$ incurred during one cycle. In
addition, we measure the Wasserstein length of the cycle, $\wassLen=\int_{0}^{\Tcyc}\dot{W}(t)\,dt$. We remind the reader that, in a system without external fluxes, $\wassLen$ is the \emph{minimal activity} required to traverse the cycle 
in concentration space. 
Our finite-time TSL for $\wassLen$~\eqref{eq:tslwassLenFT} gives the lower bound $\epTSL^{\wassLen} =2\wassLen\,\atanh({\wassLen}/{\totAct})$ on EP $\ep$ and excess EP $\epEX$. 

For Brusselator dynamics $\dtccc=(\dtcccx{1},\dtcccx{2})$, the Wasserstein speed can be written in closed form, 
\begin{align}
\dot{W}=\vert\dtcccx{1}+\dtcccx{2}\vert+\vert\dtcccx{2}\vert.\label{eq:taxi}
\end{align}
To see why, 
note that only the (coarse-grained) reactions $X_{1}\rightleftharpoons X_{2}$
affect the concentration of $X_{2}$, so either the forward or reverse direction must have a flux of at least $\vert\dtcccx{2}\vert$ (which one depends on the sign of $\dtcccx{2}$). 
One of the two remaining reactions
must have a minimal flux $\vert\dtcccx{1}+\dtcccx{2}\vert$ to induce the
correct evolution of $X_{1}$. Eq.~\eqref{eq:taxi} implies that the Brusselator's
Wasserstein length $\wassLen$ 
is given by the ``taxicab-geometry'' in
coordinates $(\dtcccx{1}+\dtcccx{2},\dtcccx{2})$. 

To illustrate this, Figure~\ref{fig:bruss-new}(f)
plots the three cycles from Figure~\ref{fig:bruss-new}(b)-(d) in
this coordinate system. In this space, the cycle's Wasserstein length $\wassLen$ is 
equal to the perimeter of the minimal bounding box. 
Comparing Figure~\ref{fig:bruss-new}(b)-(d) and (f), we see
that larger $\kThree$ leads to shorter cycle times and smaller Wasserstein lengths.

We illustrate our TSL $\epTSL^{\wassLen}$ in Figure~\ref{fig:bruss-new}(g) for
 $\kThree\in[\kThreeMin,\kThreeMax]$. The bound is relatively
tight, capturing about a third of the excess EP at $\kThree\approx\kThreeMin$ 
and about half around $\kThree\approx\kThreeMax$. EP is many orders
of magnitude larger than excess EP (see inset
in semi-logarithmic scale) and it does not lead to a useful TSL.

Finally, we use Eq.~\eqref{eq:timelimit} to bound the minimal time
needed to complete a cycle as
\begin{align}
\Tcyc \ge\frac{\wassLen}{\avgAct}\coth\frac{\epEX}{2\wassLen} \ge\frac{\wassLen}{\avgAct}\coth\frac{\ep}{2\wassLen}\ge\frac{\wassLen}{\avgAct}.\label{eq:bruss-tsl2}
\end{align}
The last bound $\Tcyc\ge \wassLen/ \avgAct$ holds even in the absolutely irreversible
regime where $\eprEX$ diverges. These bounds are illustrated in
Figure~\ref{fig:bruss-new}(h) for $\kThree\in[\kThreeMin,\kThreeMax]$.
The excess EP bound is relatively tight across
the range of $\kThree$ values. The EP-based bound and absolute irreversibility
bound 
are essentially equivalent, since EP
is very large for this system.

\begin{figure*}[!t]
\includegraphics[width=.23\textwidth]{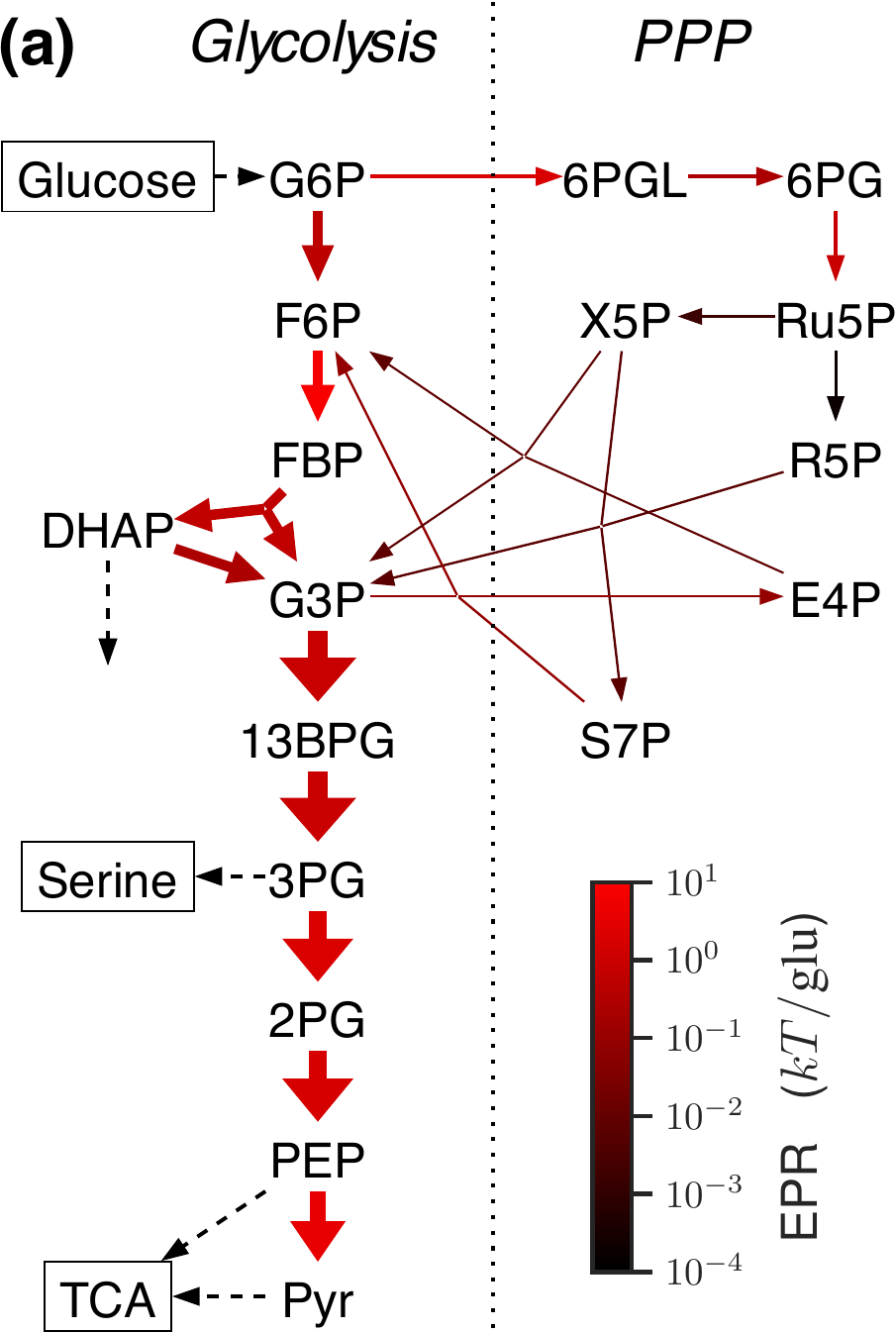}
\hspace{10pt}
\includegraphics[width=.187\textwidth]{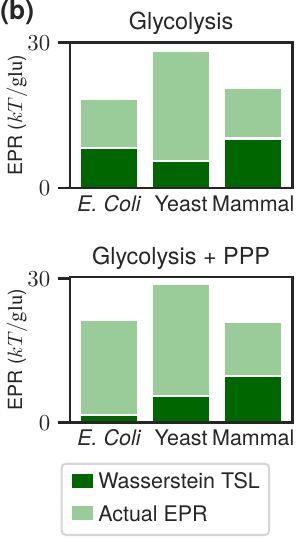}
\hspace{15pt}
\includegraphics[width=.23\textwidth]{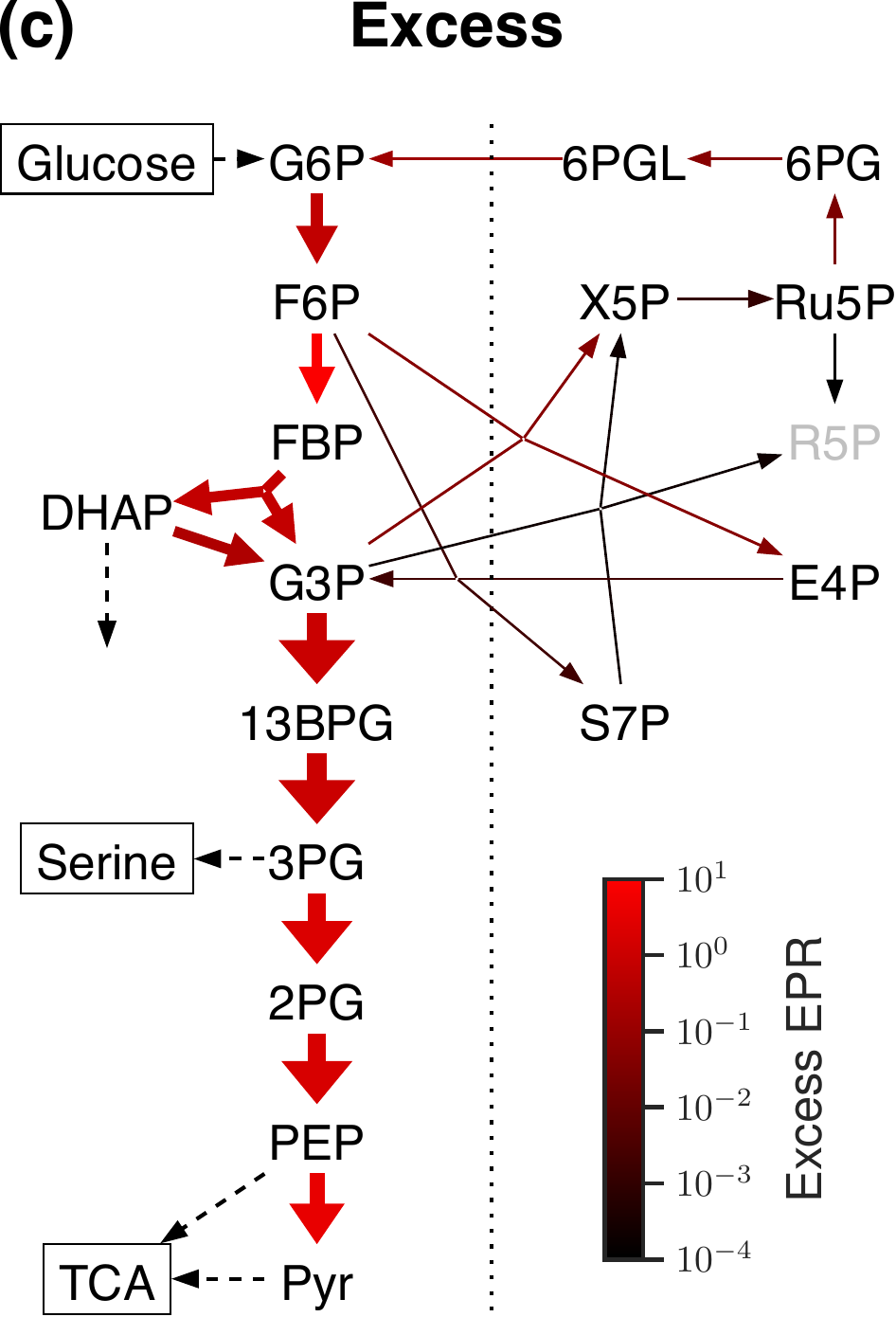}
\hspace{10pt}
\includegraphics[width=.23\textwidth]{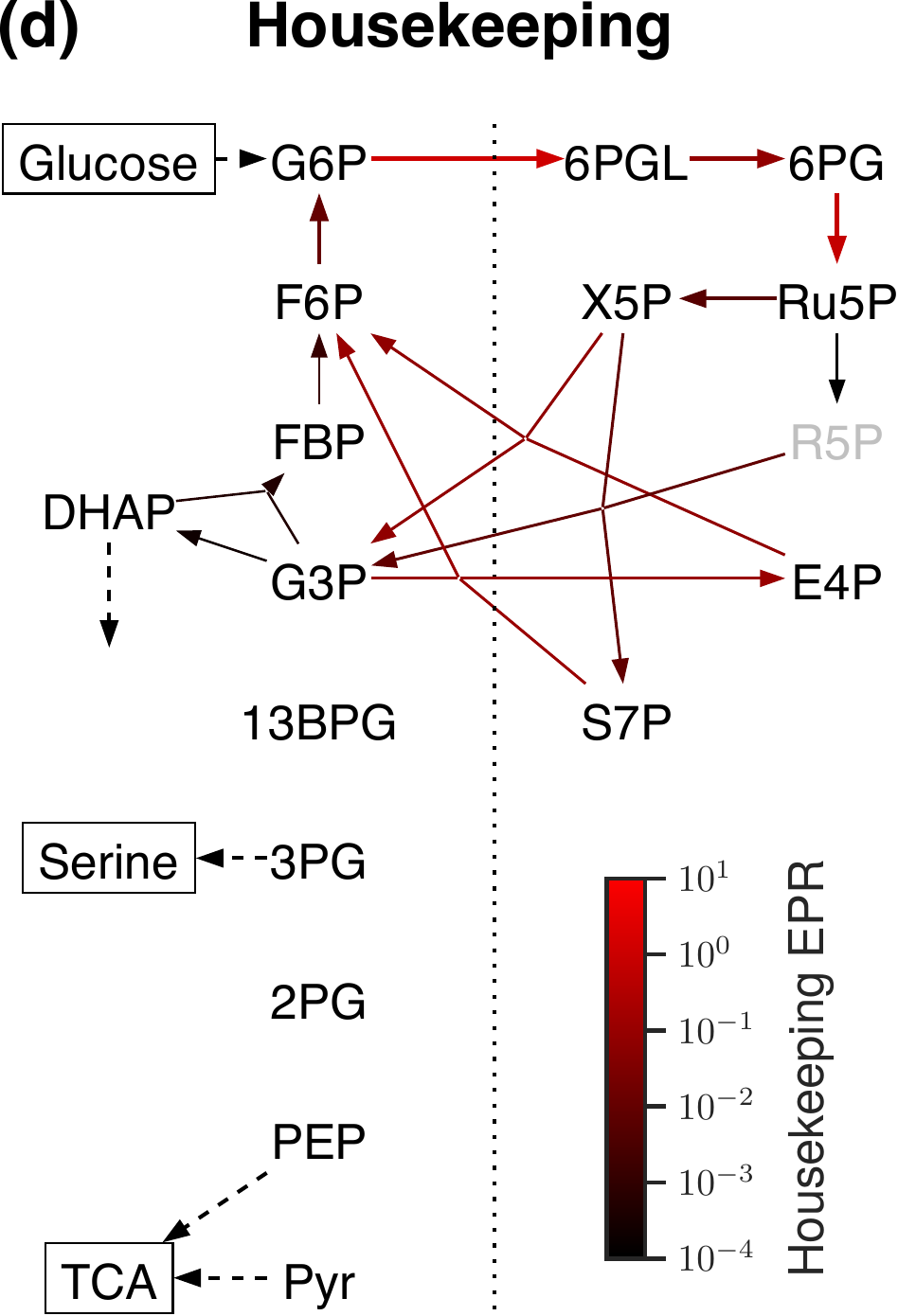}
\caption{\newstuff{\textbf{Metabolic networks for \emph{E. coli}, yeast, and mammalian cells}, based on data from Ref.~\cite{park2016metabolite}. 
\textbf{(a)} Net fluxes and dissipation for individual reactions in \emph{E. coli} for glycolysis and the pentose phosphate pathway (PPP) pathways. For each reversible reaction, arrow width indicates net flux ($\mathcal{J}_\rr=\jjrr-j_{\negedge}$) and color indicates reaction-level EPR $\sigma^{(\rr)}_\text{rev}$ from  Eq.~\eqref{eq:epr-rr}. EPR is in units of $kT$ per incoming glucose molecule. 
\textbf{(b)} Our TSL $\epr\ge 2\dot{W}\atanh({\dot{W}}/{\Act})$ bounds EPR needed to sustain external fluxes in steady state. 
 \textbf{(c)}-\textbf{(d)} The original network does not have housekeeping EPR. To illustrate our excess/housekeeping decomposition, we chemostat ribose-5-phosphate (R5P; gray), as might represent cellular homeostasis of R5P for nucleotide demand. This creates an emergent housekeeping cycle that runs through PPP and back to upper glycolysis. 
Here we show the excess/housekeeping contributions to net fluxes~\eqref{eq:jjex} and EPR~\eqref{eq:eprexhk-rr} at the level of individual reversible reactions. Arrow width indicates net flux ($\mathcal{J}_\rr=\jjrr-j_{\negedge}$) and color indicates reaction-level EPR $\sigma^{(\rr)}_\text{rev}$ from  Eq.~\eqref{eq:epr-rr}.
}}
\label{fig:metabolism}
\end{figure*}

\begin{newstuffenv}
\subsection{Metabolic networks}
\label{sec:metabolic-example}

In our last example, we illustrate our approach on real-world metabolic networks from three species: \emph{E. coli} bacteria, yeast, and a mammalian cell line. Using our TSL, we demonstrate that
central metabolism operates with high efficiency, given constraints imposed by its overall activity and stoichiometry. We also show that our excess/housekeeping decomposition is able to identify fluxes and dissipation associated with futile metabolic cycles.

As standard, especially in research on flux balance analysis~\cite{orth2010flux}, we model the metabolic networks as open deterministic CRNs in steady state.  We focus on two core pathways of central metabolism: glycolysis, responsible for breaking down glucose into pyruvate while producing ATP, and the pentose phosphate pathway (PPP), responsible for  production of NADPH and ribose-5-phosphate (used downstream for nucleotide synthesis). Central metabolism is highly conserved across different branches of life, and so the pathways in each of the three biological species involve the same $\numstate=18$ metabolites and $\numedge=34$ one-way reactions (17 reversible reactions). For illustration, we show the \emph{E. coli} network in Figure~\ref{fig:metabolism}(a), with glycolysis reactions on the left column and PPP reactions on the right column.

To determine flux and thermodynamic parameters, we use published data from isotopic labeling experiments~\cite{park2016metabolite}. In these experiments, each of the three cell types was grown on a nutrient-rich medium with high glucose. Following Ref.~\cite{park2016metabolite}, we treat the first reaction in glycolysis (phosphorylation of glucose to G6P) as external. Also, we do not explicitly represent internal co-factors and small molecules (ATP, ADP, Pi, CO2, etc.), assuming that they are chemostatted at constant concentrations by cellular homeostasis. For details about the data set, see \smref{metabolic}.

In addition to metabolic reactions, the network is also subject to external fluxes of nutrients and metabolites from the environment, the most important of which is the inflow of phosphorylated glucose (G6P), outflow of phosphoenolpyruvic acid (PEP) and pyruvate to the citric acid cycle, outflow of 3-Phosphoglyceric acid (3PG) (used for serine production), and outflow of dihydroxyacetone phosphate (DHAP). External fluxes are set to balance net production, $\II=\BBdiv\jj$, ensuring steady-state conditions.

Some of the flux and thermodynamic data are visualized in the \emph{E. coli} network in  Figure~\ref{fig:metabolism}(a). The direction and width of arrows indicate the direction and net flux of different reactions, color indicates the reaction EPR (summed for both forward and reverse reactions). Fluxes are normalized relative to the rate of glucose uptake, thus EPR values should be understood in units of $k T / \text{glucose}$. The yeast and mammalian networks (not shown), are qualitatively similar to the \emph{E. coli} network, though they exhibit less flux through the PPP pathway.

The stoichiometric matrix has rank $17=m/2$, hence according to algebraic condition~\eqref{eq:rankprop}, the forces are conservative and housekeeping EPR vanishes. This is reasonable from a biological perspective, since glycolytic metabolism appears to avoid futile cycles~\cite{russell1995energetics}. Although the forces are conservative, the steady state is nonequilibrium due to external fluxes.

Next, we use our TSL from Eq.~\eqref{eq:tslW0} to bound the EPR:
\begin{align}
\epr\ge2\dot{W}\,\atanh\frac{\dot{W}}{\Act} \,.
\label{eq:mettsl}
\end{align}
The Wasserstein speed $\dot{W}$~\eqref{eq:wassdef} quantifies the minimal activity required to balance external fluxes $\II=\BBdiv\jj$, given the system's fixed stoichiometry. It can be understood as a topology-dependent measure of the intensity of the external fluxes.

Results for the three biological species (\emph{E. coli}, yeast, and mammalian) are shown in Figure~\ref{fig:metabolism}(b). In the top row, we compute the TSL only for the glycolysis pathway. 
The dark green bars show the minimal bound~\eqref{eq:mettsl}, while the lighter green bars show the actual EPR. We observe that glycolysis is remarkably efficient: its dissipation is close to the fundamental minimum permitted by its stoichiometry and overall activity (reactions per unit time). \emph{E. coli} and mammalian cells achieve nearly $\approx$50\% efficiency, while yeast achieves $\approx$20\% efficiency. This accords with biochemical studies, which have shown that glycolysis pathway appears highly optimized for thermodynamic performance~\cite{helnrlch1997theoretical,melendez1997theoretical}. 

In the bottom row of Figure~\ref{fig:metabolism}(b), we compute the TSL when including both the glycolysis and PPP pathways. For yeast and mammalian cells, the results are essentially the same as without PPP (top row), because these networks have small flux through the PPP pathway in this data. The \emph{E. coli} has  significant flux through the PPP pathway. For this reason, when accounting for PPP, \emph{E. coli} is further away from the fundamental bound set by the TSL, indicating that the PPP pathway is less thermodynamically efficient. Arguably, this is not surprising, since the main function of PPP is synthesis of precursor metabolites rather than energy conversion. 

In our final analysis, we demonstrate our excess/housekeeping decomposition. As mentioned, the metabolic networks considered above  have maximum rank and thus no housekeeping EPR. In biological terms, their stoichiometric matrix does not permit ``futile cycles'' (sequences of reactions that contribute to EPR but have no net effect on metabolites). However, the definition of futile cycle depends on which metabolites are  tracked internally, and ``emergent cycles''~\cite{polettini2014irreversible} can appear when metabolites are removed from the stoichiometric matrix and treated as chemostatted. 

To illustrate this phenomenon, we removed ribose-5-phosphate (R5P) from the stoichiometric matrix of the \emph{E. coli} network. Biologically, this may represent chemostatting of R5P by homeostatic mechanisms that buffer nucleotide demand. This reduces the rank of the stoichiometric matrix to 16, leading to an emergent cycle and a non-trivial excess/housekeeping decomposition. Figure~\ref{fig:metabolism}(c)-(d) shows excess/housekeeping fluxes and dissipation at the level of individual reversible reactions (see Eqs.~\eqref{eq:jjex}~and~\eqref{eq:eprexhk-rr}). 
Thus, by treating R5P as chemostatted, we reveal a maintenance cycle that circulates carbon through the  PPP back to upper glycolysis. This cycle is captured by the housekeeping fluxes $\jj_\hk$ shown in Figure~\ref{fig:metabolism}(d).

\end{newstuffenv}

\global\long\def\varpotential{\mathcal{L}}%
\global\long\def\hess{\text{hess}}%
\global\long\def\potons{\pot^{\ons}}%
\global\long\def\jjons{\jj^{\ons}}%
\global\long\def\jjonsR{\jjR^{\ons}}%
\global\long\def\jjonsrr{j_{\rr}^{\ons}}%
\global\long\def\pothesseq{\pot^{\hess}}%
\global\long\def\pothessst{\bm{\psi}^{\text{\ensuremath{\hess}}}}%
\global\long\def\jjhess{\jj^{\hess}}%
\global\long\def\jjhessR{\jjR^{\hess}}%
\global\long\def\jjhessrr{j_{\rr}^{\hess}}%
\global\long\def\eprEXhess{\sigma_{\ex}^{\hess}}%
\global\long\def\eprHKhess{\sigma_{\hk}^{\hess}}%
\global\long\def\mmCGF{\Phi}%
\global\long\def\jjSymm{\jj^{s}}%
\global\long\def\jjSymmrr{j_{\rr}^{s}}%
\global\long\def\jjSymmrrR{\tilde{j}_{\rr}^{s}}%
\renewcommand{\arraystretch}{1.3} %
\begin{table}[b]
\newstuff{\begin{tabular}{|>{\centering}p{2.2cm}|>{\centering}p{1.05cm}|>{\centering}p{1.8cm}|>{\centering}p{1.95cm}|>{\centering}p{.75cm}|}
\hline 
 & Hatano-\\Sasa & Euclidean-\\Onsager~\cite{kohei2022} & Hessian\\geometry~\cite{Kobayashi2022} & Ours\tabularnewline
\hline 
\hline 
Variational definition & & $\checkmark$ & $\checkmark$ & $\checkmark$\tabularnewline
\hline 
Large deviations & $\checkmark$ & & $\checkmark$ & $\checkmark$\tabularnewline
\hline 
Thermodynamic inference & & & & $\checkmark$\tabularnewline
\hline 
Minimum EP principle & & $\checkmark$ & & \tabularnewline
\hline 
Additive invariance & & & & $\checkmark$\tabularnewline
\hline 
Coarse-graining & MJPs & & & $\checkmark$\tabularnewline
\hline 
\end{tabular}
}
\caption{\newstuff{Comparison between our approach and several other generalized potentials
and excess/housekeeping decompositions. See Section \ref{sec:comparison2}
for details. }}\label{tab:comparison}
\end{table}

\begin{newstuffenv}
\section{Comparison to previous approaches}
\label{sec:comparison2}

Several other generalizations of the free energy potential and excess/housekeeping
decompositions have been proposed, including in some recent publications.
Here, we compare our approach to several existing proposals. In the
first subsection, we compare to other ``variational'' proposals, which define
the generalized potential and EPR decomposition via an optimization
problem. In the second subsection, we compare to the steady-state approach, as previously reviewed in Section~\ref{subsec:ss}.

For purposes of comparison, we will draw attention to several important properties (summarized
in Table \ref{tab:comparison}):
\begin{enumerate}
\item \emph{Variational definition}: are the quantities defined in a variational
way, via the optimization of some objective function? Such definitions typically guarantee existence of a well-defined potential and decomposition in  a general class of systems, and they allow derivations of TURs and related thermodynamic bounds.

\item \emph{Large deviations}: do the quantities have an interpretation
in terms of a large deviations principle? This provides  physical meaning in terms of fluctuation statistics.

\item \emph{Thermodynamic inference}: do the quantities permit thermodynamic
inference using local-in-time statistics? This allows  estimation and validation from real-world experimental data.

\item \emph{Minimum EP principle}: does the variational principle 
identify minimal EP required to achieve some desired state
evolution? This is relevant for thermodynamic optimal control, %
i.e., implementing a given evolution while minimizing entropy production.

\item \emph{Additive invariance}: are the definitions invariant under
joining of independent systems? This  enforces  the fundamental physical principle of extensivity.

\item \emph{Coarse-graining}: are the definitions invariant under merging of reactions with the same stoichiometry? This guarantees that the potential and excess EPR do not depend on detailed information about which reactions mediate state changes, and has practical benefits (see Section~\ref{subsec:cg}). 
\end{enumerate}

\subsection{Variational approaches}
\label{sec:comparison-variational}

Our definition of the generalized potential and excess/housekeeping
decomposition is based on the variational principle~\eqref{eq:exdual}. Several
previous proposals have also defined generalized potentials and EPR
decompositions using different (though related) variational principles. 
Here we summarize and compare these approaches.

Given a system with stoichiometry $\BBbase$ and fluxes $\jj$, we
say that a generalized potential $\hat{\pot}$ is defined variationally
when it is expressed as
\begin{equation}
\hat{\pot}=\argmax_{\pot\in\potspace}\varpotential(\pot,\jj,\BBbase)\,,\label{eq:varform}
\end{equation}
where $\varpotential$ is some objective function that depends on
the ``test potential'' $\pot$, the fluxes $\jj$, and the stoichiometric
matrix $\BBbase$ (it could also depend on the state
$\pp$, though we leave this dependence implicit in our notation).
The optimal potential exists as long as $\varpotential$ satisfies
some mild regularity conditions, though it may not be unique (e.g.,
due to conserved quantities). As we will see below, in existing proposals,
the objective involves the Legendre transform of some convex function.

One appealing feature of the variational approach is that a generalized
free energy can be defined for any system, whether stochastic or deterministic,
as long as one can define an appropriate objective  $\varpotential$.
Of course, different objectives produce different results,
and the choice of the objective must be justified by physical, operational,
or formal considerations. 

In our approach, the generalized potential $\potopt$ is defined using
the following objective,
\begin{equation}
\potopt=\argmax_{\pot\in\potspace}\big[-\jj^{\top}\BBgrad\pot-\mmCGF(\jj,\negBBgrad\pot)\big],\label{eq:psiopt}
\end{equation}
where for convenience we introduced the function $\mmCGF(\jj,\ee):=\jj^{\top}(e^{-\ee}-\oneone)$.
The excess EPR is the maximum value of the objective, $\eprEX=-\jj^{\top}\BBgrad\potopt-\mmCGF(\jj,\negBBgrad\potopt)$. 
The optimality condition, $\BBdiv\jj=\BBdiv(\jjR\circ e^{\negBBgrad\potopt})$,
implies that $\potopt$ exponentially tilts the reverse
fluxes so as to recover the actual net production.

As discussed above, $\mmCGF$ is the cumulant generating
function (CGF) of short-time dynamical fluctuations, and the objective~\eqref{eq:psiopt} is defined in terms of its Legendre transform. 
For this reason, our definitions 
have a natural interpretation in terms of large deviations: $\eprEX$
is the rate function that controls irreversibility of the state evolution,
and the potential $\potopt$ is the ``most irreversible'' state
observable. Our definitions are directly accessible to thermodynamic
inference. In particular, local-in-time statistics provide a tight
lower bound on the excess EPR via the TUR~\eqref{eq:estEPR}.

Several other variational definitions have been advanced in the literature.
For instance, the present authors previously proposed the Euclidean-Onsager
potential~\citep{kohei2022}, defined as
\begin{align}
\potons & =\argmax_{\pot\in\potspace}\big[-2\jj^{\top}\BBgrad\pot-\pot^{\top}\BBdiv L(\jj)\BBgrad\pot\big]\label{eq:optons}
\end{align}
Here we introduced the diagonal matrix $L_{\rr\rr}:=\frac{1}{2}(\jjrr-\jjRrr)/\ln(\jjrr/\jjRrr)$ which encodes reaction-level Onsager coefficients 
that map forces to net fluxes. The associated excess EPR $\eprEXons$
is defined as the maximum of the objective~\eqref{eq:optons}. 
Eq.~\eqref{eq:optons} involves the Legendre transform of the
quadratic function $\pot^{\top}\BBdiv L(\jj)\BBgrad\pot$.
In \smref{ons}, we discuss this definition in more depth, and we derive the inequality %
$\eprEX\le\eprEXons$.

The Euclidean-Onsager approach does not have a direct relation to
large deviations and, outside the linear response regime, it is not
easily accessible to thermodynamic inference. Nonetheless, it has
a natural interpretation in terms of a minimum EP principle.
In particular, it was shown that $\eprEXons$ is the minimum EPR incurred
by any fluxes $\jj^{\prime}$ that have the same net production ($\BBdiv\jj^{\prime}=\BBdiv\jj$) and the same Onsager coefficients ($L(\jj^{\prime})=L(\jj)$) 
as $\jj$~\cite{kohei2022}. Moreover, 
these optimal fluxes have conservative forces $\negBBgrad\potons$.

Another generalized potential was proposed in Ref.~\cite{Kobayashi2022}. It is  helpful to introduce the ``symmetrized'' fluxes $\jjSymmrr:=\surd({\jjrr\jjRrr})$,
which are equilibrium fluxes that have the same ``frenetic activity''
as the actual fluxes,  $\surd({\jjSymmrr\jjSymmrrR})=\surd({\jjrr\jjRrr})$.
 The authors define two different potentials, which in our notation
can be written as:
\begin{align}
\pothesseq & =\argmax_{\pot\in\potspace}\Big[-\jj^{\top}\BBgrad\pot-\mmCGF(\jjSymm,-\BBgrad\pot)\Big]\label{eq:hgeq}\\
\pothessst & =\argmax_{\pot\in\potspace}\Big[-\mmCGF(\jjSymm,\ln(\jj/\jjR)+\BBgrad\pot)\Big]\label{eq:hgss}
\end{align}
where $\mmCGF$ is the short-time CGF mentioned above. The corresponding
excess EPR $\eprEXhess$ has a somewhat complicated expression; for
 details and a numerical comparison, see \smref{hesscomp}. The authors
refer to their proposal as one based on ``Hessian geometry'',
to contrast with Riemannian geometry as might be induced by a quadratic metric 
(e.g., $L(\jj)$ from Eq.~\eqref{eq:optons}).

The potentials $\pothesseq$ and $\pothessst$ can be interpreted in terms of the
family of fluxes $\ee\mapsto \jjSymm\circ e^{\ee/2}$, characterized by  varying thermodynamic forces $\ee$ and fixed frenetic activity $\jjSymm$. Within this family, the forces
$-\BBgrad\pothesseq$ induce the correct state dynamics, $\BBdiv(\jjSymm\circ e^{-\BBgrad\pothesseq/2})=\BBdiv\jj$, while the forces $\ln(\jj/\jjR)+\BBgrad\pothessst$ make the fluxes
stationary, $\BBdiv(\jjSymm \circ e^{(\ln(\jj/\jjR)+\BBgrad\pothessst)/2})=\zz$. We
note that in the Euclidean-Onsager approach, with its simpler quadratic
structure, the potential $\potons$ plays both of
these dynamical roles.

The Hessian-geometry approach is closely
related to research on large deviations of CRNs~\cite{mielke_relation_2014,mielkeNonequilibriumThermodynamicalPrinciples2017,kaiserCanonicalStructureOrthogonality2018,renger2018gradient,rengerOrthogonalityFluxesGeneral2020,pattersonVariationalStructuresGradient2023}, and the two objectives~\eqref{eq:hgeq}-\eqref{eq:hgss} involve Legendre transforms
of the CGF. However, the CGFs are evaluated
under the ``reference'' equilibrium fluxes $\jjSymm$, rather than
the actual fluxes $\jj$. For this reason, 
this approach is not directly amenable to thermodynamic inference
using local-in-time statistics. Also, to our knowledge, the
excess EPR $\eprEXhess$ does not have an interpretation as
a minimum EPR expression.

To summarize, we have shown that our approach is closely related to local-in-time statistics and thermodynamic inference.
The Euclidean-Onsager approach is related to the minimum EP principle
with prescribed Onsager coefficients. The Hessian approach is related to dynamical large deviations of equilibrium systems
with prescribed frenetic activity.

Interestingly, there is another aspect which distinguishes our approach from the others: 
among the objectives listed above, ours~\eqref{eq:psiopt} is the only
one that is linear in the fluxes $\jj$. This formal property has important physical consequences. Consider two
systems characterized by fluxes $\jj$ and $\jj^{\prime}$ that have
the same stoichiometry $\BBbase$ and the same generalized potential
$\potopt$. For example, these might represent reactions occurring
within two different regions of a reactor volume. Then, when the systems
are coarse-grained into a single system with fluxes $\jj+\jj^{\prime}$,
our definition is unique in guaranteeing that the generalized
potential $\potopt$ remains unchanged and the excess EPR adds extensively,
$\eprEX(\jj+\jj^{\prime})=\eprEX(\jj)+\eprEX(\jj^{\prime})$. This
invariance mirrors that of conservative systems: when systems with
the same conservative forces are combined, the forces do not change and 
the EPR adds extensively. This additive invariance also underlies
the coarse-graining property, discussed in Section~\ref{subsec:cg}, 
which states that our definitions are invariant when merging reactions
with the same stoichiometry.

All of the above proposals become equivalent
in the linear-response regime near equilibrium. Specifically, when
$\jj\approx\jj^{\eq}$ (where $\jj^{\eq}=\jjR^{\eq}$ are some equilibrium
fluxes),  all of the objectives reduce to the quadratic form of Eq.~\eqref{eq:ovv}. 
Thus, all objectives share the
same properties and interpretations (large deviations, thermodynamic
inference, minimum EPR, additive invariance) near equilibrium, but
become distinct in the far-from-equilibrium regime. 

Finally, we may 
consider Fokker-Planck dynamics as the continuum limit of MJPs in
the linear response regime~\citep[Appendix~B]{kohei2022}. In this limit, the dynamics converge to the
 continuity equation $\partial_{t}\rho_{t}=-\divop \bm{\mathrm{j}}$, and the generalized potentials converge to the potential proposed by Maes and Netočný (MN) 
for Fokker-Planck systems~\citep{maes2014nonequilibrium}. As described in \smref{mn}, it has a variational characterization~\cite{dechant2022geometric},
\begin{align}
\varphi^{\text{MN}}=\argmax_{\varphi:\mathbb{R}^{k}\to\mathbb{R}}\int\big( 2\,\bm{\mathrm{j}}\cdot\grad\varphi-\rho_t \left\Vert \grad\varphi\right\Vert ^{2} \big) d\bm{r}\,,
\label{eq:varMN}
\end{align}
analogous to Eq.~\eqref{eq:ovv}. 
The variational definitions in  Eqs.~\eqref{eq:psiopt}-\eqref{eq:hgss} provide different generalizations of the Maes-Netočný decomposition to far-from-equilibrium discrete
systems.

\subsection{Steady-state approach}

\label{subsec:hs-comparison}

\global\long\def\HHhs{G}%

We now briefly compare our proposal to the steady-state approach, as previously discussed in Section~\ref{subsec:ss}. 
For simplicity, we focus on the case of MJPs and complex-balanced
CRNs, where the steady-state potential has the simple form $\potss=\grad_{\pp}D(\pp\Vert\ppss)=\ln(\pp/\ppss)$
and the excess/housekeeping decomposition is the Hatano-Sasa (HS)
decomposition.

As we show in \smref{hs}, the HS housekeeping EPR can be expressed in the notation of Eq.~\eqref{eq:ddd3} as
\begin{align}
\eprHKhs=\DDexpFam(\ff\Vert\negBBgrad\potss),
\end{align}
This has a simple connection to our information-geometric interpretation, as visualized in 
Figure~\ref{fig:pyth}(b): the HS housekeeping EPR is 
represented by a line from the actual forces $\ff$ to the 
point $\negBBgrad\potss$ on the conservative manifold. Since our
housekeeping EPR satisfies $\eprHK=\min_{\potlag}\DDexpFam(\ff\Vert\negBBgrad\potlag)$, 
it is always smaller than the HS housekeeping EPR: 
\begin{align}
\eprHK\le\eprHKhs\qquad\eprEX\ge\eprEXhs.\label{eq:hscomparison}
\end{align}
The difference $\eprEX-\eprEXhs\ge0$
has been previously termed ``coupling EPR'' in Fokker-Planck systems~\cite{dechant2022geometricCoupling}, where it quantifies the gap between the HS and Maes-Netočný~\cite{maes2014nonequilibrium} definitions of housekeeping EPR.

In terms of the properties listed in Table~\ref{tab:comparison}, 
the steady-state potential is closely related to the large deviations
 of steady-state fluctuations (see Section~\ref{subsec:ss}). However, for nonstationary systems, it is not directly accessible to thermodynamic inference using
local-in-time statistics. In addition, the HS excess
EPR does not satisfy a minimum EPR principle. 

The steady-state potential does not satisfy ``additive invariance'', since combining systems with the same steady-state potential $\potss$ does not necessarily preserve $\potss$. However, in MJPs, it obeys the coarse-graining condition discussed in Section~\ref{subsec:cg}: $\potss$ and $\eprEXhs$ are invariant under merging of transitions with same stoichiometry (i.e., same $i\to j$) mediated by different reservoirs. For MJPs where the coarse-grained fluxes have a conservative form,  as in Eq.~\eqref{eq:linf} 
and the two-level MJP discussed in Section~\ref{subsec:2level}, our potential $\potopt$ and EPR decomposition is the same one as the HS one.

Finally, we consider the linear-response regime, for simplicity focusing on the case of MJPs. 
Near steady state $\ppp\approx \pppss$, 
we may approximate $\potssx=\ln(\pppx{\xx}/\pppssx{\xx})\approx(\pppx{\xx}-\pppssx{\xx})/\pppssx{\xx}$.
Multiplying both sides by $R_{ji}\pppssx{\xx}$ and summing gives
\begin{align}
\dtppp\approx-\HHhs(\pppss)\potss\,,\label{eq:onsHS}
\end{align}
where $\HHhs_{ji}(\pppss):=-R_{ji}\pppssx{\xx}$ acts as the HS mobility matrix. This may be compared to our linear-response equation, Eq.~\eqref{eq:onsevol0}. 
However, in nonconservative systems, the matrix $\HHhs$ is generally not symmetric,
thus Eq.~\eqref{eq:onsHS} does not satisfy Onsager's reciprocal
relations. Interestingly, our steady-state mobility matrix is the additive symmetrization
of the HS one, $\HH_{\sss}(\pppss)=(\HHhs(\pppss)+{\HHhs}(\pppss)^{\top})/2$, see Eq.~\eqref{eq:mjpH}. As expected,
the two agree for conservative systems.

Finally, Mandal and Jarzynski~\cite{mandal2016analysis} analyzed the linear-response
regime of the HS decomposition for MJPs. They showed that the HS excess
EPR defines a Riemannian geometry over thermodynamic states. However,
the HS friction tensor is not local-in-time, but rather determined
by fluctuations under the stationary process~\cite[Eq.~(24)]{mandal2016analysis}. 
\end{newstuffenv}

\section{Discussion}

\label{sec:Discussion}

In this paper, we proposed a generalized free energy and excess/housekeeping decomposition for nonconservative systems. We demonstrated that our approach is
applicable to a broad
class of stochastic and deterministic systems. We demonstrate that our definitions are amenable to thermodynamic inference, and that they lead to useful bounds such as thermodynamic uncertainty relations (TURs) and thermodynamic speed limits (TSLs). 

In conservative systems, the nonequilibrium free energy can be understood from two different perspectives. The first perspective is state-based (static): the nonequilibrium free energy is related to deviation between the actual state and the equilibrium state, and it controls properties like equilibrium fluctuations and extractable work in nonequilibrium states. The second perspective is dynamical (flux-based): the nonequilibrium free energy gives rise to conservative thermodynamic forces, and it controls properties like dynamical fluctuations and gradient flow dynamics. This dynamical perspective can be traced back to the visionary work of Onsager~\cite{onsager1931reciprocal1,onsager1931reciprocal2,onsager1953fluctuations}.

These two perspectives lead to different generalizations to nonconservative systems. Until now, most research has considered the first (static) perspective by defining the generalized potential in terms of the deviation between the actual state and (nonequilibrium) steady state.

We proposed an alternative approach based on the dynamical perspective. Here, the generalized potential is characterized by offering the ``conservative approximation'' to the forces. From the perspective of large deviations, the generalized potential is the most irreversible state observable, and the excess EPR is defined as its 
degree of irreversibility.

We mention several possible directions for future work.

\newstuff{First, to our knowledge, our analysis of metabolic networks represents the first application of a TSL to a complex biological reaction network. A promising direction for further research is the application of theoretical results from nonequilibrium thermodynamics to large-scale metabolic modeling and flux-balance analysis~\cite{wachtel2022free,gagrani2025thermodynamic,Sugie2025}.}

Second, as we discuss near Eq.~\eqref{eq:eprLR}, in the linear-response regime, our excess EPR defines a Riemannian geometry over the set of thermodynamic states. It would be interesting to explore thermodynamic length and optimal protocols using this geometry in %
nonconservative systems.

Third, in this paper, we explored thermodynamic bounds, such as TURs and TSLs, for the excess EPR. However, we also showed that the housekeeping EPR, the non-excess contribution to dissipation, quantifies the nonconservative nature of the forces. 
Future work may investigate the housekeeping contribution in more depth, e.g., from the perspective of large deviations, TURs and TSLs, as well as decompositions into elementary cycles~\citep{kohei2022}.

Fourth, here we considered the excess/housekeeping decomposition applied to the EPR at a given instant in time. However, the variational principle that defines excess EPR may also be considered for time-extended stochastic processes, where entropy production is quantified as the relative entropy between trajectory distributions. This may lead to fluctuation theorems for excess/housekeeping EPR and generalizations to non-Markovian stochastic processes. There are also interesting connections to other trajectory-level variational principles, such as Schrödinger-bridge problems~\citep{baradat2020minimizing,movilla2024inferring} and Maximum Caliber inference~\citep{dixit2018perspective}. 

Finally, it is interesting to consider our generalized potential and information-geometric decomposition in other types of systems, including reaction-diffusion~\citep{nagayama2023geometric}, hydrodynamic~\citep{yoshimura2023geometric},
and quantum~\citep{yoshimura2024force} systems, which have recently been explored using Euclidean geometry.

\vspace{5pt}

\begin{acknowledgments}
\newstuff{The authors thank Masafumi Oizumi and Praful Gagrani for fruitful discussions.} 
A.~D.~is
supported by JSPS KAKENHI Grants No.~19H05795, and No.~22K13974.
K.~Y.~is supported by Grant-in-Aid for JSPS Fellows (Grant No.~22J21619).
S.~I. is supported by JSPS KAKENHI Grants No.~21H01560, No.~22H01141, No.~23H00467, and No.~24H00834, JST ERATO Grant No.~JPMJER2302, and UTEC-UTokyo FSI Research Grant Program. A.
K. received funding from the European Union’s Horizon 2020 research
and innovation programme under the Marie Sklodowska-Curie Grant Agreement
No. 101068029. 
\end{acknowledgments}

\appendix

\begin{newstuffenv} 

\section{Nonlinear MJPs}
\label{sec:nonlinear_mjps}

In this appendix, we discuss applications of our approach to nonlinear MJPs. Nonlinear MJPs are stochastic master equations in which the transition
rates may depend on the probability distribution. In terms of the MJP notation of Eq.~\eqref{eq:g2}, this means that the rates $\Rija$ may depend on $\ppp$. Nonlinear MJPs are often used in physics~\cite{malek-mansour_master_1975,nicolis_self-organization_1977,korbel2021stochastic} and biology~\cite{frank2013strongly,allahverdyan2011chatelier,patterson2020probabilistic} to model many-body systems (many particles, regions, species, etc.) with mean-field interactions~\cite{feng1992solutions,patterson2020probabilistic,kolokoltsov2011mean}.

We illustrate our formalism on a classic model by Malek-Mansour and Nicolis of local fluctuations in a reaction-diffusion system~\citep{malek-mansour_master_1975}. Let us consider a system that contains a single chemical species, and let $\xx \in\{0,1,2,\dots\}$
indicate the number of particles of that species within a given small region. It is assumed that the region is locally well-mixed, that it undergoes the same fluctuations as its environment, and that it is statistically independent of its environment. Then, the probability distribution $\ppp$ of $\xx$ evolves according to
a birth-death process \citep{malek-mansour_master_1975},
\begin{align*}
\dtpppx{\xx} & =R_{\xx}(\ppp)+\mathscr{D}\langle \xx\rangle_{\ppp}(\pppx{\xx-1} - \pppx{\xx})+\mathscr{D}((\xx+1)\pppx{\xx+1} - \xx \pppx{\xx}). 
\end{align*}
(The term $\mathscr{D}\langle \xx\rangle_{\ppp}\pppx{\xx-1}$ is omitted
at the boundary $\xx=0$.) The first term $R_{\xx}(\ppp)$ is the contribution
due to local reactions, the second term due to particle exchange
from the environment, and the third term due to particle
exchange to the environment. $\mathscr{D}$ is an effective
diffusion frequency and $\langle \xx\rangle_{\ppp}$ is the expected
particle count under distribution $\ppp$ (i.e., particle density
in the environment). Because the transition rate $\mathscr{D}\langle \xx\rangle_{\ppp}$
depends on $\ppp$, this process is a nonlinear
MJP. 

This system can be put in the form of the discrete
continuity equation~\eqref{eq:cont-equation} by introducing an appropriate
incidence matrix $\BBbase$ and probability fluxes. In particular, the fluxes due to
exchanges with the environment are defined as 
\begin{align}
j_{i\to i+1}(\ppp) & =\mathscr{D}\langle i\rangle_{\ppp}\pppx{\xx}\nonumber \\
j_{i+1\to i}(\ppp) & =\mathscr{D}(i+1)\pppx{\xx+1}\label{eq:jf}
\end{align}
The force associated with gaining one particle from
the reservoir can be written as a sum of three terms,
\[
f_{i\to i+1}=\ln\frac{j_{i\to i+1}}{j_{i+1\to i}}=\ln\frac{\pppx{\xx}}{\pppx{\xx+1} }+\ln\frac{1}{\xx+1}-\ln\frac{1}{\langle \xx\rangle_{\ppp}},
\]
representing the change of the region's statistical
entropy, its internal entropy, and the environment's entropy. When 
the reaction term $R_{i}(\ppp)$ arises from a chemical master equation,
it can also be expressed in terms of an incidence matrix and reaction fluxes.

Using these fluxes, EPR is defined as in the main text, Eq.~\eqref{eq:eprn2}. The generalized free
energy and the excess/housekeeping decomposition are also defined as in the main text, Eqs.~\eqref{eq:exdual} and~\eqref{eq:poisson-1}. As usual, the excess term
quantifies the nonstationary contribution, while the housekeeping
term quantifies the nonconservative contribution. 

To understand the meaning of conservative forces, it is useful to note that any Poisson distribution ($\pppx{\xx}^{*}=e^{-\lambda}\lambda^{\xx}/\xx!$), 
satisfies detailed balance for the exchange fluxes, 
\begin{align*}
j_{\xx\to \xx+1}(\ppp^{*})=j_{\xx+1\to \xx}(\ppp^{*}),
\end{align*}
irrespective of the mean $\lambda$. Similarly, the forces for the exchange transitions
are conservative,
\begin{align}
 f_{i\to i+1}=\phi_{i}^{*}-\phi_{i+1}^{*}\,,
\end{align}
for the potential $\phi_{i}^{*}=\ln(\pppx{\xx}/\pppx{\xx}^{*})$. Therefore, the overall system is conservative, and the housekeeping EPR vanishes, as long as
the reaction fluxes that specify $R_{i}(\ppp)$ obey detailed balance
for some Poisson equilibrium distribution.

\end{newstuffenv}

\section{Systems with odd variables}

\label{app:odd} 
\global\long\def\oddconj{\epsilon}%
\global\long\def\timeinterval{dt}%
\global\long\def\Reieja{R_{\oddconj i\oddconj j}^{\bath}}%
\global\long\def\Rejeia{R_{\oddconj j\oddconj i}^{\bath}}%

\subsection{Entropy production rate}

We show how our approach applies to MJPs with odd variables,
such as velocity or momentum, whose sign changes under time
reversal. 

We first write the expression of EPR. Consider an MJP coupled to a
single heat bath that evolves over a small time interval $[t,t+\timeinterval]$.
The conditional probability that the system is in state $\yy$ at
time $t+dt$, given state $\xx$ at time $t$, is
\[
T_{\yy\vert\xx}=\timeinterval\,R_{\yy\xx}+O(\timeinterval^{2}).
\]
The EP is the relative entropy between the forward and backward joint 
distributions, 
\begin{align}
\ep=D\big(\pppx{\xx}T_{\yy\vert\xx}\Vert\pppx{\yy}T_{\oddconj\xx\vert\oddconj\yy}\big),\label{eq:derivapp}
\end{align}
where $\oddconj\xx$ indicates the conjugation of microstate $\xx$
(odd-parity variables flipped in sign). The conjugation of the reverse
conditional probability $T_{\oddconj\xx\vert\oddconj\yy}$ follows from the principle of local detailed balance
for systems with odd variables (see Refs.~\citep{spinney2012entropy,spinney2012nonequilibrium,lee2013fluctuation},
also Section~5.3.4 in~\citep{gardinerHandbookStochasticMethods2004}).
We may write Eq.~\eqref{eq:derivapp} more explicitly as 
\[
\ep=\underbrace{\sum_{\xx\ne\yy}\pppx{\xx}T_{\yy\vert\xx}\ln\frac{\pppx{\xx}T_{\yy\vert\xx}}{\pppx{\yy}T_{\oddconj\xx\vert\oddconj\yy}}}_{\text{Transitions}}+\underbrace{\sum_{\xx}\pppx{\xx}T_{\xx\vert\xx}\ln\frac{\pppx{\xx}T_{\xx\vert\xx}}{\pppx{\xx}T_{\oddconj\xx\vert\oddconj\xx}}}_{\text{Diagonals}}.
\]
The second term  is the contribution to
EP due to different escape rates under the forward and reverse
dynamics. This contribution only appears in systems with odd variables.

The EPR is the time derivative of EP, $\epr=d_t \ep$.
With a bit of algebra, this derivative can be found as 
\begin{align}
\epr=\sum_{\yy\ne\xx}\Big(\pppx{\xx}R_{\yy\xx}\ln\frac{\pppx{\xx}R_{\yy\xx}}{\pppx{\yy}R_{\oddconj\xx\oddconj\yy}}-\pppx{\xx}R_{\yy\xx}+\pppx{\yy}R_{\oddconj\xx\oddconj\yy}\Big).\label{eq:eprodd-1}
\end{align}
For derivations, see~\citep[Eq.~(4.10),][]{spinney2012use} or~\citep[Eq.~(23),][]{liu2012splitting}. 
For a system coupled to multiple reservoirs indexed by $\bath$, this may be generalized as
\begin{align}
\epr=\sum_{\yy\ne\xx,\bath}\Big(\pppx{\xx}\Rjia\ln\frac{\pppx{\xx}\Rjia}{\pppx{\yy}\Reieja}-\pppx{\xx}\Rjia+\pppx{\yy}\Reieja\Big).\label{eq:eprodd}
\end{align}

The EPR can be expressed as a relative entropy between flux vectors.
We define a reaction $\rr$ for each one-way transition $(\xx\to\yy,\bath)$ 
with flux $\jjrr=\pppx{\xx}\Rjia$ and reverse
flux $\jjRrr=\pppx{\yy}\Reieja$. Importantly, unlike in systems without odd variables, the reverse flux does not have to correspond to the forward flux of any reaction. The force
across reaction $\rr$ is
\begin{align}
\ffrr=\ln\frac{\jjrr}{\jjRrr}=\ln\frac{\pppx{\xx}\Rjia}{\pppx{\yy}\Reieja},\label{eq:forceodd}
\end{align}
as in Eq.~\eqref{eq:forcedef}. Finally, the EPR~\eqref{eq:eprodd}
can be written as the relative entropy between forward and reverse
fluxes, as in Eq.~\eqref{eq:eprn}:
\[
\epr=\DD(\jj\Vert\jjR).
\]
Note that Eq.~\eqref{eq:eprodd} does not have the simple ``flux-force''
form $\epr=\sum_{\rr}\jjrr\ffrr$, as it does in systems without odd
variables. This reflects the irreversibility due to different forward and reverse escape rates.

\subsection{Generalized free energy and excess/housekeeping decomposition}

Many of our results continue to hold for MJPs with odd variables,
including the definition of the generalized potential and the excess/housekeeping
decomposition. One important caveat is that we do not
simplify our expressions by using Eq.~\eqref{eq:antisymmetry},
the relationship between forward and reverse fluxes that holds without odd variables. Without using this equation, excess EPR~\eqref{eq:exdual} is defined as
\begin{align}
\eprEX=\max_{\pot\in\potspace}\,\Big[-\pot^\top \BBdiv \jj-\jjR^{\top}(e^{\negBBgrad\pot}-\oneone)\Big],\label{eq:eprn-odd}
\end{align}
which is the analogue of Eq.~\eqref{eq:epr00}. The optimality condition that defines
$\potopt$ is given by 
\begin{align}
 \BBdiv\jj =\BBdiv(\jjR\circ e^{-\BBgrad\potopt}),\label{eq:poisson-odd}
\end{align}
rather than Eq.~\eqref{eq:poisson-1}. The expression of excess EPR
as information-theoretic optimal transport, as in Eq.~\eqref{eq:exdef},
remains unchanged. In the special case where Eq.~\eqref{eq:antisymmetry} holds, Eq.~\eqref{eq:eprn-odd} reduces to Eq.~\eqref{eq:exdual}, and Eq.~\eqref{eq:poisson-odd}
reduces to Eq.~\eqref{eq:poisson-1},

Some results must be qualified in the presence of odd variables.
For instance, in MJPs with odd variables, excess EPR does not necessarily vanish in stationarity,
unless the steady state is symmetric under conjugation: $\pppssx{\xx}=\pppssx{\oddconj\xx}$.
Similarly, we do not have the inequality between our excess EPR and
HS excess EPR, Eq.~\eqref{eq:hscomparison}, unless the steady-state
distribution is symmetric under conjugation. Finally, our thermodynamic
speed limits do not hold in general for systems with odd variables.
These differences arise because, for systems with odd variables and asymmetric escape rates,
the steady state may be nonequilibrium even when the thermodynamic
forces are conservative. The steady state is equilibrium if the forces are conservative
and the steady state is symmetric under conjugation of odd variables.
(See Ref.~\citep{lee2013fluctuation} for further discussion.)

Odd variables are problematic for the HS decomposition, e.g., 
they can lead to negative values of HS housekeeping EPR~\citep{spinney2012nonequilibrium,ford2012entropy,lee2013fluctuation}.
To our knowledge, no universally-applicable housekeeping/excess decomposition has been previously
proposed for systems with odd variables. 

\subsection{Example: particle on a ring}

\label{app:odd-example}

\global\long\def\paramJ{\eta}%
\global\long\def\locs{k}%

We provide an example to illustrate our excess/housekeeping
decomposition in the presence of odd variables. 
We also compare to the HS decomposition, where the housekeeping EPR can
take unphysical negative values~\citep{lee2013fluctuation,spinney2012nonequilibrium,ford2012entropy}.

\begin{figure*}
\centering

\begin{minipage}[b]{0.45\textwidth}%
\vspace{5pt}\includegraphics[width=1\textwidth]{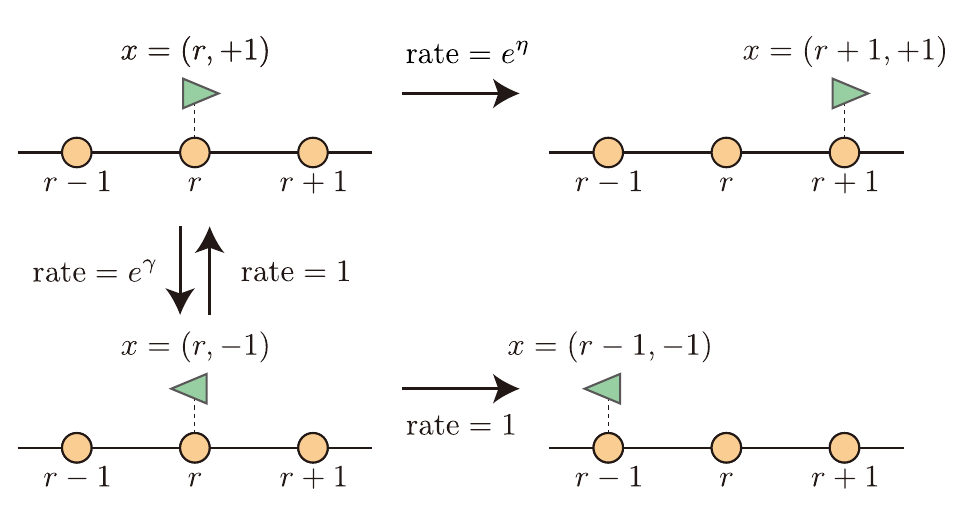}\vspace{30pt}%
\end{minipage}\qquad
\begin{minipage}[b]{0.45\textwidth}%
\includegraphics[width=1\textwidth]{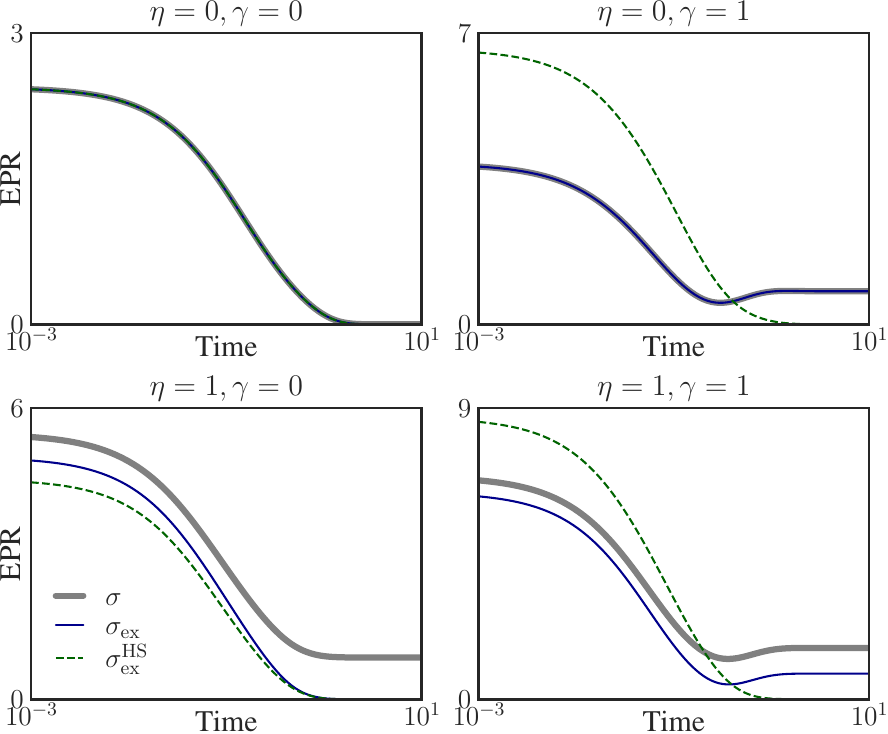}%
\end{minipage}\caption{\protect\emph{Left}: a discrete system with odd variables,
consisting of a particle on a ring with position $r\in\{1,\dots,\protect\locs\}$
and odd velocity $v\in\{-1,+1\}$~\citep{lee2013fluctuation,spinney2012nonequilibrium,ford2012entropy}.
\emph{Right}: time courses of EPR $\protect\epr$, our excess EPR $\protect\eprEX$,
and HS excess EPR $\protect\eprEXhs$ for this model. $\protect\paramJ\protect\ne0$
means forces are nonconservative, $\gamma\protect\ne0$ means steady-state
distribution is not symmetric under conjugation of odd variables.
HS decomposition can give unphysical values ($\protect\eprEXhs>\protect\epr,\protect\eprHKhs<0$)
when $\gamma\protect\ne0$.}\label{fig:odd}
\end{figure*}

We use an existing model from the literature~\citep{lee2013fluctuation,spinney2012nonequilibrium,ford2012entropy},
illustrated in Figure~\ref{fig:odd} (left). There is a particle on
a ring with $\locs$ locations; the particle also has an odd ``velocity'' degree
of freedom, indicating whether it is moving clockwise or counter-clockwise. The system's microstate is specified by $\xx=(r,v)$,
where $r\in\{1,\dots,\locs\}$ is the position of the particle on
the ring and $v\in\{{-1},+1\}$ is the velocity. Conjugation involves
flipping the sign of the velocity variables,
\[
\oddconj(r,v)=(r,-v).
\]
The rate matrix contains two types of jumps: around the ring $(r,v)\to(r+v,v)$ and velocity flips $(r,v)\to(r,-v)$. 

The transition rates are parameterized as
\begin{align*}
R_{(r+v,v)\leftarrow(r,v)} & =e^{\delta_{v,1}\paramJ}\\
R_{(r,-v)\leftarrow(r,v)} & =e^{\delta_{v,1}\gamma}.
\end{align*}
In words, the particle moves in the direction of its velocity with
rate $e^{\paramJ}$ when $v=+1$ and rate $1$ when $v=-1$; the velocity
flips with rate $e^{\gamma}$ when $v=+1$ and rate 1 when $v=-1$. 
The steady state is given by 
\begin{align}
\pppssx{r,v}=\frac{\delta_{v,1}+\delta_{v,-1}e^{\gamma}}{\locs(e^{\gamma}+1)}.\label{eq:odd-ss}
\end{align}
The parameter $\paramJ$ controls the strength of driving around the ring, leading to nonconservative
forces when $\paramJ\ne0$. The parameter $\gamma$ controls the breaking
of symmetry of velocity flips, leading to a steady-state distribution
that is asymmetric under conjugation ($\pppssx{r,v}\ne\pppssx{r,-v}$)
when $\gamma\ne0$. The steady state is  equilibrium only when $\paramJ=0$
and $\gamma=0$. 
Using Eq.~\eqref{eq:forceodd}, the forces for the two types of
transitions are 
\begin{alignat}{1}
f_{(r+v,v)\leftarrow(r,v)} 
 & =\ln\frac{\pppx{r,v} R_{(r+v,v)\leftarrow(r,v)}}{\pppx{r+v,v} R_{(r,-v)\leftarrow(r+v,-v)}}\nonumber \\
 & =\ln\frac{\pppx{r,v}}{\pppx{r+v,v}}+\delta_{v,1}\paramJ-\delta_{-v,1}\paramJ\nonumber \\
 & =\ln\frac{\pppx{r,v}}{\pppx{r+v,v}}+v\paramJ\\
f_{(r,-v)\leftarrow(r,v)} 
 & =\ln\frac{\pppx{r,v}R_{(r,-v)\leftarrow(r,v)}}{\pppx{r,-v}R_{(r,-v)\leftarrow(r,v)}}=\ln\frac{\pppx{r,v}}{\pppx{r,-v}}
\end{alignat}

In Figure~\ref{fig:odd} (right), we visualize the time-dependent
values of EPR $\epr$, our excess EPR $\eprEX$, and the HS excess
EPR $\eprEXhs$. We consider a system with $\locs=4$ positions and
the non-steady-state initial distribution $\pppx{r,v}\propto10\delta_{r,0}\delta_{v,1}+1$.
We consider four parameter values.

In the first condition (top left), $\paramJ=0$ and $\gamma=0$, so all transitions
are symmetric. Here the forces are conservative, $f=\negBBgrad\pot$
for $\pot=\ln\ppp$ (up to a constant) and the steady-state distribution is symmetric
under conjugation. The steady state is in equilibrium
and $\epr=\eprEX=\eprEXhs$ at all times.

In the second condition (top right), $\paramJ=0$ and $\gamma=1$, so velocity
flips $(r,-1)\to(r,+1)$ occur more frequently than $(r,+1)\to(r,-1)$.
The forces are conservative, $f=\negBBgrad\pot$ for $\pot=\ln\ppp$,
but the steady state is not symmetric under conjugation.
For this reason, the steady state is nonequilibrium ($\epr>0$ in steady state).
Since the forces are conservative, our housekeeping EPR vanishes and
$\epr=\eprEX$ at all times. The HS decomposition gives different
numerical values, and it can produce unphysical negative values ($\eprEXhs>\epr,\eprHKhs<0$).

In the third condition (bottom left), $\paramJ=1$ and $\gamma=0$, so movements
along the ring with positive velocity are faster than those with negative
velocity. The steady state is symmetric under time-reversal but the
forces $f_{(r+v,v)\leftarrow(r,v)}$ are not conservative, so the
steady state is nonequilibrium. Our decomposition and HS decomposition
both obey $0\le\eprEX\le\epr$ and $0\le\eprEXhs\le\epr$. We verify that, in systems with time-reversal-symmetric steady states,
$\eprEX\ge\eprEXhs$ and $\eprEX=\eprEXhs=0$ in steady state.

In the fourth condition, $\paramJ=1$ and $\gamma=1$, so the forces
are not conservative and the steady state is not symmetric under conjugation. The HS decomposition again gives unphysical values
$\eprEXhs>\epr,\eprHKhs<0$. Under our decomposition, neither $\eprEX$
nor $\eprHK$ vanish in steady state.

\bibliography{writeup}

\begin{thebibliography}{164}
\bibitem [{not()}]{notehydro}%
  \BibitemOpen
  \href@noop {} {}\bibinfo {note} {For physical relevance, we should assume
  that short time intervals are longer than the hydrodynamic timescale, so that
  assumptions of Markovian and overdamped dynamics are justified.}\BibitemShut
  {Stop}%
\end{thebibliography}

\ifarxiv

\vfill
\clearpage
\setcounter{equation}{0} 
\counterwithout{equation}{section}
\renewcommand{\theequation}{S\arabic{equation}}

\setcounter{page}{1} %
\renewcommand{\thepage}{S\arabic{page}} %

\renewcommand{\thefigure}{S\arabic{figure}} 
\renewcommand{\appendixname}{}
\setcounter{section}{0} 
\setcounter{figure}{0} 
\makeatletter
\renewcommand{\thesection}{SM\arabic{section}}
\renewcommand{\thesubsection}{\thesection.\arabic{subsection}}
\renewcommand{\p@subsection}{}
\makeatother

\onecolumngrid

\begin{center}
 \vspace*{1cm}
 {\huge \textbf{Supplemental Material}}
 \vspace{1cm}
\end{center}

\twocolumngrid

\section{Generalized free energy with conservation laws}
\label{app:sm1}

Our variational principle~\eqref{eq:exdual} uniquely identifies the optimal conservative force $\BBgrad \potopt$. The free energy potential $\potopt$ is uniquely defined up to the nullspace of $\BBgrad$. The degeneracy within this nullspace represents different choices of conserved quantities.

In some cases, it may be useful to identify the precise potential $\potopt$ that is consistent with the conserved quantities of the current state $\pp$. The conserved quantities of $\pp$ may be obtained as $\nproj \pp$, 
where $\nproj$ is the projector onto the nullspace of $\BBgrad$,
\begin{align}
 \nproj=I-\BBgrad^+\BBgrad\,,
\end{align}
and $\BBgrad^+$ is the pseudo-inverse. 
For example, for an MJP with an irreducible rate matrix, the nullspace of $\BBgrad$ is spanned by the vector $\oneone$ (representing conservation of probability) and $\nproj \ppp$ represents the conserved quantity $\oneone^\top \ppp$ (total probability). 

Recall that, in the case of conservative systems, the free energy potential is $\poteq = \ln(\pp/\ppeq)$, and the equilibrium state is recovered from $\pp$ and $\poteq$ as $\ppeq = \pp \circ e^{-\poteq}$. There may be different possible equilibrium states, representing different choices of conserved quantities. The physically relevant equilibrium state has the same conserved quantities as $\pp$, thus $\nproj\pp=\nproj \ppeq=\nproj (\pp \circ e^{-\poteq})$.

It is natural to generalize this to nonconservative systems by defining $\pp^\star := \pp \circ e^{-\potopt}$ as the ``pseudo-canonical'' state associated with $\pp$ and $\potopt$. Then, $\potopt$ may be chosen so that the pseudo-canonical state has the same conserved quantities as the actual state $\pp$, 
\begin{align}
\nproj \pp=\nproj \pp^\star\equiv \nproj (\pp\circ e^{-\potopt})\,.\label{eq:conslaw}
\end{align}
For the example of the irreducible MJP where $\pp\equiv \ppp$ is a probability distribution, this would guarantee that %
$\ppp^\star$ is also a normalized probability distribution,
\begin{align}
\oneone^\top \ppp^\star \equiv \oneone^\top(\ppp\circ e^{-\potopt}) = \oneone^\top \ppp= 1.
\label{eq:normPSS}
\end{align}

Here we show how to find the generalized potential $\potopt$ that optimizes our variational principle~\eqref{eq:exdual} while satisfying the conservation law~\eqref{eq:conslaw}. To do so, we 
introduce the following information-theoretic variational principle:
\begin{align}
 \bm{y}^\star &= \argmin_{\bm{y} \in \distspace} D(\bm{y}\Vert \pp \circ e^{-\potopt})\;\suchthat\; \nproj \bm{y} =\nproj \pp \,.\label{eq:bccdd1}
\end{align}
Here $\potopt$ is any optimal potential found by optimizing Eq.~\eqref{eq:exdual} and $D$ is the generalized relative entropy~\eqref{eq:klstate}. 
This is a convex optimization problem that can be solved using standard numerical methods.

The generalized free energy potential that optimizes Eq.~\eqref{eq:exdual} and satisfies the conservation laws~\eqref{eq:conslaw} is 
\begin{align}
 {\pot}^{\star\star} = -\ln\frac{\bm{y}^\star}{\pp}\,.
 \label{eq:sm033}
\end{align}
To see why, write 
 the Lagrangian of the optimization~\eqref{eq:bccdd1} as
\begin{align*}
\sum_i \Big(y_i \ln \frac{y_i}{\ppx{i} e^{-\potoptx}} - y_i + \ppx{i} e^{-\potoptx}\Big) + \bm{\lambda}^\top(\nproj \bm{y} - \nproj\pp)\,.
\end{align*}
Taking derivatives with respect to each $y_i$, setting to zero, and rearranging gives
\begin{align}
 -\ln\frac{\bm{y}^\star}{\pp}= \potopt + \nproj^\top \bm{\lambda}^\star, 
 \label{eq:sm034}
\end{align}
where the value of the optimal Lagrange multipliers $\bm{\lambda}^\star$ is determined by the constraint $\nproj \bm{y} =\nproj \pp$. 
Comparing Eqs.~\eqref{eq:sm033} and \eqref{eq:sm034}, we see that $ {\pot}^{\star\star}=-\ln(\bm{y}^\star/\pp)$ differs from $\potopt$ by a vector in the image of $\nproj^\top = \nproj$, i.e., a null vector of $\BBgrad$, therefore ${\pot}^{\star\star}$ achieves the same objective in the variational principle~\eqref{eq:exdual} as $\potopt$. Also, by construction, $\bm{y}^\star = \pp \circ e^{- {\pot}^{\star\star}}$ satisfies $\nproj \bm{y}^\star = \nproj \pp$, which is the conservation law~\eqref{eq:conslaw}. We see that the optimization~\eqref{eq:bccdd1} identifies the physically relevant pseudo-canonical state.

\section{Relation to variational principle from Ref.~\citep{shiraishi2019information}}

\label{app:sm3}

In previous work, Shiraishi and Saito~\citep{shiraishi2019information} proposed 
a variational principle for EPR in passive MJPs. They
showed that 
\begin{align}
\epr=\max_{\qq}\,\Big[-\frac{d}{dt}D(\ppp(t)\Vert\qq(-t))\Big].\label{eq:shi-gen}
\end{align}
where $D$ is the relative entropy between probability distributions
and the maximization is over all probability distributions
$\qq$. The notation $\qq(-t)$ indicates that this distribution
evolves backwards in time under the reverse rates, 
\begin{align}
-\frac{d}{dt}\qqx{\xx}(-t)=\sum_{\yy(\ne\xx),\bath}(\qqx{\yy}(-t)\Rija-\qqx{\xx}(-t)\Rjia).\label{eq:gdd2}
\end{align}

Here we show that this variational principle is a special case of
the variational principle for excess EPR~\eqref{eq:exdual}, in this
way recovering Eq.~\eqref{eq:shi-gen-1} in the main text. We first rewrite the optimization~\eqref{eq:shi-gen-1} as
\begin{equation}
\max_{\qq}\Big[\sum_{\xx}\dtpppx{\xx}\ln\frac{\qqx{\xx}}{\pppx{\xx}}+\sum_{\xx}\pppx{\xx}\frac{d}{dt}\ln\qqx{\xx}(-t)\Big],\label{eq:vcxv3}
\end{equation}
\newstuff{where we use that $\BBdiv \jj=\dtppp$ for an MJP, since MJPs do not have external flows.} 
Next, we define a reaction $\rr$ for each transition $(\yy\to\xx,\bath)$,
with forward and reverse fluxes $\jjrr=\pppx{\yy}\Rija$ and $\jjRrr=\pppx{\xx}\Rjia$.
We then write the variational principle~\eqref{eq:exdual} as 
\begin{align*}
\eprEX & =\max_{\pot\in\mathbb{R}^{\numstate}}\Big[-\sum_{\xx}\dtpppx{\xx}\potx-\sum_{\xx\ne\yy,\bath}\pppx{\yy}\Rija(e^{\potx-\potxx{\yy}}-1)\Big].
\end{align*}
We change the optimization variable from potentials to probability
distributions $\qq$ via $\ln\qqx{\xx}=\ln\pppx{\xx}-\potx+\text{const}$ (the additive constant cancels). 
Using this replacement, we can write 
\begin{align}
\eprEX= & \max_{\qq}\Big[\sum_{\xx}\dtpppx{\xx}\ln\frac{\qqx{\xx}}{\pppx{\xx}}-\sum_{\xx\ne\yy,\bath}\pppx{\yy}\Rija\Big(\frac{\pppx{\xx}}{\qqx{\xx}}\frac{\qqx{\yy}}{\pppx{\yy}}-1\Big)\Big]\label{eq:adsf}
\end{align}
We rewrite the second sum as 
\begin{align}
 & \sum_{\yy\ne\xx,\bath}\Big(\frac{\pppx{\xx}}{\qqx{\xx}}\qqx{\yy}\Rija-\pppx{\yy}\Rija\Big)\\
 & =\sum_{\yy\ne\xx,\bath}\Big(\frac{\pppx{\xx}}{\qqx{\xx}}\qqx{\yy}\Rija-\pppx{\xx}\Rjia\Big)\\
 & =\sum_{\xx}\frac{\pppx{\xx}}{\qqx{\xx}}\sum_{\yy(\ne\xx),\bath}\big(\qqx{\yy}\Rija-\qqx{\xx}\Rjia\big)\nonumber \\
 & =-\sum_{\xx}\frac{\pppx{\xx}}{\qqx{\xx}}\frac{d}{dt}\qqx{\xx}(-t)=-\sum_{\xx}\pppx{\xx}\frac{d}{dt}\ln\qqx{\xx}(-t),\label{eq:app-rev-evo}
\end{align}
where in the last line we used the definition~\eqref{eq:gdd2}. Combining
shows the equivalence to~\eqref{eq:vcxv3}.

\allowdisplaybreaks
\section{Simplified TUR for excess EPR}

\label{app:tur}

\begin{newstuffenv}
In this section, we derive a simplified ``thermodynamic uncertainty relation'' (TUR) for excess EPR. This bound is less tight than the higher-order TUR~\eqref{eq:estEPR}. However, unlike the higher-order TUR, which requires knowledge of all cumulants, the simplified bound involves only a few simple statistics, and it may be practical in cases where measuring all cumulants is difficult.

The simplified TUR bounds excess EPR in terms of three quantities, defined for any state observable $\pot\in\potspace$:
\begin{enumerate}
\item The expected change of the observable due to net production: $K^{(1)}_{\pot} := \jj^\top \BBgrad \pot$.
\item The maximum change of the observable across any reaction, $\Vert \BBgrad\pot\Vert_\infty:=\max_\rr \vert \BBgrad\pot\vert_\rr$. In the literature, this quantity is sometimes called the discrete Lipschitz constant. 
\item The ``activity of the observable'' $\Act_{\pot}:=\sum_\rr\jjrr \vert \BBgrad\pot\vert_\rr$, the rate of absolute (negative and positive)
changes of the observable across all reactions.
\end{enumerate}
$\Act_{\pot}$ and $\Vert \BBgrad\pot\Vert_\infty$ provide two ways to quantify the scale of dynamical fluctuations of $\pot$. They both vanish if and only if $\pot$ is a conserved quantity ($\BBgrad \pot=\zz$).

In a stochastic system, all quantities can be easily estimated from experimental two-point measurements of $\pot$ at times $t$ and $t+dt$. Writing a dataset of $N$ measurements as $\{(\phi_{t}^{(i)},\phi_{t+dt}^{(i)}): i=1\dots N\}$, we may estimate the quantities as \begin{align*}
 K^{(1)}_{\pot}&\approx N^{-1} \sum_i (\phi_{t+dt}^{(i)}-\phi_{t}^{(i)})/dt\\
 \Act_{\pot}&\approx N^{-1} \sum_i \vert \phi_{t+dt}^{(i)}-\phi_{t}^{(i)}\vert/dt\\
 \Vert \BBgrad\pot\Vert_\infty&\approx \max_i \vert \phi_{t+dt}^{(i)}-\phi_{t}^{(i)}\vert
\end{align*}

Our simplified TUR bounds the excess EPR as:
\begin{align}
 \eprEX \ge \frac{2K_{\pot}^{(1)} }{\Vert \BBgrad\pot\Vert_\infty}\atanh\frac{ K_{\pot}^{(1)} }{\Act_{\pot}}
 \,. \label{eq:tur1}
\end{align}
A further bound can be derived based on the activity $\Act= \Vert \jj\Vert_1$ (overall number of reactions per unit time). We note that 
$\Act_{\pot}\le \Act \Vert \BBgrad\pot\Vert_\infty$, then plug into Eq.~\eqref{eq:tur1} to give 
\begin{align}
 \eprEX \ge \frac{2K_{\pot}^{(1)} }{\Vert \BBgrad\pot\Vert_\infty}\atanh\frac{ K_{\pot}^{(1)} }{\Act \Vert \BBgrad\pot\Vert_\infty}
 \,. \label{eq:tur1-act}
\end{align}

To derive Eq.~\eqref{eq:tur1}, we begin by rewriting Eq.~\eqref{eq:exdual}, the variational expression for excess EPR, as 
\begin{align}
\eprEX & =\max_{\lambda\in\mathbb{R},\pot\in\potspace}\Big[-\lambda\chngObs-\sum_{\rr}\jjrr(e^{\lambda[\BBgrad\pot]_{\rr}}-1)\Big]\label{eq:legendre-tur}
\end{align}
Here we used that the optimization is invariant to scaling of $\pot\mapsto \lambda\pot$. 
Next, we will use the bounds 
\begin{align*}
e^{\alpha x}-1 & \le xe^{\alpha}-\left|x\right| & & \text{ for }\alpha\in\mathbb{R},x\in[0,1]\\
e^{\alpha x}-1 & \le-xe^{-\alpha}-\left|x\right| & & \text{ for }\alpha\in\mathbb{R},x\in[-1,0]
\end{align*}
Taking $x=[\BBgrad\pot]_{\rr}/\Vert \BBgrad\pot\Vert_\infty\in[-1,1]$ in each term in the sum in Eq.~\eqref{eq:legendre-tur} and rearranging gives the bound
\begin{align}
\eprEX\ge\frac{1}{\Vert \BBgrad\pot\Vert_\infty}\max_{\lambda}\Big[-\lambda\chngObs+\actOBS-\actOBSpos e^{\lambda}-\actOBSneg e^{-\lambda}\Big]\label{eq:leg3}
\end{align}
where we defined the positive $\actOBSpos$ and negative $\actOBSneg$ 
activity of the observable as 
\begin{align*}
\actOBSpos & :=\sum_{\rr:[\BBgrad\pot]_{\rr}>0}\jjrr[\BBgrad\pot]_{\rr}=\frac{1}{2}(\actOBS+\chngObs)\\
\actOBSneg & :=\sum_{\rr:[\BBgrad\pot]_{\rr}<0}\jjrr[-\BBgrad\pot]_{\rr}=\frac{1}{2}(\actOBS-\chngObs).
\end{align*}
The last optimization %
can be solved by taking derivatives,
which gives the optimal $\lambda$ as 
\[
\lambda^{*}=\ln\frac{\actOBSneg}{\actOBSpos}=\ln\frac{\actOBS-\chngObs}{\actOBS+\chngObs}=2\atanh\frac{\chngObs}{\actOBS}
\]
where we used $\actOBS=\actOBSpos+\actOBSneg$ and $\chngObs=\actOBSpos-\actOBSneg$.
Plugging into Eq.~\eqref{eq:leg3} gives Eq.~\eqref{eq:tur1}.
\end{newstuffenv}

\section{Thermodynamic speed limit (TSL)}
\label{app:tsl}

Here we derive several results relevant to our Wasserstein TSL. 

\subsection{Duality for 1-Wasserstein distance}
\newcommand{\otherpp}{\bm{y}}

\begin{newstuffenv}
The 1-Wasserstein distance between states $\pp$ and $\otherpp$ is
\begin{align}
W(\pp,\otherpp)=\min_{\bm{m}\in\fluxspace}\Vert \bm{m}\Vert _{1}\suchthat\BBdiv\bm{m}=\otherpp-\pp\label{eq:dd}
\end{align}
assuming the constraint is feasible for some $\bm{m}$. The Lagrangian formulation is 
\begin{align*}
W(\pp,\otherpp)&=\min_{\bm{m}\in\fluxspace}\max_{\pot\in\potspace} \Big[\Vert \bm{m}\Vert _{1}-\pot^\top (\BBdiv\bm{m}-(\otherpp-\pp))\Big]\\
&=\max_{\pot\in\potspace} \min_{\bm{m}\in\fluxspace} \Big[\Vert \bm{m}\Vert _{1}-\pot^\top (\BBdiv\bm{m}-(\otherpp-\pp))\Big]
\end{align*}
where we used duality to exchange $\min$ and $\max$. Note that 
\begin{align*}\min_{\bm{m}\in\fluxspace} \Big[\Vert \bm{m}\Vert _{1}-\pot^\top\BBdiv\bm{m}\Big]=\begin{cases}0 & \text{if\;}\Vert \BBgrad \pot\Vert_\infty\le 1\\-\infty &\text{otherwise}\end{cases}
\end{align*}
Thus, we may write the optimization in its dual form as
\begin{align}
W(\pp,\otherpp)&=\max_{\pot\in\potspace:\Vert \BBgrad \pot\Vert_\infty\le 1} \pot^\top (\otherpp-\pp)
\label{eq:dddual}
\end{align}

The short-time limit of the Wasserstein distance is 
\begin{align}
 \dot{W}(\bm{v}) := \lim_{dt\to 0}\frac{1}{dt} {W}(\pp,\pp+dt\,\bm{v})
\end{align}
Using Eq.~\eqref{eq:dddual}, whose objective is linear in $\otherpp-\pp$, we have
\begin{align}
 \dot{W}(\bm{v}) &=\max_{\pot\in\potspace:\Vert \BBgrad \pot\Vert_\infty\le 1} \pot^\top\bm{v} = {W}(\pp,\pp+\bm{v}) 
 \label{eq:WspeedDual}
\end{align}
Finally, combining the right side with Eq.~\eqref{eq:dd}, we arrive at the form that appears in the main text:
\begin{align}
\dot{W}(\bm{v})=\min_{\bm{m}\in\fluxspace}\Vert \bm{m}\Vert _{1}\suchthat\BBdiv\bm{m}=\bm{v}\label{eq:ddww}
\end{align}
Observe that the optimal fluxes are absolutely irreversible (either $m_\rr >0$ or $m_{\negedge}>0$ but not both), since otherwise, by removing the reversible contribution, one could decrease the activity $\Vert \bm m\Vert_1$ while still satisfying the constraint $\BBdiv\bm{m}=\bm{v}$. 
\end{newstuffenv}

\begin{newstuffenv}
 \subsection{Derivation of TSL~\eqref{eq:tslW0} and optimal fluxes}

We first rewrite Eq.~\eqref{eq:WspeedDual} as
\begin{align}
 \dot{W} &=\max_{\pot\in\potspace:\Vert \BBgrad \pot\Vert_\infty\le 1} \pot^\top\evoll \,.
 \label{eq:appwspeeddual}
\end{align}
Considering the optimal $\pot$ in the optimization, we may then apply our simplified TUR~\eqref{eq:tur1-act}:
\begin{align}
 \eprEX \ge \frac{2K_{{\pot}}^{(1)} }{\Vert \BBgrad{\pot}\Vert_\infty}\atanh\frac{ K_{{\pot}}^{(1)} }{\Act\Vert \BBgrad{\pot}\Vert_\infty}
 \,. \label{eq:tur-wass0}
\end{align}
Recall that $\Vert \BBgrad{\pot}\Vert_\infty\le 1$ by construction, and that $K_{{\pot}}^{(1)}=\jj^\top \BBgrad {\pot}=\dot{W}$. Plugging in gives the lower bound
\begin{align}
 \eprEX \ge 2\dot{W} \atanh\frac{\dot{W}}{\Act}
 \,. \label{eq:tur-wass}
\end{align}

Next, we use the known TSL for total EPR~\cite{nagayama2025infinite}:
\begin{align}
\min_{\bm{g}\in\mathcal{S}}\epr(\bm{g})= 2\dot{W} \atanh\frac{\dot{W}}{\Act}\,,
\label{eq:wassbndepr}
\end{align}
where $\mathcal{S}=\{\bm{g} \in \fluxspace:\BBdiv\bm{g}=\evoll,\left\Vert \bm{g}\right\Vert _{1}=\Act\}$ is the set of fluxes with net production $\evoll$ and dynamical activity $\Act$. 
Recall the definition~\eqref{eq:tsldef} of $\eprTSL:=\min_{\bm{g}\in\mathcal{S}}\eprEX(\bm{g})$. Since $\epr \ge \eprEX$ always, we may combine with Eq.~\eqref{eq:wassbndepr} to write
\begin{align}
 \eprTSL\le 2\dot{W} \atanh\frac{\dot{W} }{\Act}\,.
 \label{eq:tur-wass3}
\end{align}
Combining Eqs.~\eqref{eq:tur-wass} and \eqref{eq:tur-wass3} implies $\eprTSL=2\dot{W} \atanh({\dot{W}}/{\Act})$, as in Eq.~\eqref{eq:tslW0}.

We now provide the form of the optimal fluxes $\jj^\star$ in Eq.~\eqref{eq:wassbndepr} and demonstrate that they have conservative forces. Thus, for these fluxes, we have
\begin{align*}
 \eprEX(\jj^\star)=\epr(\jj^\star) = \eprTSL = 2\dot{W} \atanh\frac{\dot{W} }{\Act}
\end{align*}

Given a system with net production $\evoll$ and activity $\Act$, let $\dot{W}$ be the Wasserstein speed, $\bm g$ the optimal fluxes in the primal formulation~\eqref{eq:ddww}, and $\pot^\prime$ the optimal potential in the dual formulation~\eqref{eq:appwspeeddual}. Recall that the  optimal fluxes $\bm g$ in Eq.~\eqref{eq:ddww} obey $\BBdiv \bm g=\evoll$ and are absolutely irreversible, so 
$\sum_{\rr}\vert g_\rr - g_{\negedge}\vert=2\Vert \bm{g}\Vert_1 =2\dot{W}$ (the factor 2 arises because each one-way reaction $\rr$ is counted twice in the first sum). 

To derive the optimal fluxes $\jj^\star$, we use the argument found in Appendix~K in Ref.~\cite{nagayama2025infinite}. We first define another potential,
\begin{align}
 \potopt=-2\pot^\prime\atanh\frac{\dot{W} }{\Act}
\end{align}
and the following fluxes,
\begin{align}
\jjrr^\star =\frac{(g_\rr -g_{\negedge})e^{[\negBBgrad \potopt]_\rr}}{e^{[\negBBgrad \potopt]_\rr}-1}\qquad \jjRrr^\star =\frac{g_\rr-g_{\negedge}}{e^{[\negBBgrad \potopt]_\rr}-1}\label{eq:optfluxx}
\end{align}

We show that the fluxes $\jj^\star$ have net production $\BBdiv \jj^\star = \evoll$, activity $\Act$, and EPR $\epr = 2\dot{W}\atanh({\dot{W}}/{\Act})$, thus they are optimal in Eq.~\eqref{eq:wassbndepr}.  The net production obeys
\begin{align*}
 \BBdiv\jj^{\star}&=\frac{1}{2}\BBdiv(\jj^{\star}-\jjR^{\star}) \\&=\frac{1}{2}\BBdiv(\bm{g}-\tilde{\bm{g}})=\BBdiv\bm{g}=\evoll%
\end{align*}
In the second line, we used  definition~\eqref{eq:optfluxx} and $\BBdiv\bm{g}=\evoll$. 

To calculate the dynamical activity, we consider the activity contributed by each pair of nonzero fluxes $(\jjrr,\jjRrr)$:
\begin{align*}
\jjrr^\star+\jjRrr^\star&= \frac{e^{[\negBBgrad\potopt]_\rr}+1}{e^{[\negBBgrad\potopt]_\rr}-1}(g_{\rr}-g_{\negedge})\\
 & =\frac{g_{\rr}-g_{\negedge}}{\tanh([\negBBgrad\potopt]_\rr/2)}\\
 & =\frac{g_{\rr}-g_{\negedge}}{\tanh\left(\atanh(\dot{W}/\Act)[\BBgrad \pot^\prime]_\rr\right)}\\
 & =\frac{g_{\rr}-g_{\negedge}}{\tanh\left(\atanh(\dot{W}/ \Act)\frac{g_{\rr}-g_{\negedge}}{|g_{\rr}-g_{\negedge}|}\right)}\\
 & =\frac{g_{\rr}-g_{\negedge}}{\frac{\dot{W}}{ \Act}\frac{g_{\rr}-g_{\negedge}}{|g_{\rr}-g_{\negedge}|}}=\frac{ \Act}{\dot{W}}|g_{\rr}-g_{\negedge}| %
\end{align*}
In the fourth line, we used the known property~\cite[Eq.~(G3)]{nagayama2025infinite}
\begin{align}
 [\BBgrad \pot^\prime]_\rr=(g_{\rr}-g_{\negedge})/{|g_{\rr}-g_{\negedge}|}\label{eq:fjsdn}
\end{align}
In the last line, we used that $\tanh(\atanh(x)\gamma)=\gamma x$ for $\gamma\in\{-1,1\}$. Since $\sum_{(\rr,\negedge)}|g_{\rr}-g_{\negedge}|=\dot{W}$ (summed  across reversible reaction pairs), the activity of $\jj^\star$ is equal to $\Act$. 

Finally, we compute the EPR as
\begin{align*}
\epr(\jj^{\star}) =-\jj^{\star\top}\BBgrad\potopt =-\bm{g}^{\top}\BBgrad\potopt
\end{align*}
where we used $\BBdiv\jj^{\star}=\BBdiv\bm{g}$. Using the definition of $\potopt$,
\begin{align*}
 \epr(\jj^{\star})& =\Big(2\atanh\frac{\dot{W}}{ \Act}\Big) \bm{g}^{\top}\BBgrad\pot^{\prime}\\
 & =\Big(\atanh\frac{\dot{W}}{ \Act}\Big)(\bm{g}-\tilde{\bm{g}})^{\top}\BBgrad\pot^{\prime}\\
 & =\Big(\atanh\frac{\dot{W}}{ \Act}\Big)\vert\bm{g}-\tilde{\bm{g}}\vert^{\top}\oneone\\
 & =2\dot{W}\atanh\frac{\dot{W}}{\Act}
\end{align*}
where we used Eq.~\eqref{eq:fjsdn} in the penultimate line.
\end{newstuffenv}

\section{Comparison to other decompositions}

\label{app:sm6}

Here we compare our approach to several excess/housekeeping decompositions
that have been previously proposed in the literature.

\subsection{Euclidean-Onsager decomposition}

\label{app:ons}

In this paper, we considered the excess/housekeeping decomposition
based on relative entropy between flux vectors. In our previous work~\cite{kohei2022}, we explored a
 decomposition based on a Euclidean distance, as was also mentioned in Section~\ref{sec:comparison-variational} in the main text.

In particular, the distance between forces $\mathscr{D}$, defined
using relative entropy in Eq.~\eqref{eq:ddd3}, was instead defined
using a generalized squared Euclidean norm,
\begin{align}
\mathscr{D}^{\prime}(\ee\Vert\ee^{\prime}):=\left\Vert \ee-\ee^{\prime}\right\Vert _{L}^{2}:=(\ee-\ee^{\prime})^{\top}L(\ee-\ee^{\prime}),\label{eq:distons}
\end{align}
where $L\in\mathbb{R}_{+}^{\numedge\times\numedge}$ is a diagonal
matrix with entries $L_{\rr\rr}=\frac{1}{2}(\jjrr-\jjRrr)/\ffrr$. $L$ \newstuff{has been previously called the ``nonequilibrium conductance matrix''~\cite{vroylandt2018degree,raux2024thermodynamic}}, and it can be
understood as an Onsager-type matrix that specifies a linear relationship
between forces and net fluxes, $\jj-\jjR=L\ff$, thus we may term
the decomposition as the \emph{Euclidean-Onsager decomposition}. Note
that the factor 1/2 appears in our definition of $L$, but not in
Ref.~\citep{kohei2022}, due to a minor change of convention (in
this paper we consider reversible reactions as two separate reactions).

Using Eq.~\eqref{eq:distons}, the overall EPR can be written as
$\epr=\left\Vert \ff-\zz\right\Vert _{L}^{2}$. The housekeeping
EPR and generalized potential may be defined via a Euclidean projection,
\begin{equation}
\eprHKons=\min_{\pot\in\potspace}\left\Vert \ff-(\negBBgrad\pot)\right\Vert _{L}^{2}\equiv\left\Vert \ff-(\negBBgrad\potons)\right\Vert _{L}^{2}\,.\label{eq:onshk}
\end{equation}
This is the Euclidean analogue of Eq.~\eqref{eq:hkdef-1}. Also, the (Euclidean version of the) Pythagorean relation~\eqref{eq:pyth} applies in a similar 
way. We note that Eq.~\eqref{eq:optons} in the main text can be derived from Eq.~\eqref{eq:onshk} by expanding, dropping the term $\ff^\top L \ff$ (which does not depend on $\pot$) and using $\BBdiv L \ff =\BBdiv \jj$.

As discussed in Ref.~\citep{kohei2022}, the Euclidean-Onsager excess
EPR can be physically interpreted as the minimal EPR achievable by
manipulating forces while keeping the Onsager-type coefficients
$L$ fixed. Our information-geometric decomposition is not interpreted
in terms of a minimal EPR principle, but rather in terms of dynamical
large deviations and thermodynamic uncertainty relations. The information-geometric
approach is arguably more natural in the far-from-equilibrium regime,
where the relationship between forces and net fluxes becomes nonlinear.
Unlike the bounds derived in Ref.~\citep{kohei2022}, our approach
leads to speed limits that may be tight far-from-equilibrium and far-from-stationarity.

We now show that the two approaches agree to third order in the linear-response
regime near equilibrium. We also show that our excess EPR is always
smaller than the Euclidean-Onsager excess EPR.

Recall that for systems without odd variables, each reaction $\rr$
is paired with a unique reverse reaction $\negedge$ such that $\ffrrR=-\ffrr$.
Consider the relative entropy between the forward fluxes $\jj=\jjR\circ e^{\ff}$
and any other $\jjR\circ e^{\ee}$, where $\ee$ is anti-symmetric
($\eerr=-\eerrR$): 
\begin{align*}
\DDexpFam(\ff\Vert\ee) & =\sum_{\rr}\jjrr(e^{-(\ffrr-\eerr)}+(\ffrr-\eerr)-1)\\
 & =\sum_{\rr}\jjRrr(e^{\ffrr-\eerr}-(\ffrr-\eerr)-1).
\end{align*}
In the second line, we used anti-symmetry of $\ff$ and $\ee$. Combining
these expressions, and using $\jjrr=e^{\ffrr}\jjRrr$, gives 
\begin{align*}
\DDexpFam(\ff\Vert\ee) & =\frac{1}{2}\sum_{\rr}\jjRrr\big((e^{\ffrr-\eerr}-(\ffrr-\eerr)-1)\\
 & \qquad\quad+e^{\ffrr}(e^{-(\ffrr-\eerr)}+(\ffrr-\eerr)-1)\big).
\end{align*}
We now rewrite the right hand side as 
\begin{align}
\DDexpFam(\ff\Vert\ee) & =\frac{1}{2}\sum_{\rr}\jjRrr\big(h(\ffrr-\eerr,\ffrr)+\frac{e^{\ffrr}-1}{\ffrr}(\ffrr-\eerr)^{2}\big)\nonumber \\
 & =\frac{1}{2}\sum_{\rr}\jjRrr h(\ffrr-\eerr,\ffrr)+\left\Vert \ff-\ee\right\Vert _{L}^{2},\!\!\!\!\!\!\label{eq:d1}
\end{align}
where for convenience we defined the following function: 
\[
h(a,b)=\Big[\frac{(e^{a}-a-1)+e^{b}(e^{-a}+a-1)}{a^{2}}-\frac{e^{b}-1}{b}\Big]a^{2}.
\]
It can be verified 
that the term inside the brackets is always nonnegative. Thus, $h$ is nonnegative and  $\DDexpFam(\ff\Vert\ee)\ge\left\Vert \ff-\ee\right\Vert _{L}^{2}$
given Eq.~\eqref{eq:d1}. Finally, since $\ee=\negBBgrad\pot$ is
anti-symmetric, we arrive at the inequality between the information-geometric
and Euclidean-Onsager housekeeping EPR:
\begin{equation}
\eprHK=\min_{\pot}\DDexpFam(\ff\Vert\negBBgrad\pot)\ge\min_{\pot}\left\Vert \ff-(\negBBgrad\pot)\right\Vert _{L}^{2}=\eprHKons.\label{eq:onsineq}
\end{equation}
For a numerical comparison, see Section~\ref{app:numericalcomp}.

We now consider the limit in which the two decompositions agree. Using
the derivations above, we have the bounds 
\begin{align}
0\le\eprHK-\eprHKons\le\frac{1}{2}\sum_{\rr}\jjRrr\,h(\ffrr+[\BBgrad\potons]_{\rr},\ffrr).\label{eq:d2-1}
\end{align}
The function $h(a,b)$ vanishes to first order around $a=b$ and $a=0$
(in general, $h(a,b)$ is symmetric under the transformation $a\mapsto b-a$).
Given Eq.~\eqref{eq:d2-1}, considering $a=b$ implies that $\eprHK$
and $\eprHKons$ agree to first order around $\BBgrad\potons=0$
(steady state) while considering $a=0$ implies that they agree to first order
around $\ff=\negBBgrad\potons$ (conservative forces in a passive
system). We can ask if they also agree to second order there. A Taylor
expansion of $h(a,b)$ shows that second order terms do not vanish
except in the limit $b\to0$. Given Eq.~\eqref{eq:d2-1},
this is the equilibrium limit $f_{\rr}\to0$, where the 
force across each reaction vanishes. Note that 
\[
\alpha\left\Vert \ff\right\Vert ^{2}\ge\left\Vert \ff\right\Vert _{L}^{2}\ge\left\Vert \ff+\BBgrad\potons\right\Vert _{L}^{2}\ge \beta \left\Vert \ff+\BBgrad\potons\right\Vert ^{2}
\]
where $\alpha=\max(\jjrr+\jjRrr)/2$, $\beta =\min_{\rr}\sqrt{\jjrr\jjRrr}$,
and $\Vert\cdot\Vert$ is the usual Euclidean norm. The middle inequality comes from Eq.~\eqref{eq:onshk}, the others from bounds between the
arithmetic, geometric, and logarithmic mean. Thus, if $f_{\rr}\to0$,
then $\ffrr+[\BBgrad\potons]_{\rr}$, the first argument
of $h$ in Eq.~\eqref{eq:d2-1}, also vanishes. We expand $h$
in each argument and rearrange to give 
\begin{align}
h(\gamma,f)=\frac{1}{12}\gamma^{2}\left(\gamma-f\right)^{2}+\mathcal{O}(\epsilon^{5})=\mathcal{O}(\epsilon^{4})
\end{align}
for $f,\gamma\sim\epsilon$. Plugging into Eq.~\eqref{eq:d2-1}
shows that $\eprHK$ and $\eprHKons$ agree to third order in the
equilibrium limit.

\subsection{Comparison with Hessian decomposition~\NoCaseChange{\citep{Kobayashi2022}} and Euclidean-Onsager decomposition~\NoCaseChange{\citep{kohei2022}}}
\label{app:hesscomp}

\global\long\def\netforce{\bm{\FreeEnergy}}%
\global\long\def\netforcerr{\FreeEnergy_{r}}%
\global\long\def\barBBdiv{\breve{\BBbase}^{\top}}%
\global\long\def\barBBgrad{\breve{\BBbase}}%
\global\long\def\netflux{\bm{\mathcal{J}}}%
\global\long\def\netjR{\mathcal{J}_{r}}%

\label{app:numericalcomp} We numerically compare the excess/housekeeping
decomposition from this paper with two others: the Euclidean-Onsager
decomposition described in Section~\ref{app:ons}, the ``Hessian
decomposition'' recently proposed in Ref.~\citep{Kobayashi2022}
(see also Ref.~\citep{kobayashi2024information}). While an inequality
exists between the Euclidean-Onsager decomposition and our decomposition,
Eq.~\eqref{eq:onsineq}, no inequality between the Hessian decomposition
and the others has been proved analytically. Nonetheless, our numerical
results prove that they are different. 

We briefly summarized some aspects of the Hessian decomposition in Section~\ref{sec:comparison2} of the main text. 
However, to be self-contained, here we provide a more thorough summary of the Hessian decomposition
from Ref.~\citep{Kobayashi2022}, also keeping closer to the original notation. 

Consider a system without
odd variables that has reversible $\numedge$ reactions, having forward
and reverse fluxes $\jjrr$ and $\jjRrr$. We use $r\in\{1,2,\dots,M/2\}$
to label each pair of one-way reactions $\rr$ and $\negedge$, where
the forward/reverse fluxes of the pair are indicated as $\jjrrCRN=\jjrr$
and $\jjRevrrCRN=\jjRrr$. We define a vector of currents (net fluxes)
$\netflux\in\mathbb{R}^{M/2}$ as $\netjR:=\jjrrCRN-\jjRevrrCRN$,
a vector of ``frenetic activities'' $\bm{\omega}\in\mathbb{R}_{+}^{M/2}$
as $\omega_{r}:=2\sqrt{\jjrrCRN\jjRevrrCRN}$, and a vector of (half)forces
$\netforce\in\mathbb{R}^{M/2}$ as $\netforcerr=\frac{1}{2}\ln(\jjrrCRN/\jjRevrrCRN)$.
We use the notation $\barBBdiv$ to indicate the $\numstate\times\numedge/2$
matrix that only has columns for the forward reaction ($\rr$) in
each pair $(\rr,\negedge)$. $\barBBdiv$ maps currents to state
evolution: $\dtpp=\barBBdiv\netflux=\BBdiv\jj$. Note that Ref.~\citep{Kobayashi2022}
uses the convention that forces $\netforcerr=\frac{1}{2}\ln(\jjrrCRN/\jjRevrrCRN)$
are scaled by $1/2$ relative to the forces as defined in this paper,
$f_{r}=\ln(\jjrrCRN/\jjRevrrCRN)$. 

The currents can be expressed as 
\[
\netjR=\omega_{r}\sinh(\netforcerr)=%
\jjrrCRN-\jjRevrrCRN.
\]
This equation can be solved for $\netforcerr$ as $\netforcerr=\sinh^{-1}(\netjR/\omega_{r})$. 
These relations can also be derived from a canonical structure.
Define two dual convex functions which are the Legendre conjugate
of each other: for a fixed $\bm{\omega}$, the convex function 
\[
\Psi_{\omega}(\netflux'):=\sum_{r}\left[\netjR'\sinh^{-1}\frac{\netjR'}{\omega_{r}}-\omega_{r}\left[\sqrt{1+\Big(\frac{\netjR'}{\omega_{r}}\big)^{2}}-1\right]\right]
\]
is the Legendre conjugate of 
\begin{align*}
\Psi_{\omega}^{*}(\netforce')=\sum_{r}\omega_{r}\big[\cosh(\netforcerr')-1\big],
\end{align*}
and they specify the current and force across reaction $r$ as 
\begin{align*}
\netjR=\partial_{\netforcerr}\Psi_{\omega}^{*}(\netforce),\quad\netforcerr=\partial_{\netjR}\Psi_{\omega}(\netflux).
\end{align*}
Note that for any $\bm{\omega}$, these functions achieve their   minima at $\Psi_{\omega}(\zz)=\Psi_{\omega}^{*}(\zz)=0$.

\begin{figure}
\includegraphics[width=1\columnwidth]{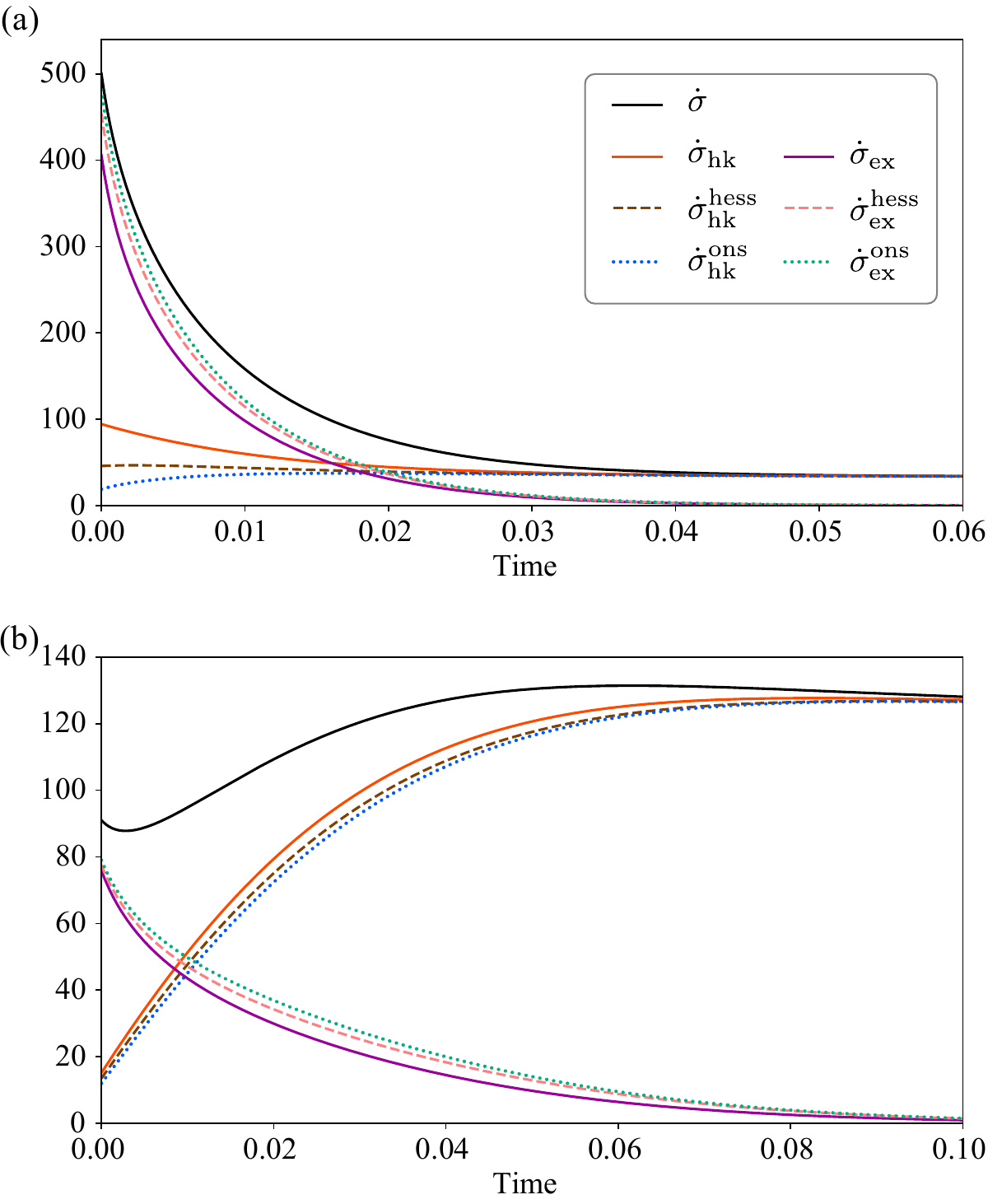} \caption{Comparison of three EPR decompositions. We calculate EPRs of the chemical
reaction network in Eq.~\eqref{eq:appcrn} for two distinct rate
constants (detailed values are given in the text).}
\label{fig:compareeprs} 
\end{figure}

In general, a convex function $\varphi(\bm{x})$ 
induces the Bregman divergence $D_\varphi(\bm{x},\bm{x}'):=\varphi(\bm{x})-\varphi(\bm{x}')-\langle\bm{x}-\bm{x}',\grad \varphi(\bm{x}')\rangle\geq0$,
where $\langle\cdot,\cdot\rangle$ is the normal inner product and
$\grad\varphi(\bm{x})=(\partial_{x_{1}}\varphi(\bm{x}),\partial_{x_{2}}\varphi(\bm{x}),\dots)^{\top}$ is the gradient vector 
of the function~\citep{amari2016information}. For a fixed $\bm{\omega}$,
we can define the Bregman divergences $D_{\omega}$ and the dual one
$D_{\omega}^{*}$ by 
\begin{align*}
&D_{\omega}(\netflux',\netflux''):= \\
&\qquad \Psi_{\omega}(\netflux')-\Psi_{\omega}(\netflux'')-\langle\netflux'-\netflux'',\grad\Psi_{\omega}(\netflux'')\rangle\\
&D_{\omega}^{*}(\netforce',\netforce''):= \\
&\qquad \Psi_{\omega}^{*}(\netforce') - \Psi_{\omega}^{*}(\netforce'') - \langle\netforce'-\netforce'',\grad \Psi_{\omega}^{*}(\netforce'')\rangle
\end{align*}
As a general property of Bregman divergences and the Legendre transformation,
we have 
\begin{align}
D_{\omega}(\netflux',\netflux'')=D_{\omega}^{*}(\netforce'',\netforce')
\end{align}
when $(\netflux',\netforce')$ and $(\netflux'',\netforce'')$ are
Legendre dual coordinates. In this situation, we also have 
\begin{align}
D_{\omega}(\netflux',\netflux'')=\Psi_{\omega}(\netflux')+\Psi_{\omega}^{*}(\netforce'')-\langle\netflux',\netforce''\rangle,
\end{align}
which leads to 
\begin{align}
\epr=\langle\netflux,\netforce\rangle=\Psi_{\omega}(\netflux)+\Psi_{\omega}^{*}(\netforce).
\end{align}
In general, these Bregman divergences cannot be expressed
directly in terms of the relative entropy, because the current $\netjR$ can be negative.
Therefore, they do not relate to the EPR in the same direct way
as the relative entropy 
$\mathcal{D}$ in Eq.~\eqref{eq:eprn}.

The Hessian decomposition~\citep{Kobayashi2022} is defined by using
two special points: $(\netflux_{\eq},\netforce_{\eq})$, which represent
conservative currents/forces, and $(\netflux_{\sss},\netforce_{\sss})$,
which represent steady-state currents/forces. Given these two pairs
of currents/forces, we have 
\begin{align}
\eprEX^{\text{hess}}:=\Psi_{\omega}(\netflux_{\eq})+D_{\omega}^{*}(\netforce,\netforce_{\sss}),\\
\eprHK^{\text{hess}}:=\Psi_{\omega}^{*}(\netforce_{\sss})+D_{\omega}(\netflux,\netflux_{\eq}).
\end{align}
To explain how $(\netflux_{\eq},\netforce_{\eq})$ and $(\netflux_{\sss},\netforce_{\sss})$
are determined, we define two kinds of sets. $\mathcal{P}(\netflux')$
is the set of currents that produce the same dynamics as $\netflux'$,
\begin{align}
\mathcal{P}(\netflux'):=\{\netflux''\in\mathbb{R}^{M/2}\mid\barBBdiv\netflux''=\barBBdiv\netflux'\}.
\end{align}
$\mathcal{M}_{\omega}(\netforce')$ is defined as
the set of currents that are induced by $\netforce'$ plus some conservative
force, 
\begin{align}
\mathcal{M}_{\omega}(\netforce'):=\{\grad\Psi_{\omega}^{*}(\netforce'')\mid\netforce''\in\netforce'+\mathrm{im}\,\barBBgrad\},
\end{align}
where $\netforce'+\mathrm{im}\,\barBBgrad:=\{\netforce'+\barBBgrad\pot\mid\pot\in\mathbb{R}^{N}\}$.
Then, $\netflux_{\eq}$ and $\netflux_{\sss}$ are given as unique
intersections as 
\begin{align}
\netflux_{\eq}:=\mathcal{P}(\netflux)\cap\mathcal{M}_{\omega}(\zz),\;\;\netflux_{\sss}:=\mathcal{P}(\zz)\cap\mathcal{M}_{\omega}(\netforce),
\end{align}
while the corresponding forces $\netforce_{\eq}$ and $\netforce_{\sss}$
are given by $\grad\Psi_{\omega}(\netflux_{\eq})$ and $\grad\Psi_{\omega}(\netflux_{\sss})$.
Therefore, holding the frenetic activity fixed, $\netflux_{\eq}$
is the current induced by a conservative force which recovers the
original dynamics, while $\netflux_{\sss}$ is the steady-state current
given by the force that has the same nonconservative contribution
as the actual force.

We note the variational characterizations of $(\netflux_{\eq},\netforce_{\eq})$
and $(\netflux_{\sss},\netforce_{\sss})$, which simplify numerical
calculation of the decomposition. $\netflux_{\eq}$ is given
by 
\begin{align}
\netflux_{\eq}=\argmin_{\netflux'\in\mathcal{P}(\netflux)}\Psi_{\omega}(\netflux') .
\end{align}
\newstuff{In the main text, this variational principle appeared in its dual form in Eq.~\eqref{eq:hgeq}, though in slightly different notation.} $\netforce_{\sss}$ is obtained as 
\begin{align}
\netforce_{\sss}=\argmin_{\netforce'\in\netforce+\mathrm{im}\,\barBBgrad}\Psi_{\omega}^{*}(\netforce').
\end{align}
\newstuff{In the main text, this variational expression appeared as Eq.~\eqref{eq:hgss}, though in slightly different notation.}

Next, we consider a specific CRN as an example. In Ref.~\citep{Kobayashi2022},
the authors discuss the reaction network 
\begin{align}
2X\underset{k_{1}^{\leftarrow}}{\overset{k_{1}^{\rightarrow}}{\rightleftarrows}}2Y\underset{k_{2}^{\leftarrow}}{\overset{k_{2}^{\rightarrow}}{\rightleftarrows}}X+Y\underset{k_{3}^{\leftarrow}}{\overset{k_{3}^{\rightarrow}}{\rightleftarrows}}2X,\label{eq:appcrn}
\end{align}
with mass-action kinetics. 
We calculate excess and housekeeping EPR for our decomposition ($\eprHK,\eprEX$),
the Euclidean-Onsager decomposition ($\eprHKons,\eprEXons$), and the Hessian
decomposition ($\eprHKhess,\eprEXhess$). We use the same
parameters as Ref.~\citep{Kobayashi2022}. Concretely, we use the
rate constants $k_{1}^{\rightarrow}=1/2,k_{1}^{\leftarrow}=2,k_{2}^{\rightarrow}=4,k_{2}^{\leftarrow}=47/4,k_{3}^{\rightarrow}=\sqrt{2},$
and $k_{3}^{\leftarrow}=15/2+2\sqrt{2}$ to obtain (a) in Figure~\ref{fig:compareeprs},
or $k_{1}^{\rightarrow}=1/2,k_{1}^{\leftarrow}=2,k_{2}^{\rightarrow}=1/17,k_{2}^{\leftarrow}=85/8,k_{3}^{\rightarrow}=273/68,$
and $k_{3}^{\leftarrow}=137/68$ to obtain (b). The three decompositions
are exhibited in Figure~\ref{fig:compareeprs}, which reproduces numerical
results obtained in Ref.~\citep{Kobayashi2022}. The inequality $\eprEX\leq\eprEXons$
is also verified. In addition, we observe numerically that $\eprEX\leq\eprEXhess \leq\eprEXons$,
although we have not proved analytically that these inequalities hold
in general.

\begin{newstuffenv}

\subsection{Maes-Netočný decomposition}
\label{app:mn}

Maes and Netočný (MN) proposed an excess/housekeeping decomposition
for Langevin systems~\citep{maes2014nonequilibrium}. In later work,
the MN decomposition was shown to have a geometric interpretation
in terms of Euclidean projections and optimal transport~\citep{dechant2022geometric}. 
Recently, it has also been generalized to hydrodynamic systems~\citep{yoshimuraTwoApplicationsStochastic2024}.
Finally, as mentioned in the main text, the MN decomposition is recovered by the Euclidean-Onsager
decomposition in the continuum limit of an MJP~\citep[Appendix~B]{kohei2022}.
The continuum limit corresponds to the linear-response regime near equilibrium.

The variational expression~\eqref{eq:varMN} can be derived in the following way. First, we write the MN excess EPR as
\begin{align}
\eprEX^{\text{MN}}&=\max_{\varphi:\mathbb{R}^{k}\to\mathbb{R}}\int\big( 2\,\bm{\mathrm{j}}\cdot\grad\varphi-\rho_t \left\Vert \grad\varphi\right\Vert ^{2} \big) d\bm{r} \label{eq:varMN2}
\end{align}
such that the optimizer is the potential that appears in Eq.~\eqref{eq:varMN}. Without loss of generality, we may add an inner optimization over scalar multipliers $\lambda$,
\begin{align*}
\eprEX^{\text{MN}} &=\max_{\varphi:\mathbb{R}^{k}\to\mathbb{R}}\max_{\lambda \in\mathbb{R}}\int\big( 2\lambda\,\bm{\mathrm{j}}\cdot\grad\varphi-\lambda^2\rho_t \left\Vert \grad\varphi\right\Vert ^{2} \big) d\bm{r} 
\end{align*}
The inner optimization can be solved to give $\lambda^*=({\int \bm{\mathrm{j}}\cdot\grad\varphi\, d\bm{r} })/({\int \rho_t \left\Vert \grad\varphi\right\Vert ^{2} \, d\bm{r}})$. Plugging in and simplifying allows us to rewrite Eq.~\eqref{eq:varMN2} in an equivalent form,
\begin{align}
    \eprEX^{\text{MN}} = \max_{\varphi:\mathbb{R}^{k}\to\mathbb{R}} \frac{(\int \bm{\mathrm{j}}\cdot\grad\varphi\, d\bm{r} )^2}{\int \rho_t \left\Vert \grad\varphi\right\Vert ^{2} \, d\bm{r} }
\end{align}
This expression for the MN excess EPR appeared as Eq.~(13) in Ref.~\cite{dechant2022geometric}.
\end{newstuffenv}

\subsection{Hatano-Sasa (HS) decomposition}

\label{app:hs}

We derive an inequality between our excess/housekeeping
EPR and the HS excess/housekeeping EPR:
\begin{align}
\eprHK\le\eprHKhs\qquad\eprEX\ge\eprEXhs\label{eq:appbndHS}
\end{align}
Our derivations apply to (1) MJPs without odd variables, (2) MJPs
with odd variables and time-symmetric steady states, and (3) deterministic
CRNs with complex balance and mass-action kinetics. In all cases,
we show that the HS housekeeping EPR can be written as the generalized
relative entropy, 
\begin{align}
\eprHKhs=\DDexpFam(\ff\Vert\negBBgrad\pot^{\sss}),\label{eq:df2}
\end{align}
where $\potssx:=\ln(\ppx{\xx}/\ppssx{\xx})$ is defined via the steady-state
distribution $\ppss$. Since our housekeeping EPR satisfies the variational
principle in~\eqref{eq:hkdef}, Eq.~\eqref{eq:df2} implies Eq.~\eqref{eq:appbndHS}.

We first consider the simplest case, MJPs without odd variables. The
HS excess and housekeeping terms are given by~\citep{esposito2007entropy,esposito2010threefaces}
\begin{align}
\eprEXhs & =\sum_{\yy\ne\xx,\bath}\pppx{\xx}\Rjia\ln\frac{\pppx{\xx}\pppssx{\yy}}{\pppx{\yy}\pppssx{\xx}}\label{eq:hs01}\\
\eprHKhs & =\epr-\eprEXhs=\sum_{\yy\ne\xx,\bath}\pppx{\xx}\Rjia\ln\frac{\pppx{\xx}\Rjia}{\pppx{\xx}\Rija\pppssx{\yy}/\pppssx{\xx}}.\label{eq:hs01b}
\end{align}
Within our exponential family Eq.~\eqref{eq:expfam}, the conservative
force $\negBBgrad\pot^{\mathrm{ss}}$ specifies the fluxes 
\begin{align*}
[\jjR\circ e^{\negBBgrad\potss}]_{\xx\to\yy;\bath} & =\pppx{\yy}\Rija e^{\ln(\pppx{\xx}/\pppssx{\xx})-\ln(\pppx{\yy}/\pppssx{\yy})}=\pppx{\xx}\Rija\frac{\pppssx{\yy}}{\pppssx{\xx}}
\end{align*}
In the notation~\eqref{eq:ddd3}, this leads to Eq.~\eqref{eq:df2},
\begin{align}
 & \DDexpFam(\ff\Vert\negBBgrad\pot^{\sss})\nonumber\\
 & =\sum_{\yy\ne\xx,\bath}\Big(\pppx{\xx}\Rjia\ln\frac{\pppx{\xx}\Rjia}{\pppx{\xx}\Rija\pppssx{\yy}/\pppssx{\xx}}-\pppx{\xx}\Rjia+\pppx{\xx}\Rija\frac{\pppssx{\yy}}{\ppssx{\xx}}\Big)\nonumber \\
 & =\sum_{\yy\ne\xx,\bath}\pppx{\xx}\Rjia\ln\frac{\pppx{\xx}\Rjia}{\pppx{\xx}\Rija\pppssx{\yy}/\pppssx{\xx}}=\eprHKhs,\label{eq:ddd-1}
\end{align}
where in the second line we used that 
\begin{align*}
&\sum_{\yy\ne\xx,\bath}\Big(\pppx{\xx}\Rija\frac{\pppssx{\yy}}{\pppssx{\xx}}-\pppx{\xx}\Rjia\Big)\\
&\quad=\sum_{\xx}\frac{\pppx{\xx}}{\pppssx{\xx}}\sum_{\yy(\ne\xx),\bath}\Big(\pppssx{\yy}\Rija-\pppssx{\xx}\Rjia\Big)=0,
\end{align*}
which follows since $\pppss$ is a steady-state distribution.

Next, we consider MJPs with odd variables, under the assumption that
the steady-state distribution is symmetric under conjugation of odd
variables,
\begin{equation}
\pppssx{\xx}=\pppssx{\oddconj\xx}.\label{eq:symm}
\end{equation}
(Note that this condition is violated in our example of the particle
on a ring from Appendix~\ref{app:odd-example} whenever $\gamma\ne0$.)
The EPR is given by
\begin{align}
\epr=\sum_{\yy\ne\xx,\bath}\Big(\pppx{\xx}\Rjia\ln\frac{\pppx{\xx}\Rjia}{\pppx{\yy}\Reieja}-\pppx{\xx}\Rjia+\pppx{\yy}\Reieja\Big),
\end{align}
as discussed near Eq.~\eqref{eq:eprodd} above. The HS excess EPR
is still defined as in Eq.~\eqref{eq:hs01}, while the HS housekeeping
EPR is the remainder~\citep{spinney2012nonequilibrium,lee2013fluctuation}, $\eprHKhs=\epr-\eprEXhs$. Rearranging gives
\begin{align}
\eprHKhs & =\sum_{\yy\ne\xx,\bath}\big(\pppx{\xx}\Rjia\ln\frac{\pppx{\xx}\Rjia}{\pppx{\xx}\Reieja\pppssx{\yy}/\pppssx{\xx}}-\pppx{\xx}\Rjia+\pppx{\yy}\Reieja\big).
\label{eq:cczv23}
\end{align}
In the parameterized family of fluxes, we have 
\begin{align*}
[\jjR\circ e^{\negBBgrad\potss}]_{\xx\to\yy;\bath} & =\pppx{\yy}\Reieja e^{\ln\pppx{\xx}/\pppssx{\xx}-\ln\pppx{\yy}/\pppssx{\yy}} =\pppx{\xx}\Reieja\frac{\pppssx{\yy}}{\pppssx{\xx}}
\end{align*}
This leads to the following value of the relative entropy:
\begin{multline}
\DDexpFam(\ff\Vert\negBBgrad\pot^{\sss})=\\
\sum_{\yy\ne\xx,\bath}\Big(\pppx{\xx}\Rjia\ln\frac{\pppx{\xx}\Rjia}{\pppx{\xx}\Reieja\pppssx{\yy}/\pppssx{\xx}}-\pppx{\xx}\Rjia+\pppx{\xx}\Reieja\frac{\pppssx{\yy}}{\pppssx{\xx}}\Big)\label{eq:zcxv}
\end{multline}
Finally, we have 
\begin{align*}
 & \sum_{\yy\ne\xx,\bath}\Big(\pppx{\xx}\Reieja\frac{\pppssx{\yy}}{\pppssx{\xx}}-\pppx{\yy}\Reieja\Big)\\
 & =\sum_{\yy\ne\xx,\bath}\Big(\pppx{\xx}\Reieja\frac{\pppssx{\yy}}{\pppssx{\xx}}-\pppx{\xx}\Rejeia\Big)\\
 & =\sum_{\xx}\frac{\pppx{\xx}}{\pppssx{\xx}}\sum_{\yy(\ne\xx),\bath}\Big(\Reieja\pppssx{\yy}-\Rejeia\pppssx{\xx}\Big)\\
 & =\sum_{\xx}\frac{\pppx{\xx}}{\pppssx{\xx}}\sum_{\yy(\ne\xx),\bath}\Big(\Reieja\pppssx{\oddconj\yy}-\Rejeia\pppssx{\oddconj\xx}\Big)=0,
\end{align*}
where we used the symmetry~\eqref{eq:symm}. Plugging $\sum_{\yy\ne\xx,\bath}\pppx{\xx}\Reieja\pppssx{\yy}/\pppssx{\xx}=\sum_{\yy\ne\xx,\bath}\pppx{\yy}\Reieja$
into Eq.~\eqref{eq:zcxv} shows it is the same as the expression for $\eprHKhs$ in Eq.~\eqref{eq:cczv23}, thus giving Eq.~\eqref{eq:df2}.

\global\long\def\ww{\bm{\eta}}%
\global\long\def\wwxx{{\eta}_{\xx}}%
\global\long\def\BBxxrr{\nabla_{\rr\xx}}%
\global\long\def\ssfluxrr{\mathcal{J}_{\rr}^{\sss}}%
\global\long\def\complexes{\mathscr{C}}%
Finally, we consider deterministic CRNs that obey complex balance,
which means that the net current entering and leaving each chemical
complex vanishes in steady state~\citep{feinbergFoundationsChemicalReaction2019}.
We also assume mass-action kinetics, as in Eq.~\eqref{eq:massaction}.
In that case, the HS excess and housekeeping EPR are given by~\citep{ge2016nonequilibrium,rao2016nonequilibrium}
\begin{align}
\eprEXhs & =-\sum_{\rr}\jjrr\sum_{\xx}\BBxxrr\ln\frac{\cccx{\xx}}{\cccssx{\xx}}.\label{eq:hs03}\\
\eprHKhs & =\epr-\eprEXhs=\sum_{\rr}\jjrr\Bigg(\!\ln\frac{\jjrr}{\jjRrr}+\sum_{\xx}\BBxxrr\ln\frac{\cccx{\xx}}{\cccssx{\xx}}\!\Bigg)\label{eq:hs04}
\end{align}
Using the steady-state potential $\potssx:=\ln(\cccx{\xx}/\cccssx{\xx})$, Eq.~\eqref{eq:hs04}
can be written as 
\begin{align}
 & \eprHKhs=\jj^{\top}(\ff+\BBgrad\pot^{\sss}).\label{eq:hs11}
\end{align}
We also have the expression of the relative entropy as 
\begin{align*}
\DDexpFam(\ff\Vert\negBBgrad\pot^{\sss})= & \jj^{\top}(\ff+\BBgrad\pot^{\sss})-\sum_{\rr}(\jjrr-\jjRrr e^{[\negBBgrad\pot^{\sss}]_{\rr}}).
\end{align*}
Using this result and Eq.~\eqref{eq:hs11}, we prove Eq.~\eqref{eq:df2} by showing
\begin{equation}
\sum_{\rr}(\jjrr-\jjRrr e^{[\negBBgrad\pot^{\sss}]_{\rr}})=0.\label{eq:vv}
\end{equation}

To begin, we split Eq.~\eqref{eq:vv} into
contributions from the forward and backward direction of each reversible
reaction $r$, 
\begin{align}
\sum_{r}(\jjrrCRN-\jjRevrrCRN e^{[\negBBgrad\pot^{\sss}]_{r}})+\sum_{r}(\jjRevrrCRN-\jjrrCRN e^{[{\BBgrad}\pot^{\sss}]_{r}}).\label{eq:ap2s}
\end{align}
For mass-action kinetics Eq.~\eqref{eq:massaction}, each term in
the first sum can be written as 
\begin{align*}
 & \jjrrCRN-\jjRevrrCRN e^{[\negBBgrad\pot^{\sss}]_{r}}\\
 & =k_{r}^{\rightarrow}\prod_{\xx}\cccx{\xx}^{\nu_{\xx r}}-k_{r}^{\leftarrow}\prod_{\xx}\cccx{\xx}^{\kappa_{\xx r}}\prod_{\xx}\Big(\frac{\cccx{\xx}}{\cccssx{\xx}}\Big)^{\nu_{\xx r}-\kappa_{\xx r}}\\
 & =\prod_{\xx}\Big(\frac{\cccx{\xx}}{\cccssx{\xx}}\Big)^{\nu_{\xx r}}\left(k_{r}^{\rightarrow}\prod_{\xx}(\cccssx{\xx})^{\nu_{\xx r}}-k_{r}^{\leftarrow}\prod_{\xx}(\cccssx{\xx})^{\kappa_{\xx r}}\right)\\
 & \equiv\prod_{\xx}\Big(\frac{\cccx{\xx}}{\cccssx{\xx}}\Big)^{\nu_{\xx r}}\ssfluxrr,
\end{align*}
where $\ssfluxrr$ is the current (net flux) across reversible reaction
$r$ in steady state. In a similar way, we write each term in the second
sum in Eq.~\eqref{eq:ap2s} as 
\begin{align}
 & \jjRevrrCRN-\jjrrCRN e^{[\BBgrad\pot^{\sss}]_{r}}=-\prod_{\xx}\Big(\frac{\cccx{\xx}}{\cccssx{\xx}}\Big)^{\kappa_{\xx r}}\ssfluxrr.
\end{align}
Combining, we rewrite Eq.~\eqref{eq:vv} as 
\begin{align}
\sum_{r}\left[\prod_{\xx}\Big(\frac{\cccx{\xx}}{\cccssx{\xx}}\Big)^{\nu_{\xx r}}\ssfluxrr-\prod_{\xx}\Big(\frac{\cccx{\xx}}{\cccssx{\xx}}\Big)^{\kappa_{\xx r}}\ssfluxrr\right].\label{eq:s2}
\end{align}
Now split this sum into contributions from each reactant
complex and each product complex. Let $\complexes$ indicate the set
of reactant and product complexes, where each element of $\complexes$
is a vector $\ww\in\mathbb{N}_{0}^{\numstate}$ with $\wwxx$
being the number of species $\xx$ in complex $\ww$. Let $A(\ww)=\{r:\nu_{\xx r}=\wwxx\,\forall\xx\}$
 and $B(\ww)=\{r:\kappa_{\xx r}=\wwxx\,\forall\xx\}$ 
indicate the sets of reactions that have complex $\ww$ as reactant and product, respectively. 
Then, we can rewrite
Eq.~\eqref{eq:s2} as
\begin{align}
& \sum_{\ww\in\complexes}\left[\sum_{r\in A(\ww)}\!\prod_{\xx}\Big(\frac{\cccx{\xx}}{\cccssx{\xx}}\Big)^{\nu_{\xx r}}\!\ssfluxrr-\!\!\!\sum_{r\in B(\ww)}\!\prod_{\xx}\Big(\frac{\cccx{\xx}}{\cccssx{\xx}}\Big)^{\kappa_{\xx r}}\!\ssfluxrr\right] \nonumber \\
 & =\sum_{\ww\in\complexes}\left[\sum_{r\in A(\ww)}\!\prod_{\xx}\Big(\frac{\cccx{\xx}}{\cccssx{\xx}}\Big)^{\wwxx}\!\ssfluxrr-\!\!\!\sum_{r\in B(\ww)}\!\prod_{\xx}\Big(\frac{\cccx{\xx}}{\cccssx{\xx}}\Big)^{\wwxx}\!\ssfluxrr\right] \nonumber \\
 & =\sum_{\ww\in\complexes}\prod_{\xx}\Big(\frac{\cccx{\xx}}{\cccssx{\xx}}\Big)^{\wwxx}\left[\sum_{r\in A(\ww)}\ssfluxrr-\!\!\!\sum_{r\in B(\ww)}\ssfluxrr\right]. \label{eq:cgdd3cc} 
\end{align}
By the definition of complex balance, $\sum_{r\in A(\ww)}\ssfluxrr=\sum_{r\in B(\ww)}\ssfluxrr$
for each $\ww$ ~\citep{feinbergFoundationsChemicalReaction2019}.
Therefore, Eq.~\eqref{eq:cgdd3cc}, hence also Eq.~\eqref{eq:vv}, vanishes, which implies Eq.~\eqref{eq:df2}.

\begin{newstuffenv}
\section{Details of metabolic network analysis}
\label{app:metabolic}

Our analysis of metabolic networks is based on data published by Park et al.~\cite{park2016metabolite}, which used
isotopic labeling to measure fluxes and thermodynamics of central metabolism in \emph{E. coli}, yeast, and a mammalian cell line. 

Following Ref.~\cite{park2016metabolite}, we omit the first step of glycolysis which converts glucose to phosphorylated glucose (G6P), which is catalyzed by hexokinase in yeast and mammalian cells and (primarily) by the PTS system in \emph{E. coli}. This reaction is omitted because intracellular glucose concentrations are very difficult to measure reliably when cells are grown in glucose solution~\cite{park2016metabolite}. 

To define the one-way fluxes of metabolic reactions, we first extract net reaction fluxes ($\mathcal{J}_\rr$) for glycolysis and PPP from Table 1 in the Supplemental Material (SM) of Ref.~\cite{park2016metabolite}. We also extract thermodynamic forces ($-\Delta G$) from Table 4 in SM of Ref.~\cite{park2016metabolite}. Reported values of $-\Delta G$ are converted from $kJ/\text{mol}$ to dimensionless entropy units (multiples of $k_B T$) at $T=310$~K. One-way fluxes $\jjrr$ are calculated using the net fluxes, $\mathcal{J}_\rr = \jjrr-j_{\negedge}$, and the flux-force relationship, $-\Delta G = \ln (\jjrr/j_{\negedge})$, which gives
\begin{align}\jjrr= \frac{\mathcal{J}_\rr e^{-\Delta G}}{e^{-\Delta G}-1}\qquad \quad j_{\negedge}= \frac{\mathcal{J}_\rr}{e^{-\Delta G}-1}\,.
\end{align}
Note that metabolic reactions are enzymatic and have (approximately) Michaelis-Menten kinetics, so the flux-force relationship is valid under an appropriate definition of one-way fluxes~\cite{beard2007relationship}. 

The only additional complication involves the first two steps of the PPP pathway, the G6PDH and PGL reactions. Ref.~\cite{park2016metabolite} only reports net fluxes for the first step (G6PDH).
Since G6PDH and PGL are in series without branches, we assume that PGL has the same net flux as G6PDH. In addition, thermodynamic forces are not reported for these two reactions, likely because they involve the short-lived metabolite 6-phosphogluconolactone (6PGL) whose concentration is difficult to measure. To estimate the thermodynamic force across these two reactions, 
we use standard Gibbs free energy ($-\Delta G^\circ$) values reported in Ref.~\cite{li_database_2011}. 
The standard Gibbs free energies are reported for pH 7.3; we perform a correction to account for measured cytosolic pH values~\cite{park2016metabolite}: 7.7 for \emph{E. coli}, 7.2 for yeast, and 7.2 for mammalian cells. For the concentration of 6PGL, we estimate a value of 1 mM, the same order of magnitude as used in kinetic modeling studies~\cite{messiha2014enzyme}. The resulting $-\Delta G$ values for G6PDH and PGL reactions are: -21.5 and -14 in mammalian cells, -34.5, -7.1 in yeast cells, and -22.5, -3 in \emph{E. coli} cells.

\end{newstuffenv}

\clearpage

\fi
\end{document}